\documentclass[aps,prd,twocolumn,showpacs,floats,nofootinbib,longbibliography,10pt]{revtex4-2}

\usepackage{multirow}
\usepackage{graphicx}
\usepackage{amsmath,amsfonts}
\usepackage{dcolumn}
\usepackage{hyperref}

\mathchardef\minus = "002D
\newcommand{\swY}[4][]{{}_{{}_{#2}}\!Y^{#1}_{#3}(#4)}   
\newcommand{\swSH}[5][]{{}_{{}_{#2}}S^{#1}_{#3}(#4;#5)} 
\newcommand{\swS}[5][]{{}_{{}_{#2}}S^{#1}_{#3}(#4;#5)}  
\newcommand{\swSnorm}[5][]{{}_{{}_{#2}}\bar{S}^{#1}_{#3}(#4;#5)}  
\newcommand{\swSanom}[5][]{{}_{{}_{#2}}\hat{S}^{#1}_{#3}(#4;#5)}  
\newcommand{\scA}[4][]{{}_{{}_{#2}}A^{#1}_{#3}(#4)} 
\newcommand{\scAnorm}[4][]{{}_{{}_{#2}}\bar{A}^{#1}_{#3}(#4)} 
\newcommand{\scAanom}[4][]{{}_{{}_{#2}}\hat{A}^{#1}_{#3}(#4)} 
\newcommand{\sNlm}[3][]{{}_{{}_{#2}}N^{#1}_{#3}}
\newcommand{\YSH}[3][]{C^{#1}_{#2}(#3)}   

\newcolumntype{.}{D{.}{.}{-1}}
\newcolumntype{d}[1]{D{.}{.}{#1}}

\DeclareMathOperator{\sign}{sign}

\begin{document}

\title{Understanding solutions of the angular {T}eukolsky equation in the prolate asymptotic limit}

\author{Daniel J. Vickers}\email{dnlvickers5@gmail.com}
\affiliation{Department of Physics, Wake Forest University,
		 Winston-Salem, North Carolina 27109}
\author{Gregory B. Cook}\email{cookgb@wfu.edu}
\affiliation{Department of Physics, Wake Forest University,
		 Winston-Salem, North Carolina 27109}

\date{\today}

\begin{abstract}
Solutions to the Angular Teukolsky Equation have been used to solve various applied problems in physics and are extremely important to black-hole physics, particularly in computing quasinormal modes and in the extreme-mass-ratio inspiral problem.  The eigenfunctions of this equation, known as spin-weighted spheroidal functions, are essentially generalizations of both the spin-weighted spherical harmonics and the scalar spheroidal harmonics.  While the latter functions are quite well understood analytically, the spin-weighted spheroidal harmonics are only known analytically in the spherical and oblate asymptotic limits.  Attempts to understand them in the prolate asymptotic limit have met limited success.  Here, we make use of a high-accuracy numerical solution scheme to extensively explore the space of possible prolate solutions and extract analytic asymptotic expansions for the eigenvalues in the prolate asymptotic limit.  Somewhat surprisingly, we find two classes of asymptotic behavior.  The behavior of one class, referred to as ``normal'', is in agreement with the leading-order behavior derived analytically in prior work.  The second class of solutions was not previously predicted, but solutions in this class are responsible for unexplained behavior seen in previous numerical prolate solutions during the transition to asymptotic behavior.  The behavior of solutions in this ``anomalous'' class is more complicated than that of solutions in the normal class, with the anomalous class separating into different types based on the behavior of the eigenvalues at different asymptotic orders.  We explore the question of when anomalous solutions appear and find necessary, but not sufficient conditions for their existence.  It is our hope that this extensive numerical investigation of the prolate solutions will inspire and inform new analytic investigations into these important functions. 
\end{abstract}

\maketitle

\section{Introduction}
Spin-weighted spherical harmonics $\swY{s}{\ell m}{\theta,\phi}$ are used widely in physics, providing a complete set of orthonormal basis functions that can be used to represent tensor-valued functions on the surface of a sphere\cite{boyle-2016}.  Scalar spheroidal harmonics $\swSH{0}{\ell m}{\theta,\phi}{c}$ are similar to the common scalar spherical harmonics $Y_{\ell m}(\theta,\phi)$, except that they are obtained from separation of variables in terms of spheroidal coordinates instead of spherical coordinates.  The additional argument $c$ is the oblateness parameter, which determines the oblateness of the coordinate system.  While the spin-weighted spherical harmonics $\swY{s}{\ell m}{\theta,\phi}$ and their associated separation constants can be represented as known analytic functions, closed form solutions for the spheroidal harmonics and their separation constants are only known in special cases.  Generalizing both the spin-weighted spherical harmonics and the scalar spheroidal harmonics are the spin-weighted spheroidal harmonics(SWSHs) $\swSH{s}{\ell m}{\theta,\phi}{c}$.

Introduced by Teukolsky\cite{Teukolsky},  spin-weighted spheroidal harmonics can be expressed in terms of the spin-weighted spheroidal functions(SWSFs) $\swS{s}{\ell m}{x}{c}$ by separating out the azimuthal dependence, 
\begin{align}
	\swSH{s}{\ell m}{\theta,\phi}{c}\equiv\frac1{2\pi}\swS{s}{\ell m}{\cos\theta}{c}e^{im\phi}.
\end{align}
The differential equation governing the SWSFs is
\begin{align}\label{eqn:Angular Teukolsky Equation}
\partial_x \Big[ (1-x^2)\partial_x [\swS{s}{\ell{m}}{x}{c}]\Big] 
& \nonumber \\ 
    + \bigg[(cx)^2 - 2 csx + s& + \scA{s}{\ell m}{c} 
 \\ 
      & - \frac{(m+sx)^2}{1-x^2}\bigg]\swS{s}{\ell{m}}{x}{c} = 0,
\nonumber
\end{align}
where $m$, $s$, and $\ell$ are integers or half-integers, $\ell \geq \max(|m|,|s|)$, and $x=\cos{\theta}$ with $\theta$ being the usual polar angle. Equation (\ref{eqn:Angular Teukolsky Equation}) is often referred to as the Angular Teukolsky Equation.  As a Sturm-Liouville problem, the eigenvalue $\scA{s}{\ell m}{c}$, also referred to as the  angular separation constant, is fixed by the requirement that $\swS{s}{\ell m}{x}{c}$ be finite at $|x|=1$.

Teukolsky first derived Eq.~(\ref{eqn:Angular Teukolsky Equation}) for use in relativistic physics to describe perturbations near rotating black holes. One of the first applications of Eq.~(\ref{eqn:Angular Teukolsky Equation}) was by Bardeen and Press to solve for synchrotron radiation of a point mass in the Kerr geometry\cite{Bardeen1973}. Equation~(\ref{eqn:Angular Teukolsky Equation}) has also been used in the case of point-mass perturbations orbiting a black hole to determine radiated angular momentum and energy from the system\cite{Hughes2000}.

Equation~(\ref{eqn:Angular Teukolsky Equation}) has also been applied in fields outside of relativity since the SWSHs form a natural basis in spheroidal coordinate systems. Figueiredo used $s=1$ SWSHs to solve the two-center electron problem\cite{Bartolomeu2002}. Larsson, Levitina, and Br{\"a}ndas use the scalar spheroidal harmonics to solve various problems involved in signal processing\cite{Larsson2001}.

An especially important application of SWSHs is in determining the  quasinormal modes (QNMs) and total transmission modes (TTMs) of the Kerr geometry. 
In these applications, Eq.~(\ref{eqn:Angular Teukolsky Equation}) is solved together with the radial Teukolsky equation.  See Ref. \cite{CookZalutski2014} and references within for more details. QNMs are the natural ringing modes for linear perturbations of various fields.  In particular, the spin-weight $s=\pm2$ QNMs are used to describe the gravitation-wave ring-down of a perturbed black hole and are critical to the interpretation of gravitational-wave observations. QNMs represent solutions where boundary conditions are fixed so that waves are forbidden from entering the system.  That is, waves are not permitted to move from the black-hole horizon toward spatial infinity, or to enter from spatial infinity.  On the other hand, TTMs reverse one of these two boundary conditions. Left TTMs (TTMLs) reverse the condition at the black-hole horizon.  For TTMLs, waves are forbidden from traveling into the black-hole horizon. Right TTMS (TTMRs) reverse the condition at spatial infinity and only allow waves to travel in from spatial infinity.  Both types of TTMs represent the special modes that do not reflect off of the gravitational potential, yielding modes that travel entirely away from (TTMLs) or toward (TTMRs) the black-hole horizon.

Recently, Cook, Anacharicio, and Vickers\cite{CookVickersAnacharicio} used numerical methods to explore gravitational TTMs of the Kerr geometry, identifying a new branch of the gravitational TTMs which required solutions to Eq.~(\ref{eqn:Angular Teukolsky Equation}) in the asymptotic limit $c\to-i\infty$.  Unfortunately, relatively little is known about the asymptotic limit of the SWSHs as $c\to$ complex-$\infty$.  However, their numerical solutions were sufficiently accurate that they were able to extract the first 4 terms in the asymptotic expansion for $\scA{s}{\ell m}{c}$ in the limit $c\to-i\infty$ for $s=\pm2$, and their results were in agreement with the known asymptotic limit for the angular separation constant for the scalar spheroidal harmonics\cite{Flammer,berti2005eigenvalues}.

As mentioned above, even the scalar spheroidal harmonics do not have known analytic solutions except in special cases.  Flammer\cite{Flammer} explored these functions in the limit of small $c$, and in the asymptotic limits of purely real and purely imaginary $c$.  We refer to coordinates and solutions where $c$ is purely real as oblate. Likewise, when $c$ is purely imaginary, the coordinates and solutions are called prolate.  In general, $c$ can be complex.  In this case, there is no direct connection to a spheroidal coordinate system, but we still refer to $c$ as the oblateness parameter.  Beyond the scalar case, less is known about the asymptotic limits of the SWSHs.  In the oblate case, an asymptotic expansion for $\scA{s}{\ell m}{c}$ has been derived analytically by Ottewill and Casals\cite{CasalasOblate}.  However, for the prolate case, only the leading order term in the asymptotic expansion for $\scA{s}{\ell m}{c}$ has been determined\cite{berti2005eigenvalues}.  And, as we will show in this work, even this leading order term in the prolate expansion is not valid for all asymptotic solutions.

The success in determining an asymptotic expansion for prolate $\scA{s}{\ell m}{c}$ in the special case of $s=\pm2$\cite{CookVickersAnacharicio} was made possible by the high accuracy and precision of numerical solutions of Eq.~(\ref{eqn:Angular Teukolsky Equation}) using the spectral method described in detail in Ref.~\cite{CookZalutski2014}.  In this paper, we employ this approach to explore the prolate asymptotic limit of the SWSHs for a broad range of spin weights.  We have generated extensive numerical solutions of the SWSHs for a large range of integer values for $\ell$, $m$, and $s$; and we have constructed sequences of solutions where $\ell$, $m$, and $s$ are held fixed while $c$ covers a range of purely imaginary values from the small prolate limit $c\to-i0$ to values large enough to accurately explore the asymptotic prolate limit $c\to-i\infty$.

Using this data, we have been able to construct the first few terms in the asymptotic expansion for $\scA{s}{\ell m}{c}$ in the prolate case.  However, we have found that the asymptotic behavior has a richer structure than expected. Most of the sequences we find have a leading asymptotic behavior that is in agreement with prior analytic work\cite{berti2005eigenvalues}.  We refer to such sequences with fixed values of $\ell$, $m$, and $s$ as normal sequences, and for these sequences we have determined the asymptotic expansion through the first 5 terms.  However, for certain values of $\ell$, $m$, and $s$, the leading behavior of the asymptotic expansion is not normal.  We call such sequences anomalous, and interestingly, the asymptotic behavior mirrors that of the oblate case\cite{CasalasOblate,berti2005eigenvalues} for the first few terms.  However, complicating things, the number of terms that mirror the oblate expansion varies.  We suspect that this unusual behavior in the prolate case may be related to the difficulties in analytically constructing an asymptotic expansion\cite{berti2005eigenvalues}.

The purpose of this paper is to give a detailed description of the behavior of the the SWSHs in the prolate asymptotic limit based on our numerical solutions.  The asymptotic expansions we extract from the data will be useful in their own right.  But it is also our hope that a more complete understanding of the behavior of the SWSHs in this limit will provide some insight that will aid in deriving the asymptotic limit fully analytically. We will begin in Sec.~\ref{sec:Sols of Angular Teukolsky Eqn} with a summary of of what is known analytically about Eq.~(\ref{eqn:Angular Teukolsky Equation}). Then in Sec.~\ref{sec:Num Solutions}, we will describe the general behavior of, and fits to, our numerically generated eigensolutions.  We will also explore the conditions under which anomalous solutions exist.  Finally, in Sec.~\ref{sec:conclusion}, we summarize our findings.


\section{Analytic Solutions to the Angular Teukolsky Equation}
\label{sec:Sols of Angular Teukolsky Eqn}
An excellent overview of what is known analytically about solutions to the angular Teukolsky equation can be found in Ref.~\cite{berti2005eigenvalues}.  For brevity, we review only the most important of these results.

\subsection{Symmetries of the Angular Teukolsky Equation}

The basic symmetries obeyed by the SWSFs and the separation constants follow from Eq.~(\ref{eqn:Angular Teukolsky Equation}) through three transformations: $\{s\to-s,x\to-x\}$, $\{m\to-m,x\to-x,c\to-c\}$, and complex conjugation. From these transformations, it follows that the SWSFs and the separation constants satisfy the following conditions: 
\begin{subequations}\label{eq:swSF_all_ident}
\begin{align}
\label{eq:swSF_sx_ident}
\swS{-s}{\ell{m}}{x}{c} &= (-1)^{\ell+m}\swS{s}{\ell{m}}{-x}{c}, \\
\label{eq:swSF_mxc_ident}
\swS{s}{\ell(-m)}{x}{c} &= (-1)^{\ell+s}\swS{s}{\ell{m}}{-x}{-c}, \\
\label{eq:swSF_cc_ident}
\swS[*]{s}{\ell{m}}{x}{c} &= \swS{s}{\ell{m}}{x}{c^*}, \\
\intertext{and}
\label{eq:swSF_sA_ident}
\scA{-s}{\ell{m}}{c} &= \scA{s}{\ell{m}}{c} + 2s, \\
\label{eq:swSF_mcA_ident}
\scA{s}{\ell(-m)}{c} &= \scA{s}{\ell{m}}{-c}, \\
\label{eq:swSF_cA_ident}
\scA[*]{s}{\ell{m}}{c} &= \scA{s}{\ell{m}}{c^*}.
\end{align}
\end{subequations}

\subsection{The Spherical Limit}
As the oblateness parameter $c\rightarrow0$, the SWSHs reduce to spin-weighted spherical harmonics. Thus, the angular separation constant also tends towards the eigenvalue of the spin-weighted spherical harmonics, such that
\begin{align}\label{eqn:spherical sAlm}
	\scA{s}{\ell m}{c=0} = \ell(\ell+1)-s(s+1).
\end{align}
For small but non-vanishing $c$, the separation constant can be expressed as a Taylor series\cite{PressTeukolksy-1973}:
\begin{align}
	\scA{s}{\ell{m}}{c} = \ell(\ell+1)-s(s+1) - \frac{2ms^2}{\ell(\ell+1)}c +
	\sum_{p=2}^\infty{f_pc^p},
\end{align}
where the expansion coefficients $f_p$ have been worked out and presented through order $c^6$ in Ref~\cite{berti2005eigenvalues}.

\subsection{The Oblate Asymptotic Limit}
In the limit of large $c$, the behavior of solutions to Eq.~(\ref{eqn:Angular Teukolsky Equation}) have been most thoroughly explored for the oblate case where $c$ is purely real.
 Breuer, Ryan, and Waller\cite{AnalyticOblate} performed the first analytic derivation of the asymptotic oblate case, presenting $\scA{s}{\ell m}{c}$ as a power-series expansion in $c$. Their solution is given as
\begin{widetext}
\begin{subequations}\label{eqn:oblate solution all}
\begin{align}\label{eqn:oblate solution}
	\scA{s}{\ell m}{c} =& -c^{2} + 2_{s}q_{\ell m}c 
	- \frac12\left[{_{s}q_{\ell m}}^2 - m^2+2s+1\right] 
	+ \frac1{c}A_1 + \frac1{c^2}A_2 + \frac1{c^3}A_3 
	+ \frac1{c^4}A_4 + \mathcal{O}\left(c^{-5}\right), \\
	\intertext{where}
	\label{eqn:oblate solution A1}
	A_1=&-\frac18\Bigl[
	{_{s}q_{\ell m}}^3 - m^2{_{s}q_{\ell m}} + {_{s}q_{\ell m}}
	- 2s^2({_{s}q_{\ell m}} + m)
	\Bigr], \\
	\label{eqn:oblate solution A2}
	A_2=&-\frac1{64}\Big[
	5{_{s}q_{\ell m}}^4 - {_{s}q_{\ell m}}^2\left(6m^2-10\right) + m^4 
	- 2m^2 -4s^2\left(3{_{s}q_{\ell m}}^2+4{_{s}q_{\ell m}}m+m^2+1\right)+1\Big], \\
	\label{eqn:oblate solution A3}
	A_3=&-\frac1{512}\Bigl[
	33 {_{s}q_{\ell m}}^5
	- {_{s}q_{\ell m}}^3 \left(46 m^2+92 s^2-114\right)
	- 132 {_{s}q_{\ell m}}^2m s^2 \nonumber\\
 &\hspace{43pt}
	+ {_{s}q_{\ell m}}\left(13 m^4-36 m^2 s^2-50 m^2+8 s^4-100 s^2+37\right)
	+ 4 m^3 s^2
	+ 8 m s^4
	- 52 m s^2
   \Bigr], \\
	\intertext{and}
	\label{eqn:oblate solution A4}
	A_4=&-\frac1{1024}\Bigl[
	63 {_{s}q_{\ell m}}^6
	-20 {_{s}q_{\ell m}}^4 (5 m^2+10 s^2-17)
	-292 {_{s}q_{\ell m}}^3 m s^2 \nonumber\\
 &\hspace{43pt}
	+{_{s}q_{\ell m}}^2\bigl(39 m^4+48s^4-60m^2s^2-230(m^2+2s^2)+239\bigr)
	+{_{s}q_{\ell m}}m(72 s^4+36 m^2 s^2-372s^2) \nonumber\\
 &\hspace{43pt}
	-2m^2(m^4-12 s^4-2 m^2 s^2)
	+18 m^4
	+16 s^4
	-40 m^2 s^2
	-30 m^2
	-60 s^2
	+14{}
   \Bigr].
\end{align}
\end{subequations}
\end{widetext}
Note that the term $-2s^2({_{s}q_{\ell m}} + m)$ in Eq.~(\ref{eqn:oblate solution A1}) differs from the corresponding terms in Eq.~(4.12) of Ref.~\cite{AnalyticOblate}, Eq.~(3.7) of Ref.~\cite{CasalasOblate}, and  Eq.~(2.18) of Ref.~\cite{berti2005eigenvalues} where the factor of $2$ is missing.  Similarly, the term $-4s^2\left(3{_{s}q_{\ell m}}^2+4{_{s}q_{\ell m}}m+m^2+1\right)$ in Eq.~(\ref{eqn:oblate solution A2}) differs from the corresponding terms in Eq.~(4.12) of Ref.~\cite{AnalyticOblate}, Eq.~(3.6) of Ref.~\cite{CasalasOblate}, and  Eq.~(2.17) of Ref.~\cite{berti2005eigenvalues} where the term inside the parentheses is substantially different.  These seem to be errors that have propagated unnoticed from the original work\cite{AnalyticOblate}.  We have also expressed $A_3$ in a more compact form than found in Refs.~\cite{AnalyticOblate,CasalasOblate,berti2005eigenvalues}.  The versions of $A_3$ in those references are expressed in terms of $A_1$ but are correct when the correct version of $A_1$ is used.  

We also include a term at order $c^{-4}$ in Eq.~(\ref{eqn:oblate solution}).  The coefficient $A_4$ has not previously been determined.  However, we have been able to obtain this term via fits to numerical data.  The details of the derivation of this additional term are outlined in App.~\ref{sec:Oblate appendix}.

While Breuer, Ryan, and Waller\cite{AnalyticOblate} were able to obtain the oblate solution given in Eqs.~(\ref{eqn:oblate solution all}), they were unable to determine the quantity ${_{s}q_{\ell m}}$ as a function of $\ell$, $m$, and $s$.  This was achieved by Casals and Ottewill\cite{CasalasOblate} who used a WKB approximation for the spheroidal function near the boundaries $x=\pm1$ and matched it to a solution valid in the interior.  The full process of determining ${_{s}q_{\ell m}}$ is quite complicated and we will not describe it in any detail here.  The result is
\begin{subequations}\label{eqn:sqlm all}
\begin{align}\label{eqn:sqlm}
        _{s}q_{\ell m}=&
        \begin{cases}
            L+\frac{|m+s|+|m-s|}2+1-z_0  &\ell\geq \max\left({_s\ell_m},{_{-s}\ell_m}\right) \\
            2L+\bigl|m - |s|\bigr| - |s| + 1 &\text{otherwise}
        \end{cases},\\
\intertext{where}
	\label{eqn:sqlm slm}
	{_s\ell_m}=&|m+s|+s, \\
	\label{eqn:sqlm z0}
	z_0=&
        \begin{cases}
            0 & \text{if $\ell -{_s\ell_m}$ even} \\
            1 & \text{if $\ell -{_s\ell_m}$ odd}
        \end{cases},
\end{align}
\end{subequations}
and we have used the convenient notation that
\begin{align}\label{eqn:L def}
	L=\ell - \max(|m|,|s|).
\end{align}
We note that our forms for Eqs.~(\ref{eqn:sqlm}) and (\ref{eqn:sqlm slm}) are slightly different than those given in Refs.~\cite{CasalasOblate,berti2005eigenvalues}.  We choose these definitions because they are equivalent, simpler, and less prone to misinterpretation.\footnote{The simplification to Eq.~(\ref{eqn:sqlm}) occurs because ${_{-s}\ell_m}<\ell<{_s\ell_m}$ if $s>0$ and ${_s\ell_m}<\ell<{_{-s}\ell_m}$ if $s<0$ whenever $\ell<\max({_s\ell_m},{_{-s}\ell_m})$.}

Finally, we note that Eq.~(\ref{eqn:oblate solution all}) clearly satisfies the fundamental symmetry expressed by Eqs.~(\ref{eq:swSF_sA_ident}) and (\ref{eq:swSF_cA_ident}), but it does not satisfy Eq.~(\ref{eq:swSF_mcA_ident}).  This manifests in the complicated definition of $_{s}q_{\ell m}$ in Eq.~(\ref{eqn:sqlm all}), and in particular because $_{s}q_{\ell m} \ne -_{s}q_{\ell (-m)}$.  Because of this, the oblate asymptotic expansion of Eq.~(\ref{eqn:oblate solution all}) is only valid for positive real values of $c$.  To obtain the oblate asymptotic expansion for negative real values of $c$, one must explicitly apply Eq.~(\ref{eq:swSF_mcA_ident}).

\subsection{The Prolate Asymptotic Limit}\label{sec:prolate behavior}

In the prolate case, where $c$ is purely imaginary, much less is known analytically about the SWSHs.  Only in the scalar case where $s=0$ are solutions well understood\cite{Flammer,SpheroidalFunctions-1954}.  Flammer derived an asymptotic expansion for the angular separation constant in the $s=0$ case.  The result, making heavy use of $L$ as defined in Eq.~(\ref{eqn:L def}), is
\begin{widetext}
\begin{subequations}\label{eqn:flammer solution all}
\begin{align}\label{eqn:flammer solution}
	\scA{0}{\ell m}{\pm i|c|} &= (2L+1)|c|-\frac14\left(2L(L+1)+3-4m^2\right) 
	+ \frac{B_1(L,m)}{|c|} + \frac{B_2(L,m)}{|c|^2} + \frac{B_3(L,m)}{|c|^3} +\mathcal{O}\left(|c|^{-4}\right), \\
\intertext{where}\label{eqn:flammer solution B1}
	B_1(L,m) &= -\frac1{16}(2L+1)\left(L(L+1)+3-8m^2\right), \\
	\label{eqn:flammer solution B2}
	B_2(L,m) &= -\frac1{64}\Bigl[5\left(L(L+1)(L(L+1)+7)+3\right)
	 -48m^2\left(2L(L+1)+1\right)\Bigr], \\
\intertext{and}
\label{eqn:flammer solution B3}
	B_3(L,m) &= -\frac1{256}\biggl[{\frac14}(2L+1)\left(L(L+1)\left(33L(L+1)+415\right)+453\right) \nonumber \\
	& \hspace{2in} - 8m^2(2L+1)\left(37L(L+1)+51\right)+32m^4(2L+1)\biggr].
\end{align}
\end{subequations}
\end{widetext}
We note that there is an error in Ref.~\cite{Flammer} that is repeated in Ref.~\cite{berti2005eigenvalues} in the coefficient at order $c^{-2}$.  A term of $\frac{40}{64}L^2$ has been omitted, but this term is correctly included in Abramowitz and Stegun\cite{AbramowitzStegun}, and is correctly included in Eq.~(\ref{eqn:flammer solution B2}) which is in excellent agreement with all of our $s=0$ prolate data sets.  Because the separation constant $\scA{0}{\ell m}{\pm i|c|}$ is purely real and depends only on even powers of $m$, it is convenient to express Eq.~(\ref{eqn:flammer solution}) in terms of $|c|$.  

A path to generalize this result to $s\ne0$ is well summarized by Berti, Cardoso, and Casals\cite{berti2005eigenvalues} and depends critically on determining the number of zeros of the real part of the SWSFs.  Because we will examine the behavior of the SWSFs in some detail, we include some of the relevant details from Ref.~\cite{berti2005eigenvalues}.

Their approach begins by defining a new angular function
\begin{align}\label{eqn:sylm substitution}
	\swS{s}{\ell m}{x}{c} = (1-x)^{k_{+}}(1+x)^{k_{-}}\,{_{s}y_{\ell m}(x)}.
\end{align}
With a change of variables $u=\sqrt{2|c|}x$ and substituting Eq.~(\ref{eqn:sylm substitution}) into Eq.~(\ref{eqn:Angular Teukolsky Equation}), one can determine
\begin{align}\label{eqn:sylm differential}
	0 = \biggl\{ & (2|c|-u^{2})\frac{d^{2}}{du^{2}} \nonumber\\
	& - 2\Bigl[\sqrt{2|c|}(k_{+}-k_{-}) 
	+(k_{+}+k_{-}+1)u\Bigr]\frac{d}{du} \nonumber\\
	&+ \scA{s}{\ell m}{c} + s(s+1) - (k_{+}+k_{-})(k_{+}+k_{-}+1) \nonumber \\
	&- \frac{|c|u^2}2 - i\sqrt{2|c|}su \biggr\}{_{s}y_{\ell m}}.
\end{align}

In the prolate asymptotic limit, such that $ic\rightarrow\pm\infty$, the real components of Eq.~(\ref{eqn:sylm differential}) reduce to the differential equation for the parabolic cylinder functions, $D_{L}(u)$. Therefore, Ref.~\cite{berti2005eigenvalues} determined that the real component of the inner solution $\swS[inner]{s}{\ell m}{x}{c}$ can be approximated by using ${\rm Re}(_{s}y_{\ell m}^{inner}(x))=D_L(\sqrt{2|c|}x)$, where $D_{L}$ has $L$ zero-crossings.  The domain of this inner solution contains the region $|x|<\sqrt{(2L+1)/|c|}$ within which all of the zeros of $D_L(\sqrt{2|c|}x)$ occur. Matching the number of zero crossings of ${\rm Re}(\swS{s}{\ell m}{x}{c})$ in the spherical limit to $D_L(\sqrt{2|c|}x)$ in the prolate asymptotic limit, Ref.~\cite{berti2005eigenvalues} determined the leading-order approximation for the separation constant to be
\begin{align}\label{eqn:sAlm leading order}
	\scA{s}{\ell m}{c} = (2L+1)|c| + \mathcal{O}(|c|^{0}).
\end{align}

Using a WKB-type approximation similar to that use in Ref.~\cite{CasalasOblate}, they next determined the behavior of $\swS{s}{\ell m}{x}{c}$ in the outer region near $x=\pm1$. The result is a solution which is formally divergent at $|x|=1$, but is valid arbitrarily close to the endpoints:
\begin{align}\label{eqn:WKB outer solutions}
	\swS[outer,\pm1]{s}{\ell m}{x}{c} = & (-i)^{s}2^{L+1/2}(\pm\sqrt{2|c|})^{L}e^{-3|c|/2} \nonumber \\ 
	\times(1-x^{2}&)^{-1/4}x^{L}(1+\sqrt{1-x^{2}})^{-L-1/2} \nonumber \\ 
	&\times(x-i\sqrt{1-x^{2}})^{-s}e^{|c|\sqrt{1-x^{2}}}.
\end{align}
Equation~(\ref{eqn:WKB outer solutions}) shows that the number of real zero crossings of $\swS[outer]{s}{\ell m}{x}{c}$ depends critically on $s$. Together, the inner and outer solutions give a full account of the zeros of the real component of $\swS{s}{\ell m}{x}{c}$.

As mentioned in the Introduction, Ref.~\cite{CookVickersAnacharicio} was able to extend the asymptotic prolate expansion for the separation constant to order $c^{-2}$ for the special case of $s=\pm2$.  This was achieved by fitting to numerically determined values of the separation constant that had been computed for purely imaginary values of $c$ extending well into the asymptotic regime.  These were computed while exploring a previously unknown branch of the gravitational total-transmission modes of the Kerr geometry.  Unfortunately, because they only explored the cases of $s=\pm2$ and because of the error in Eq.~(\ref{eqn:flammer solution all}) for $s=0$ that was present in the literature, Eq.~(26) of Ref.~\cite{CookVickersAnacharicio} incorrectly associated all of the $L^2$ dependence in the $|c|^{-2}$ term to the term that scales as $s^2L^2$.  Since this equation includes an error, we do not include it here.  However, the general approach of fitting high-quality numerical solution to extract the asymptotic behavior of the separation constant is sound, and we will explore this extensively below.


\section{Numerical Solutions to the Angular Teukolsky Equation}
\label{sec:Num Solutions}
\subsection{Spectral Eigenvalue Method}

There are many approaches for numerically solving the angular Teukolsky equation\cite{fackerell-1976,SasakiNakmura1982,leaver-1986,CookZalutski2014}. In this paper, we utilized the spectral decomposition method developed in Ref.~\cite{CookZalutski2014}. In this spectral method, Eq.~(\ref{eqn:Angular Teukolsky Equation}) is converted into a matrix eigenvalue problem.  This approach has several advantages over methods based on solving a continued fraction.  By finding all the eigenvalues of the spectral matrix, this method yields many solutions simultaneously whereas continued fraction methods find one eigenvalue at a time.  Also, the exponential convergence of this spectral method means that we can find any eigenvalues with exceptional accuracy.  Since all of the results in this paper are derived directly from this method, we include a detailed description.

The spectral eigenvalue method is based on expanding the SWSFs using the spin-weighted spherical functions as a basis
\begin{align}\label{eqn:linear combination}
	\swS{s}{\ell m}{x}{c}=\sum_{\acute\ell}\YSH{\acute\ell\ell m}{c}\swS{s}{\acute\ell m}{x}{0}.
\end{align}
The coefficients $\YSH{\acute\ell\ell m}{c}$ will become the components of the eigenvector of the eigensolution labeled by $\ell$ and $m$. For sufficiently large $\acute\ell$, the expansion coefficients $\YSH{\acute\ell\ell m}{c}$ enter a convergent regime and their magnitudes decrease exponentially with increasing $\acute\ell$\cite{CookZalutski2014}. This guarantees that $\swS{s}{\ell m}{x}{c}$ can be accurately approximated using a sufficient number of terms in the sum.

Making use of Eq.~(\ref{eqn:linear combination}) in Eq.~(\ref{eqn:Angular Teukolsky Equation}), Ref.~\cite{CookZalutski2014} eliminates the $x$ dependence by use of the recurrence relation\cite{BLANCO1997}
\begin{align}\label{eqn:eliminate x}
	x \swS{s}{\ell{m}}{x}{0} = 
       & \mathcal{F}_{s\ell{m}} \swS{s}{(\ell+1){m}}{x}{0} \\
& \mbox{}
       + \mathcal{G}_{s\ell{m}} \swS{s}{(\ell-1){m}}{x}{0}
       + \mathcal{H}_{s\ell{m}} \swS{s}{\ell{m}}{x}{0}, \nonumber
\end{align}
where
\begin{subequations}\label{eqn:recurssion terms FGH}
\begin{align}\label{eqn:recurssion term F}
		& \mathcal{F}_{s\ell m} = \sqrt{\frac{\left((\ell+1)^{2}-m^{2}\right)}{(2\ell + 3)(2\ell + 1)}\frac{\left((\ell +1)^{2}-s^{2}\right)}{(\ell +1)^{2}}},\\
		\label{eqn:recurssion term G}
		& \mathcal{G}_{s\ell m} = \sqrt{\frac{\left(\ell^{2}-m^{2}\right)}{\left(4\ell^{2}-1\right)}\frac{\left(\ell^{2}-s^{2}\right)}{\ell^{2}}} \text{ if } \ell\neq 0,\, 0 \text{ otherwise},\\
		&\text{and} \nonumber \\
		\label{eqn:recurssion term H}
		& \mathcal{H}_{s\ell m} = -\frac{ms}{\ell(\ell+1)} \text{ if } \ell\neq 0,\, 0 \text{ otherwise}.
\end{align}
\end{subequations}
This yields a five-term recurrence relation on $\YSH{\acute\ell\ell m}{c}$:
\begin{align}\label{eqn:cook recursion}
		0 = & -c^{2}\mathcal{A}_{s(\acute\ell-2)m}\YSH{(\acute\ell-2)\ell m}{c} - \big[c^{2}\mathcal{D}_{s(\acute\ell-1)m} \nonumber \\ 
		&-2cs\mathcal{F}_{s(\acute\ell-1)m}\big]\YSH{(\acute\ell-1)\ell m}{c} + \big[\acute\ell(\acute\ell+1) - s(s+1) \nonumber \\
		& - c^{2}\mathcal{B}_{s\acute\ell m} + 2cs\mathcal{H}_{s\acute\ell m} - \scA{s}{\ell m}{c}\big]\YSH{\acute\ell\ell m}{c} \nonumber \\
		&- \left[c^{2}\mathcal{E}_{s(\acute\ell+1)m} - 2cs\mathcal{G}_{s(\acute\ell+1)m}\right]\YSH{(\acute\ell+1)\ell m}{c} \nonumber \\
		&- c^{2}\mathcal{C}_{s(\acute\ell+2)m}\YSH{(\acute\ell+2)\ell m}{c}
\end{align} 
where
\begin{subequations}\label{eqn:recurssion terms ABCDE}
\begin{align}\label{eqn:recurssion term A}
		& \mathcal{A}_{s\ell m} = \mathcal{F}_{s\ell m}\mathcal{F}_{s(\ell+1) m},\\
		\label{eqn:recurssion term B}
		& \mathcal{B}_{s\ell m} = \mathcal{F}_{s\ell m}\mathcal{G}_{s(\ell+1) m} + \mathcal{F}_{s(\ell-1) m}\mathcal{G}_{s\ell m} + \mathcal{H}_{s\ell m}^{2},\\
		\label{eqn:recurssion term C}
		& \mathcal{C}_{s\ell m} = \mathcal{G}_{s\ell m}\mathcal{G}_{s(\ell-1)m},\\
		\label{eqn:recurssion term D}
		& \mathcal{D}_{s\ell m} = \mathcal{F}_{s\ell m}(\mathcal{H}_{s\ell m}+\mathcal{H}_{s(\ell+1) m})\text{,}\\
		&\text{and} \nonumber \\
		\label{eqn:recurssion term E}
		& \mathcal{E}_{s\ell m} = \mathcal{G}_{s\ell m}(\mathcal{H}_{s(\ell-1) m} + \mathcal{H}_{s\ell m}).
\end{align}
\end{subequations}

Equation~(\ref{eqn:cook recursion}) represents an infinite-dimensional pentadiagonal-matrix eigenvalue problem. Each matrix is constructed for fixed values of $m$, $s$, and $c$. Their exist countably infinite eigensolutions indexed by $\ell$ for each matrix. To numerically approximate the eigensolutions for each combination of $m$, $s$, and $c$, the matrix is truncated at a finite size of $N\times N$ and its eigensolutions are determined. The $N$ eigenvalues can be indexed by $\ell\in\{\ell_{min},\ldots,\ell_{N-1}\}$ where $\ell_{min}=\max(|m|,|s|)$ and $N=\ell_{N-1}-\ell_{min}+1$.  Alternatively, using the notation of Eq.~(\ref{eqn:L def}), the eigenvalues can be indexed by $L\in\{0,\ldots,N-1\}$.  Due to the exponentially decreasing magnitudes of $\YSH{\acute\ell\ell m}{c}$ in the convergent regime, the eigensolutions to each matrix are guaranteed to be accurate for $\ell_{max}\ll\ell_{N-1}$ (or $L_{max}\ll N$), where $\ell_{max}$ (or $L_{max}$) denotes the last eigenvalue that is accurately computed.  For each value of $c$ during the creation of a solution sequence, the matrix size $N$ is confirmed to be large enough that  $|\YSH{(\ell_{max}-1)\ell m}{c}|<\epsilon_s$ and $|\YSH{\ell_{max}\ell m}{c}|<\epsilon_s$ for $\ell\in\{\ell_{min},\ldots,\ell_{max}\}$, where $\epsilon_s$ is the solution accuracy criteria.

For various combination of $m$ and $s$, we attempted to generated solution sequences for values of $c=-i10^{\delta}$ where $-5\leq\delta\leq5$ in steps of $\Delta\delta=10^{-3}$.  It was frequently too expensive to reach the upper limit of $\delta=5$, but each solution was extended to large enough $\delta$ to guarantee the sequence was in the asymptotic regime.  Determining that any sequence is in the asymptotic regime is achieved by direct inspection.  As discussed in Sec.~\ref{sec:conclusion}, we find that a normal sequence (see Sec.~\ref{sec:Classes of Prolate Eigensolutions}) transitions to the asymptotic regime at the point where $|{\rm Re}(\scA{s}{\ell m}{c})|<|c|^2$ which typically occurred for $\delta<2$ for the sequences we considered.  Continuous solution sequences were constructed from the set of returned eigenvalues obtained at each value of $c$ by assuming the eigenvalue along each sequence was smooth through the first derivative.  It is important to keep in mind that the label $\ell$ (or $L$) designating a given eigensolution is not fixed in any way by Eq.~(\ref{eqn:Angular Teukolsky Equation}).  For simplicity, we assign labels to the eigensolutions based on the limit $c\to0$ via Eq.~(\ref{eqn:spherical sAlm}). The resulting sequences are fully labeled by $m$, $s$, and $\ell$ (or $m$, $s$, and $L$) and are functions of $c$.  However, as we follow a sequence from the spherical limit to the asymptotic limit, the label $\ell$ (or $L$) may loose any connection to specific properties of the eigenfunction or eigenvalue.

\subsection{Numerical Data Sets}
\subsubsection{Prolate data sets}\label{sec:prolate data sets}
Our original data sets covered an extensive range of possible values of $m$ and $s$.  Data sets were constructed for all possible combinations of $0\le m\le20$ and $0\le s\le20$, and were constructed to keep at least the first 15 values of $\ell$ ($L<15$) with a solution accuracy of $\epsilon_s=10^{-15}$ and using 24 digits of precision.

In order to confirm that numerical errors were not affecting our solutions and to explore negative values of $m$, we created a second set of solution sequences keeping the same minimum 15 eigenvalues, but with a solution accuracy of $\epsilon_s=10^{-24}$ and using 32 digits of precision.  This set of solutions covered all possible combinations of $-10\le m\le10$ and $0\le s\le10$.

Finally, to explore some interesting behavior we found in the solutions that we will discuss below, we constructed a select set of solution sequences where we kept at minimum the first 150 eigenvalues.  These sequences were also created with a solution accuracy of $\epsilon_s=10^{-24}$ and using 32 digits of precision.  Because of the high cost of constructing these sequences, we only constructed sequences for the following $\{m,s\}$ pairs: $\{3,3\}$, $\{4,4\to10\}$, $\{6,4\to9\}$, $\{7,8\to10\}$, $\{8,7\}$, $\{9,8\to10\}$, and $\{10,10\}$.

\subsubsection{Oblate data sets}
\label{sec:Oblate data sets}
While the main focus of our exploration was the behavior of prolate solutions, it became clear that we needed to confirm and extend the results of Ref~\cite{CasalasOblate} for the asymptotic behavior in the oblate case.  In addition to modifying our sequences so that $c=10^{\delta}$ where $-5\leq\delta\leq5$ in steps of $\Delta\delta=10^{-3}$, we also modified the recurrence relation of Eq.~(\ref{eqn:cook recursion}) to move the leading order $-c^2$ asymptotic behavior [see Eq.~(\ref{eqn:oblate solution})] from the eigenvalue to the matrix coefficients.  This is possible because there is no $\ell$ dependence in this term.  We created solution sequences keeping at least the first 15 eigenvalues with a solution accuracy of $\epsilon_s=10^{-20}$ and using 24 digits of precision.  This set of solutions covered all possible combinations of $-5\le m\le10$ and $-5\le s\le10$.

\subsection{Classes of Prolate Eigensolutions}
\label{sec:Classes of Prolate Eigensolutions}
\begin{figure}
    \centering
 	\includegraphics[width=\linewidth,clip]{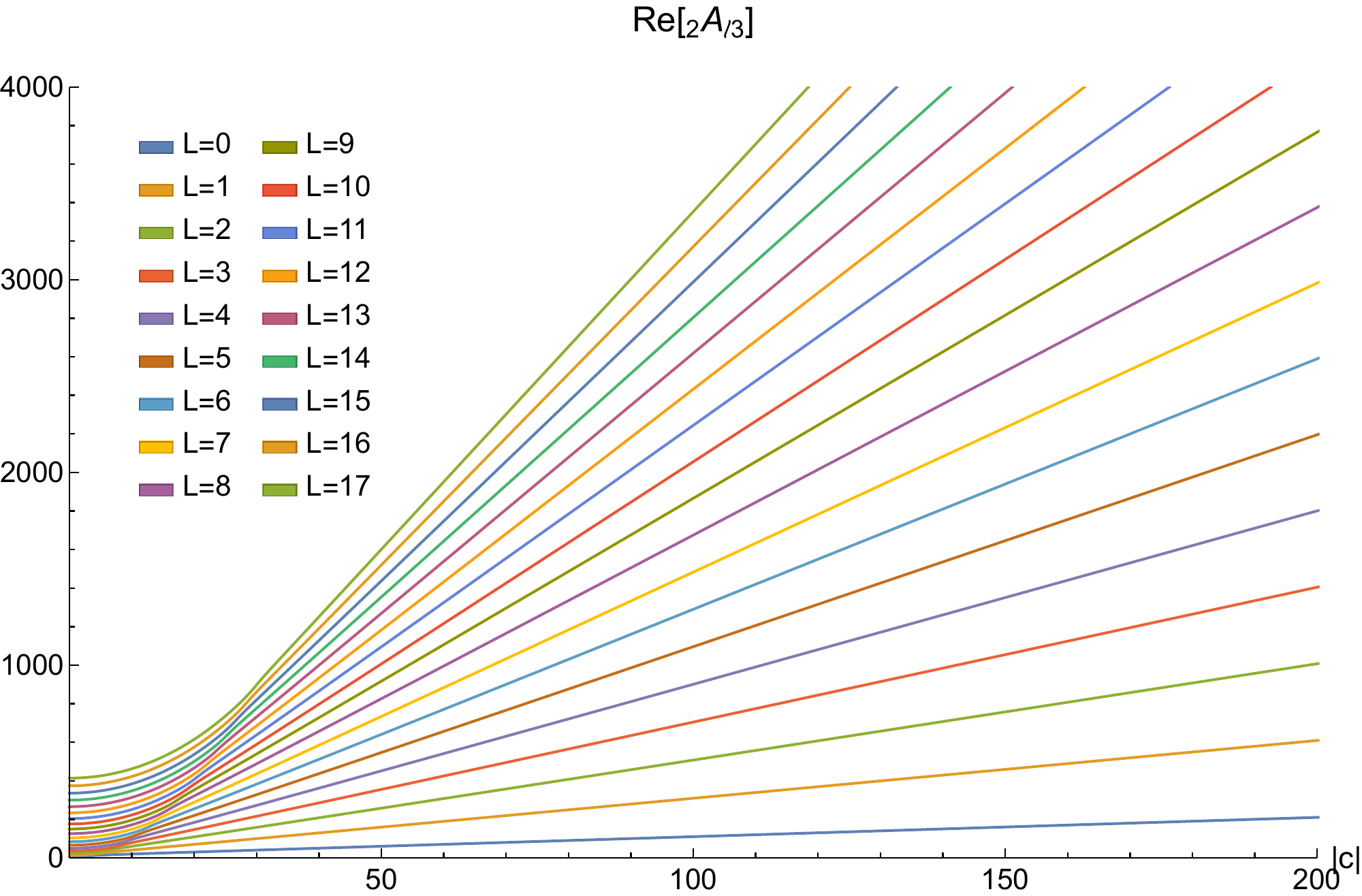}
 	\includegraphics[width=\linewidth,clip]{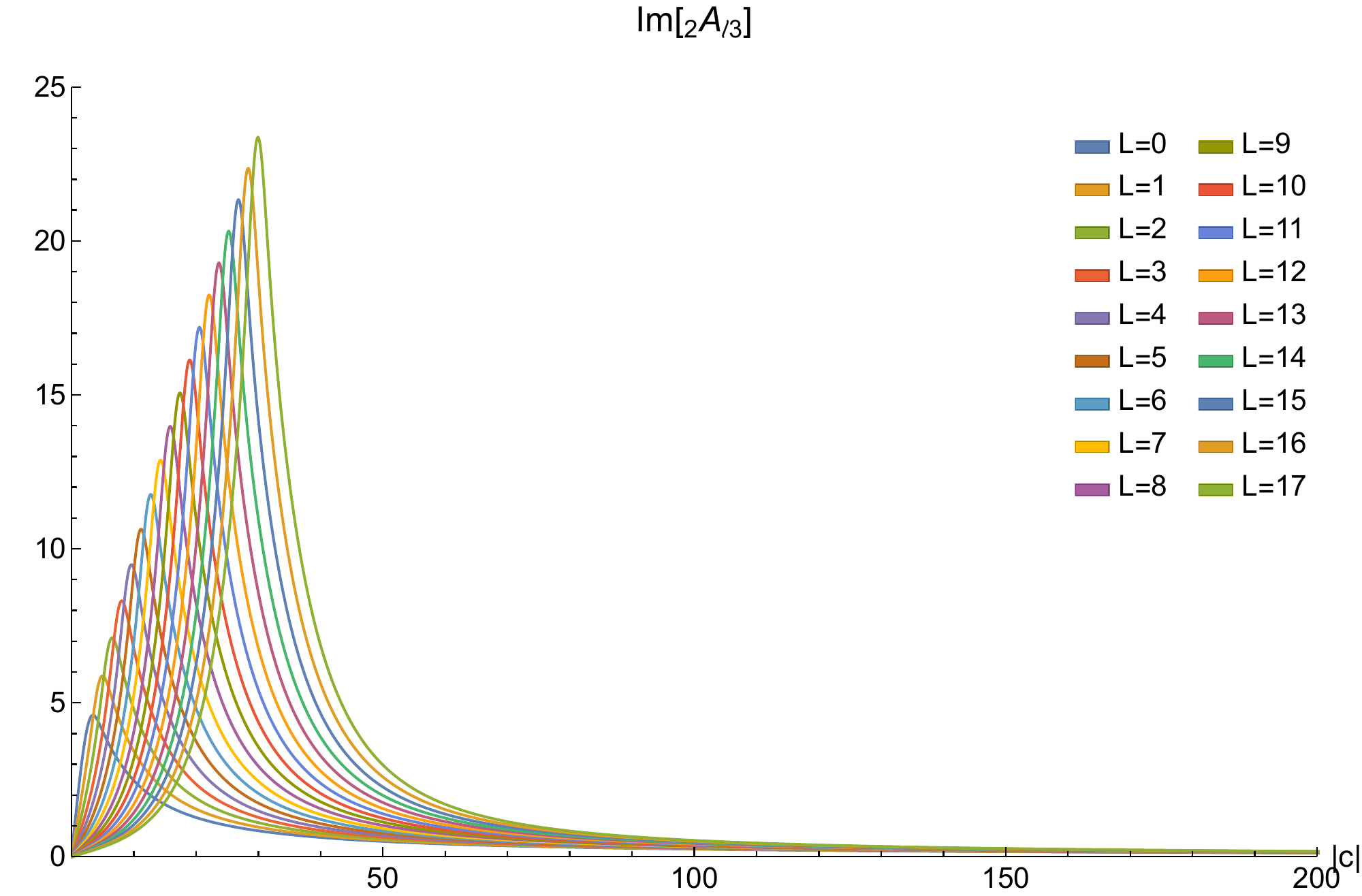}
    \caption{
    The real and imaginary components of the first 18 sequences of $\scA{2}{\ell 3}{-i|c|}$. The plot of the real components shows linear leading-order asymptotic behavior, as anticipated by Eq.~(\ref{eqn:sAlm leading order}). The imaginary component follows a leading-order asymptotic behavior of $|c|^{-1}$.}
    \label{fig:realimag sample plot}
\end{figure}
Figure~\ref{fig:realimag sample plot} shows an example set of prolate sequences where we plot the separation constant $\scA{2}{\ell 3}{-i|c|}$ with $3\le\ell\le20$ ($L\leq17$).  In the asymptotic regime, the real part of each sequence grows linearly with $|c|$ as seen in the upper plot in the figure, and the imaginary part is inversely proportional to $|c|$ as seen in the lower plot in the figure. Examination of the real part of each sequence reveals behavior that agrees with Eq.~(\ref{eqn:sAlm leading order}).  This suggests that the sequences displayed in Fig.~\ref{fig:realimag sample plot} belong to the class of solutions described by Ref.~\cite{berti2005eigenvalues}. We can further investigate the behavior of these sequences by observing the zero-crossings of the real part of the associated spin-weighted spheroidal functions.
\begin{figure}
    \centering
    \includegraphics[width=\linewidth,clip]{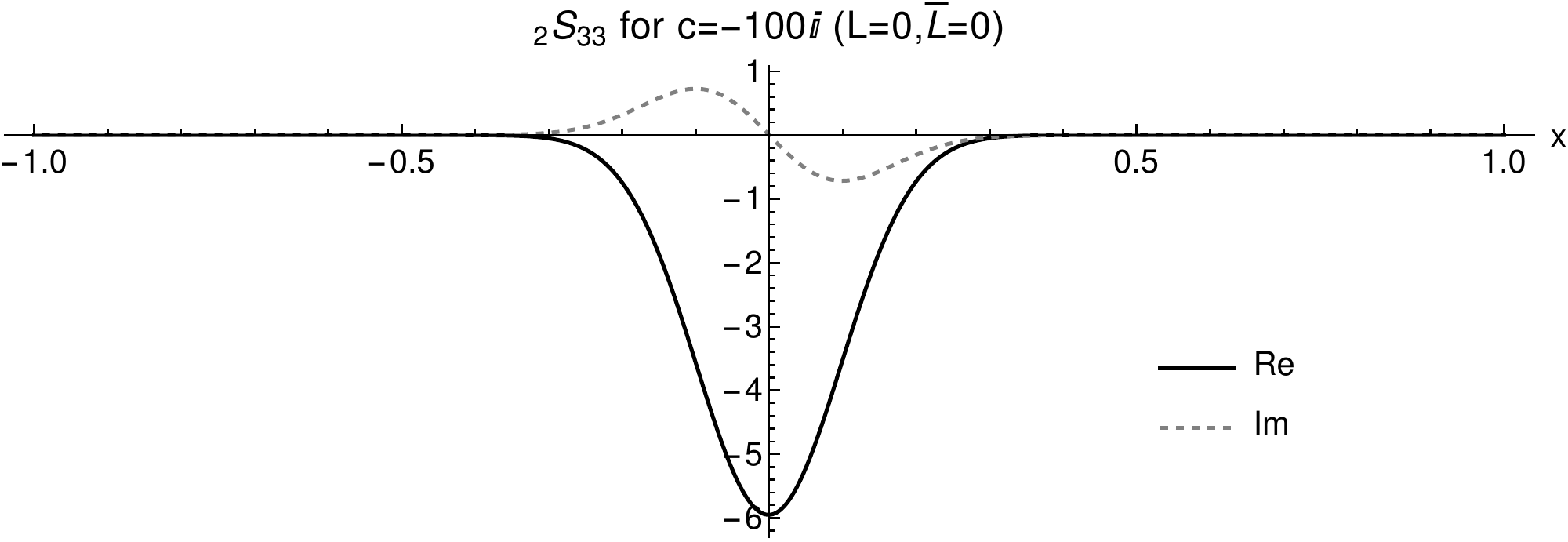}
    \includegraphics[width=\linewidth,clip]{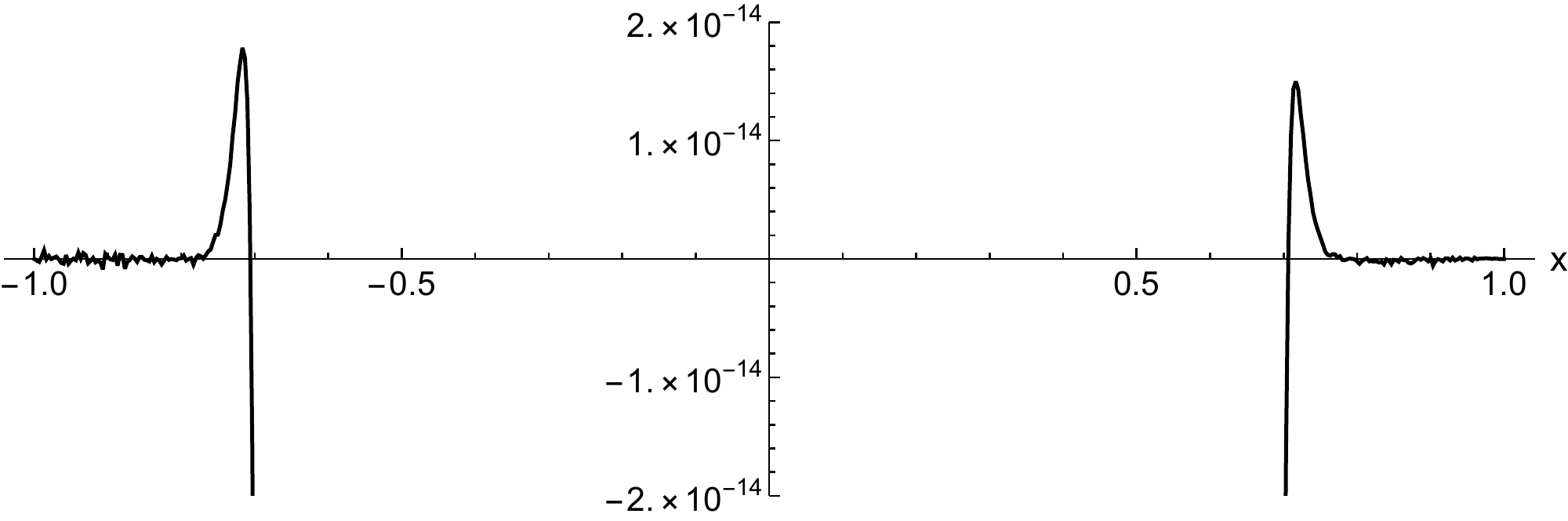}
    \caption{The real and imaginary components of $\swS{2}{33}{x}{c}$ at $c=-100i$ with an emphasis on the real zero-crossings in the outer region in the lower plot. For $s=\pm2$, one expects to see two real zero-crossings near the points of $x=\pm2^{-1/2}\approx\pm0.707$.}
    \label{fig:eigenvector 3x2x0}
\end{figure}
\begin{figure}
    \centering
    \includegraphics[width=\linewidth,clip]{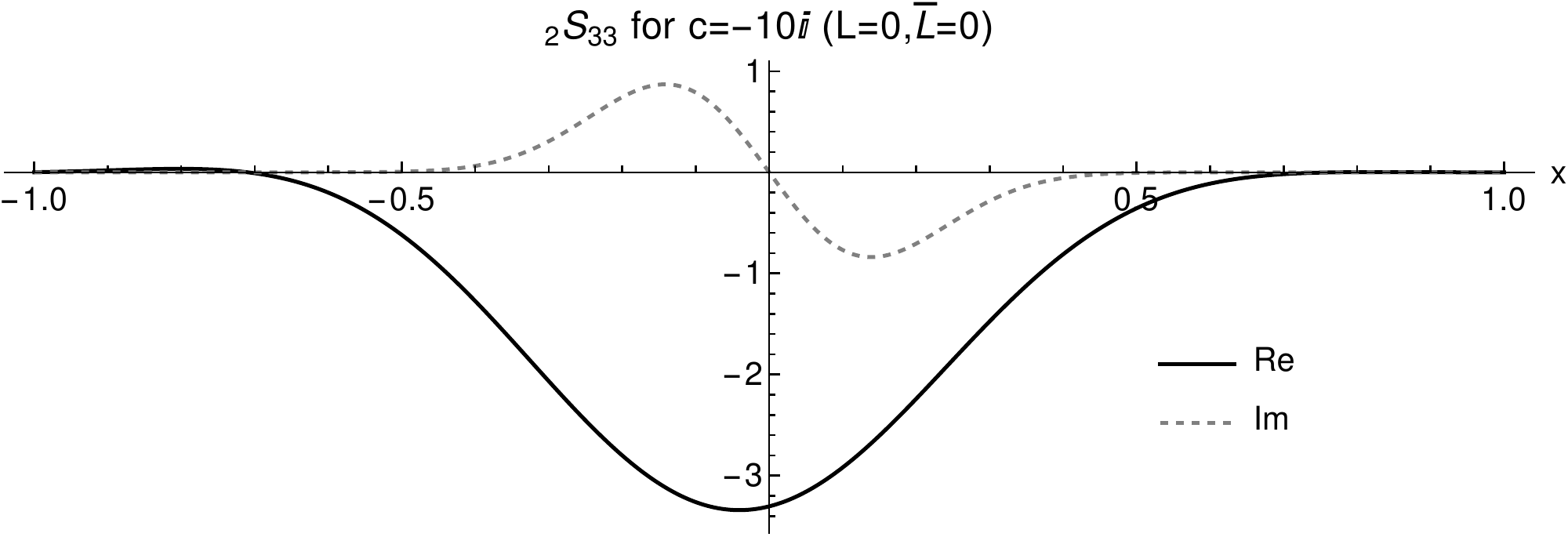}
    \includegraphics[width=\linewidth,clip]{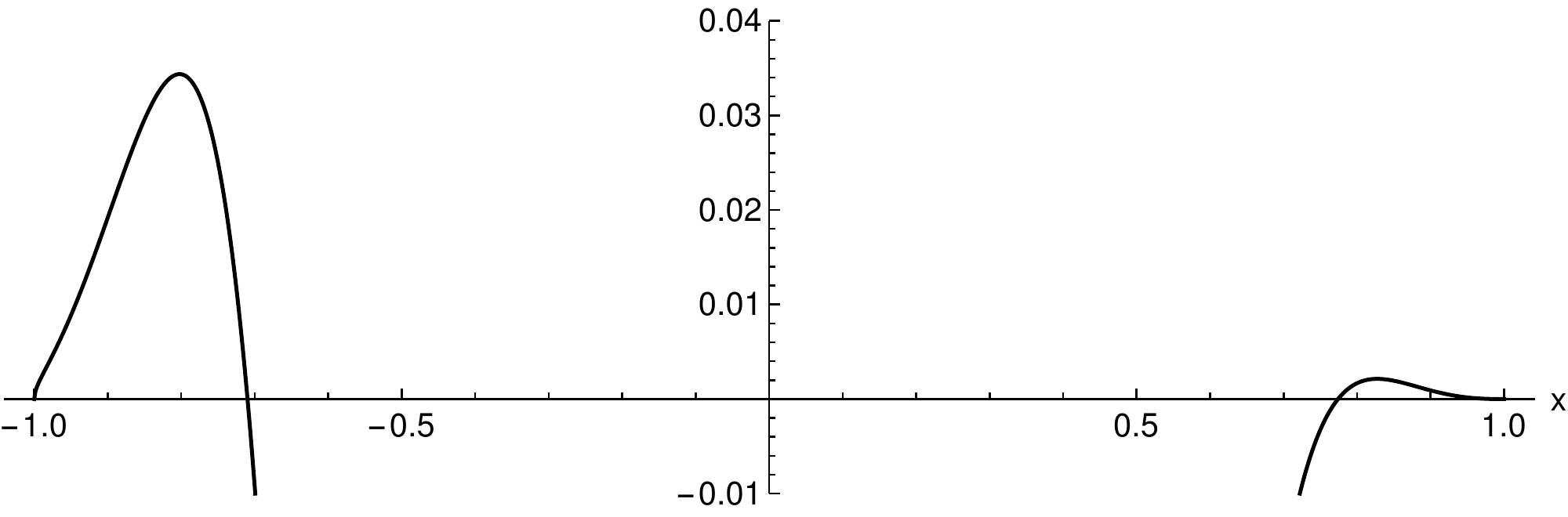}
    \caption{The real and imaginary components of $\swS{2}{33}{x}{c}$ at $c=-10i$ with an emphasis on the real zero-crossings in the outer region in the lower plot.  At $c=-10i$, the solution is only beginning to enter the asymptotic regime.  Unlike Fig.~\ref{fig:eigenvector 3x2x0}, the zero crossings are not well approximated by the roots of Eq.~(\ref{eqn:WKB outer solutions})}.
    \label{fig:eigenvector 3x2x0 wide}
\end{figure}

\begin{figure}
    \centering
    \includegraphics[width=\linewidth,clip]{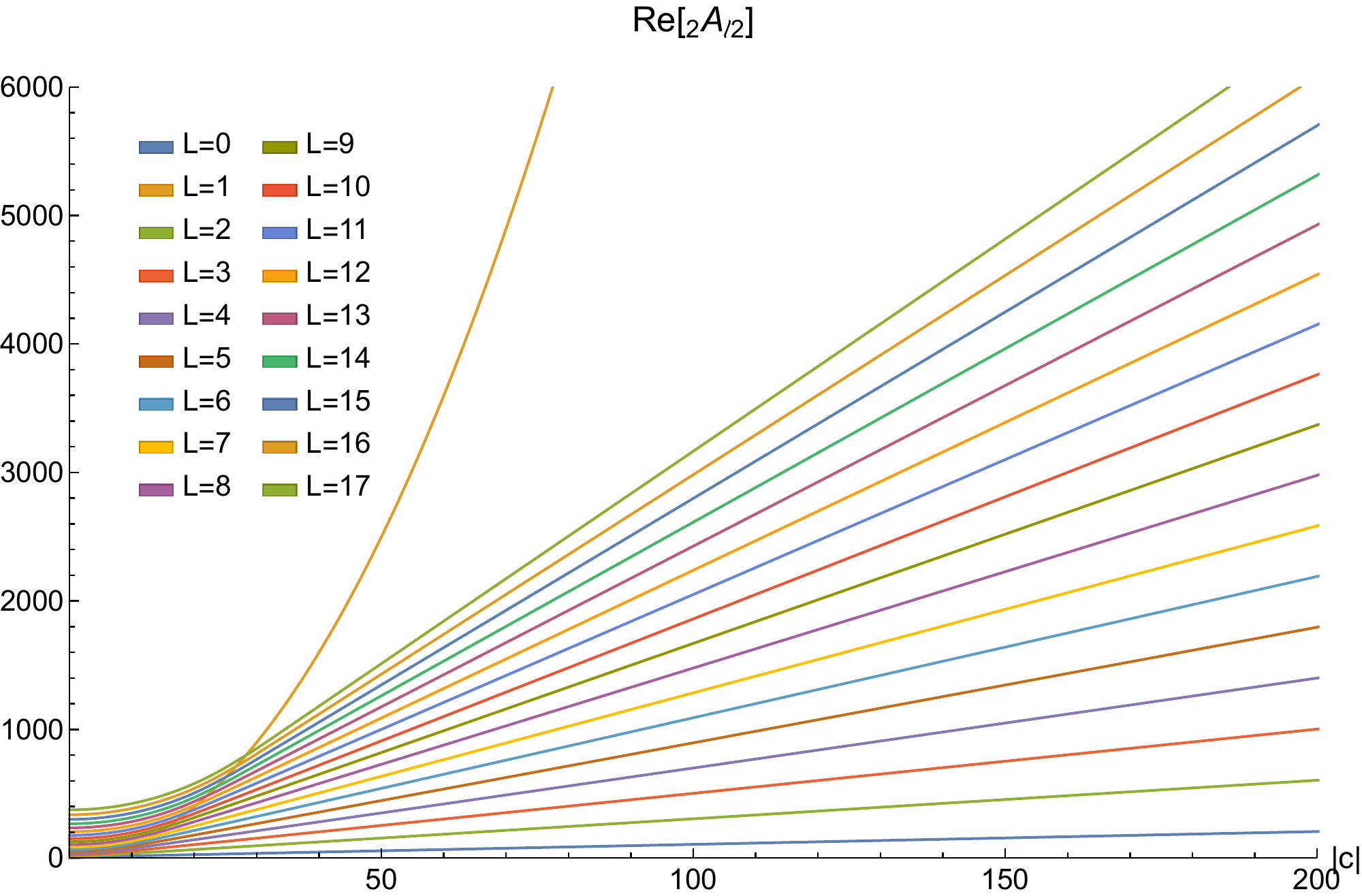}
    \includegraphics[width=\linewidth,clip]{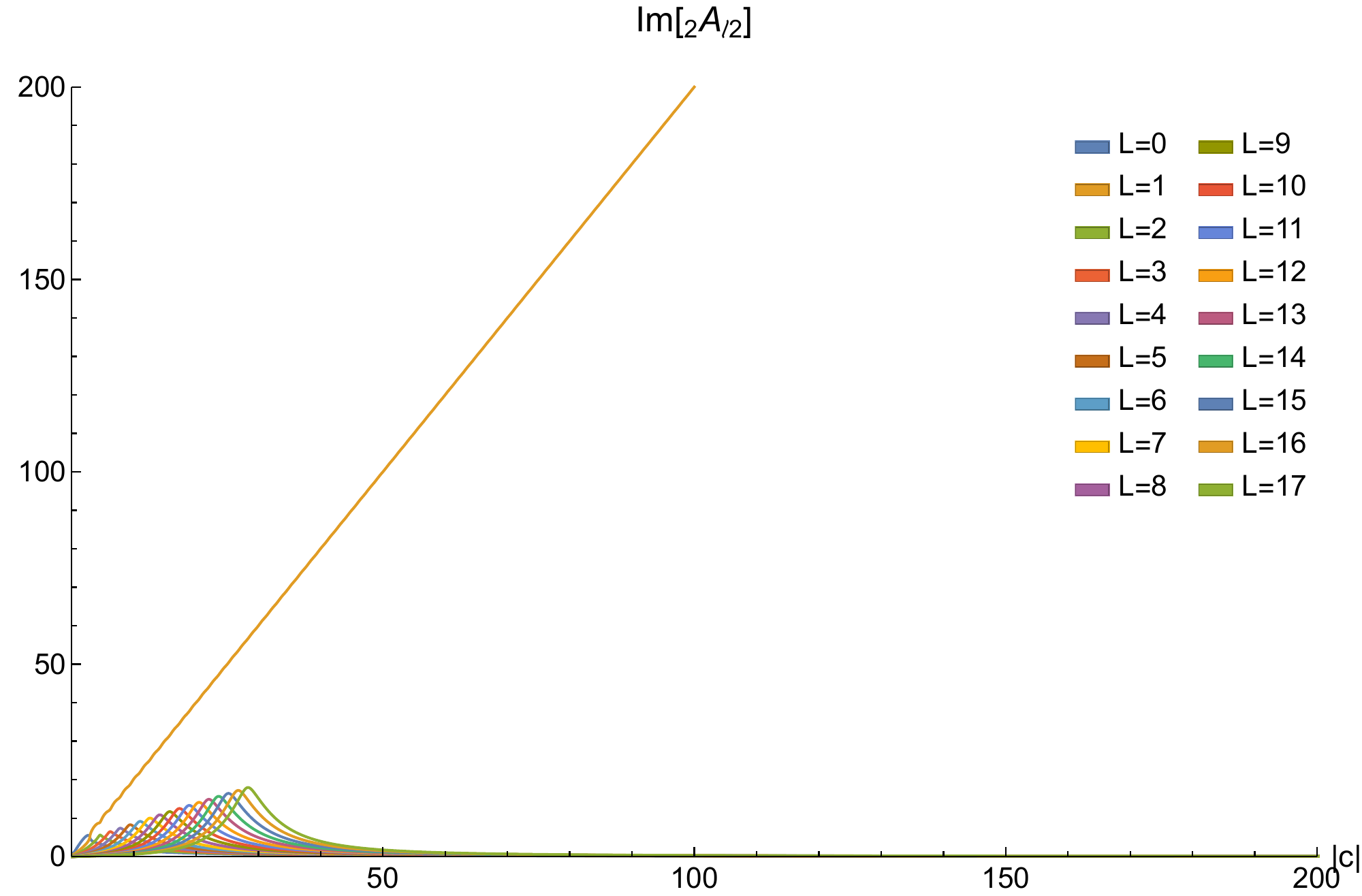}
    \caption{The real and imaginary components of the first 18  sequences of $\scA{2}{\ell 2}{-i|c|}$. Note the unusual behavior of the $L=1$ sequence, which has quadratic leading-order behavior for the real component and linear leading-order behavior for the imaginary component.}
    \label{fig:2x2 anomalous}
\end{figure}

Figure~\ref{fig:eigenvector 3x2x0} shows the real zero crossings of $\swS{2}{3 3}{x}{c}$ for $c=-100i$.  From the discussion in Sec.~\ref{sec:prolate behavior}, we expect that the $\text{Re}\big[\swS{s}{\ell m}{x}{-i|c|}\big]$ will have $L$ zero crossings in an inner region which becomes progressively narrower as $|c|$ increases. Figure~\ref{fig:eigenvector 3x2x0 wide} shows a similar plot for $c=-10i$ showing the inner region is much wider.  There should also be some number (which depends upon $s$) of additional zero crossings in the outer regions. For $s=\pm2$, Eq.~(\ref{eqn:WKB outer solutions}) gives two zero crossing at $x=\pm2^{-\frac12}$.  It can be seen in the upper plot of Fig.~\ref{fig:eigenvector 3x2x0} that $\text{Re}\big[\swS{2}{3 3}{x}{-100i}\big]$ has no real zero crossings in the inner region, appropriate for its value of $L=0$. In the lower plot, we notice that $\text{Re}\big[\swS{2}{3 3}{x}{-100i}\big]$ presents two real zero crossings in the outer regions at the prescribed locations.  In Fig.~\ref{fig:eigenvector 3x2x0 wide}, the zero crossings in the outer regions are not as well approximated by Eq.~(\ref{eqn:WKB outer solutions}) suggesting that at $c=-10i$ the solution is not yet fully in the asymptotic regime. The behavior of $\swS{2}{3 3}{x}{-100i}$ demonstrates the behavior described by the class of solutions discussed in Ref.~\cite{berti2005eigenvalues}. Most of our numerically generated prolate sequences show behavior that agrees with the predictions made in Ref.~\cite{berti2005eigenvalues}. However, there also exist a large number of individual solution sequences where neither Eq.~(\ref{eqn:sAlm leading order}) nor Eq.~(\ref{eqn:WKB outer solutions}) agree with the behavior seen in the separation constant and its associated eigenfunction. An example of one such sequence can be observed in Fig.~\ref{fig:2x2 anomalous}.

Figure~\ref{fig:2x2 anomalous} shows an example set of prolate sequences where we plot the separation constant $\scA{2}{\ell 2}{-i|c|}$ with $2\le\ell\le19$ ($L\leq17$).  While most of the sequences exhibit similar behavior to that seen in Fig.~\ref{fig:realimag sample plot}, we find that the $\scA{2}{3 {2}}{-i|c|}$ ($L=1$) eigenvalue sequence is in disagreement with the leading order behavior defined in Eq.~(\ref{eqn:sAlm leading order}). This suggests that $\scA{2}{3 {2}}{-i|c|}$ may belong to a separate, previously unknown, class of prolate solutions to Eq.~(\ref{eqn:Angular Teukolsky Equation}). An analysis of our remaining data sets demonstrates that the behavior of $\scA{2}{3 {2}}{-i|c|}$ is not unique, but is indicative of the behavior of a distinct new set of prolate solutions. Every sequence of separation constants in this new class exhibits an asymptotic leading order behavior of
\begin{align}\label{eqn:anomalous leading order}
	\text{Re}\big[\scA{s}{\ell m}{c}\big]=|c|^2+\mathcal{O}(|c|^0).
\end{align}
We will refer to all prolate eigensolutions which obey Eq.~(\ref{eqn:anomalous leading order}) as ``anomalous'' eigensolution. Eigensolutions which are in agreement with Eqs.~(\ref{eqn:sAlm leading order}) and~(\ref{eqn:WKB outer solutions}), will be referred to as ``normal'' eigensolutions.

\begin{figure}
 	\includegraphics[width=\linewidth,clip]{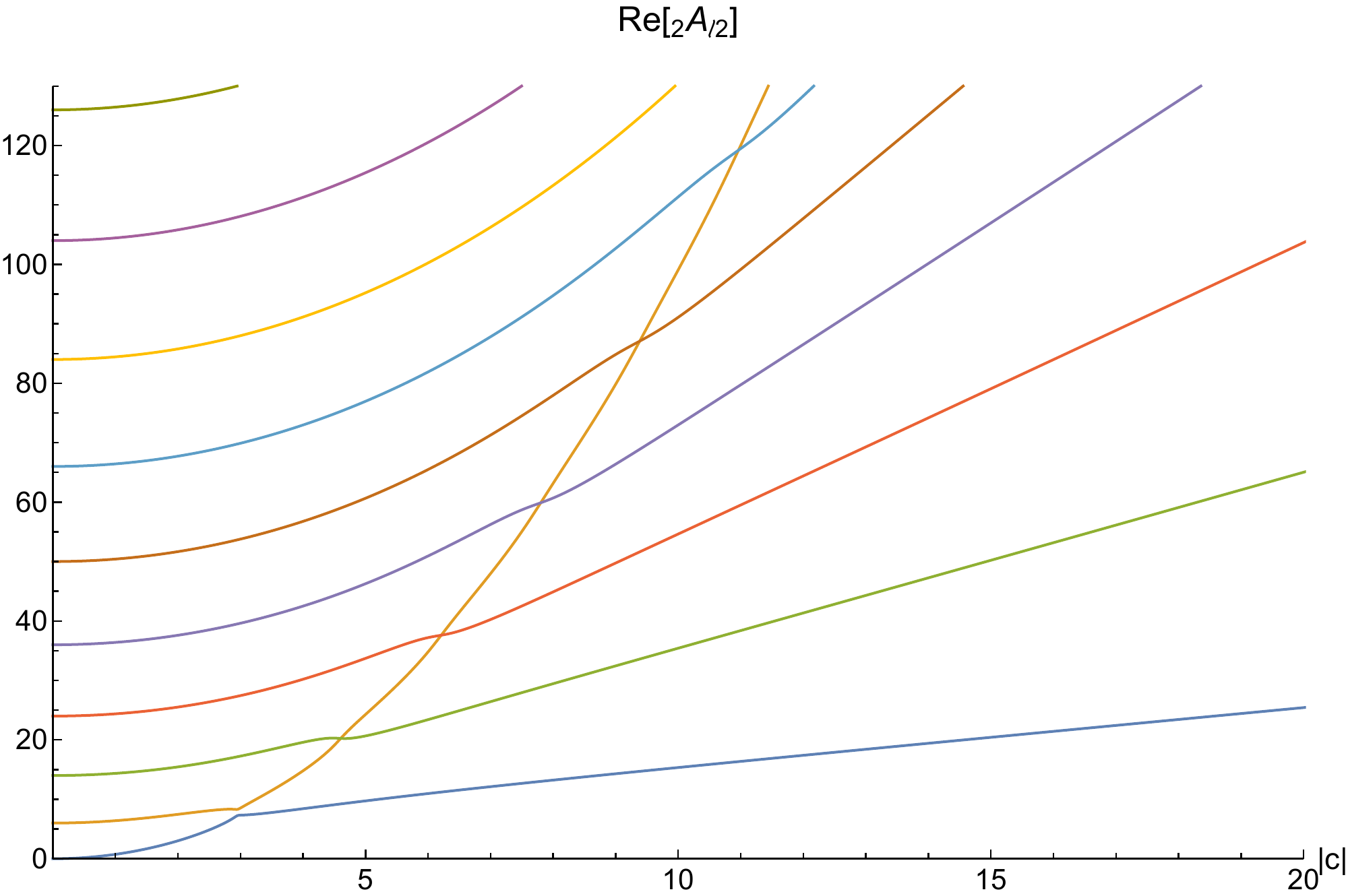}
    \includegraphics[width=\linewidth,clip]{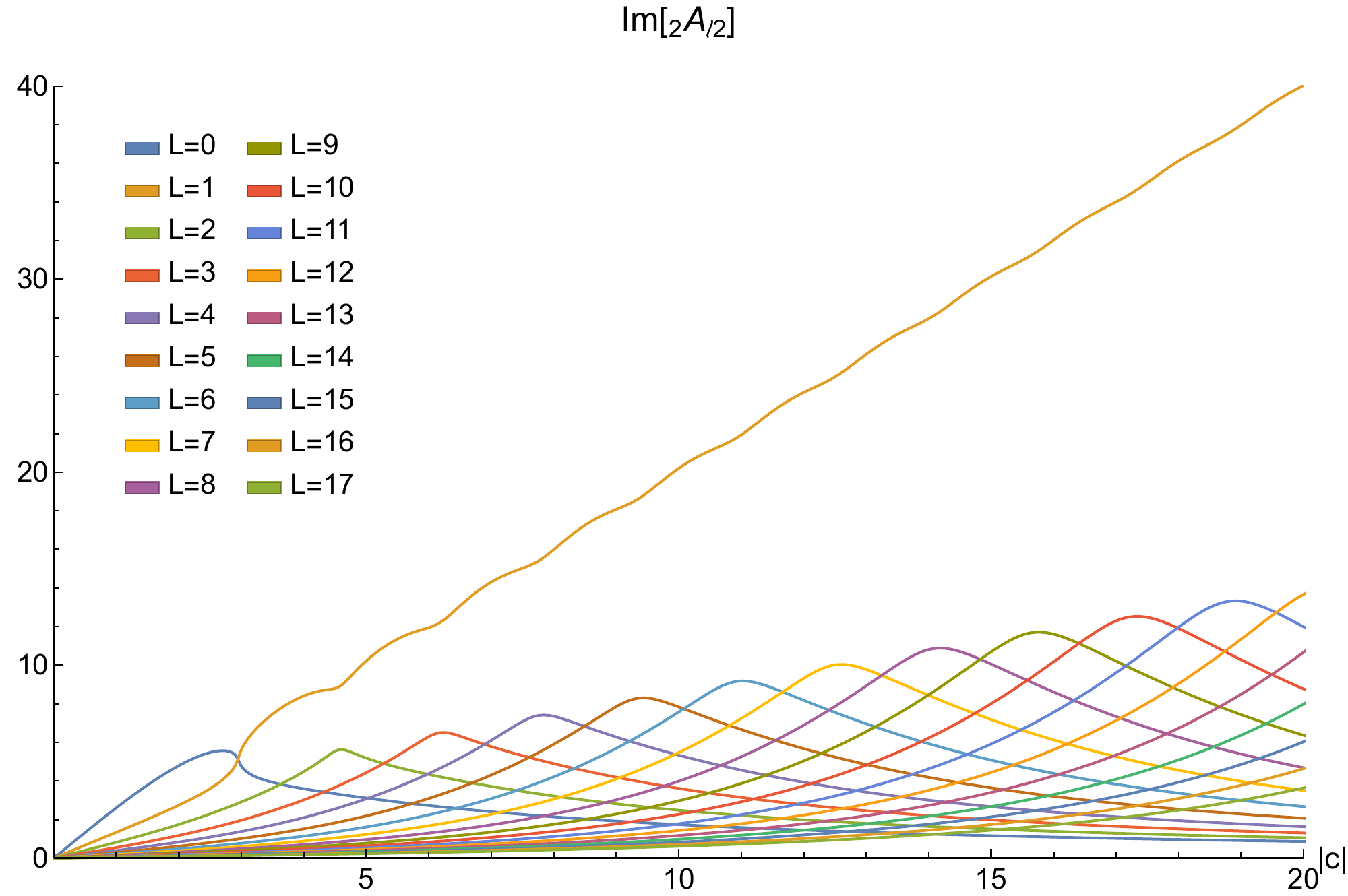}
    \caption{Close of up the same eigenvalue sequences for $\scA{2}{\ell 2}{-i|c|}$ shown in Fig.~\ref{fig:2x2 anomalous}. Note the deflection-like behavior that occurs between the real parts of the eigenvalue sequences $\scA{2}{22}{i|c|}$ and $\scA{2}{32}{i|c|}$ at $|c|\approx3$}
    \label{fig:2x2 anomalous small}
\end{figure}

Although the focus of this paper is on determining the asymptotic behavior of prolate solutions to Eq.~(\ref{eqn:Angular Teukolsky Equation}), it is also important to consider the transitional region between the asymptotic- and small-$c$ domains when anomalous solutions are present.  We can explore the behavior of sequences of the angular separation constant in this region by looking at Fig.~\ref{fig:2x2 anomalous small} which focuses on the transitional region for the sequences displayed in Fig.~\ref{fig:2x2 anomalous}.

Notice in the upper plot of Fig.~\ref{fig:2x2 anomalous small} that there exists some deflection-like behavior around $ic=3$ between the normal eigenvalue $\scA{2}{2 {2}}{-i|c|}$ ($L=0$) and the anomalous eigenvalue $\scA{2}{3 {2}}{-i|c|}$ ($L=1$). The bend in the real part of $\scA{2}{2 {2}}{-i|c|}$ was first noted in Ref.~\cite{berti2005eigenvalues} and is shown in Fig.~4 of that paper. Using a larger set of eigensolutions, we were able to determine that the bending behavior described in Ref~\cite{berti2005eigenvalues} was not unique to $\scA{2}{2 {2}}{-i|c|}$. These deflections are always found between pairs of sequences with the same values of $m$ and $s$.  In each case, the real part of an anomalous sequence's separation constant deflects away from an adjacent normal sequence of smaller $\ell$.  In the lower plot of Fig.~\ref{fig:2x2 anomalous small}, we see that the imaginary part of the $L=1$ anomalous sequence crosses the $L=0$ normal sequence near $ic=3$.  But, we note that at this crossing the real part of the separation constants deflect so that there is no degeneracy in the eigenvalues.  Subsequent to the deflection in the real part of the $L=1$ anomalous $\scA{2}{3 {2}}{-i|c|}$ sequence, we see that it crosses all of the normal sequences with $L>1$.  But, while the real parts of the sequences cross, the lower plot of Fig.~\ref{fig:2x2 anomalous small} shows that the imaginary parts do not cross and again there are no degeneracies.  More examples of similar bending behavior can be found in App.~\ref{sec:additional anom figs append}.

\begin{figure}
\includegraphics[width=\linewidth,clip]{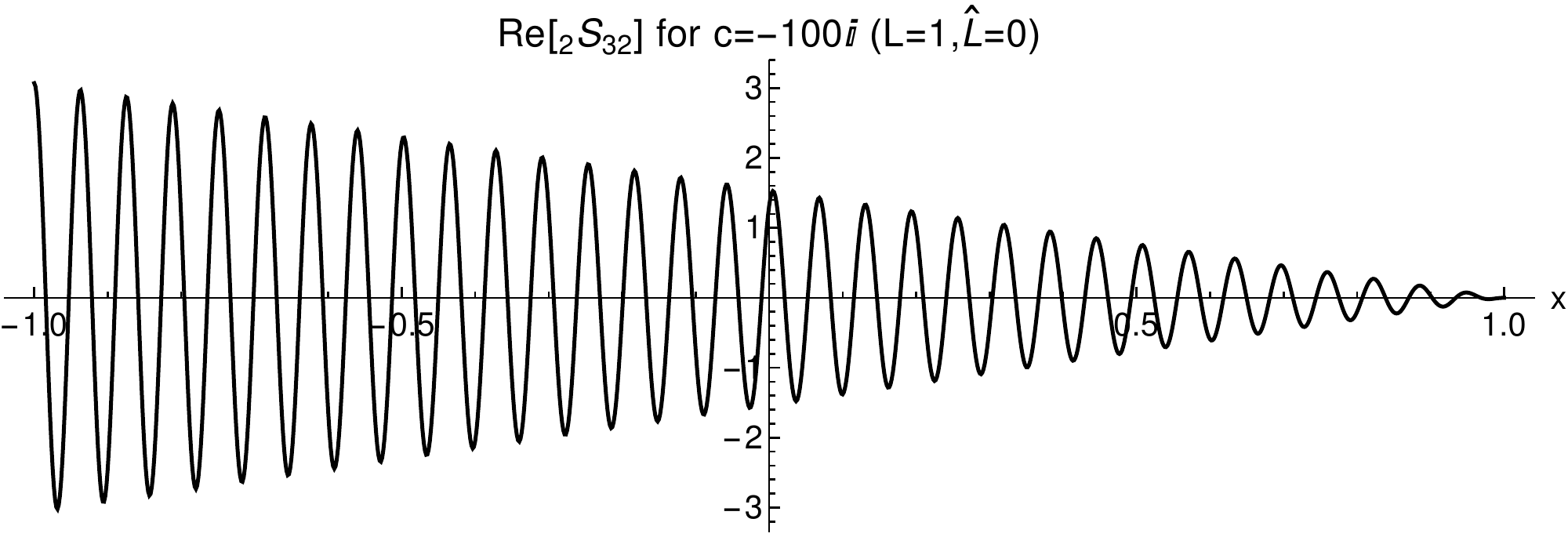}
\includegraphics[width=\linewidth,clip]{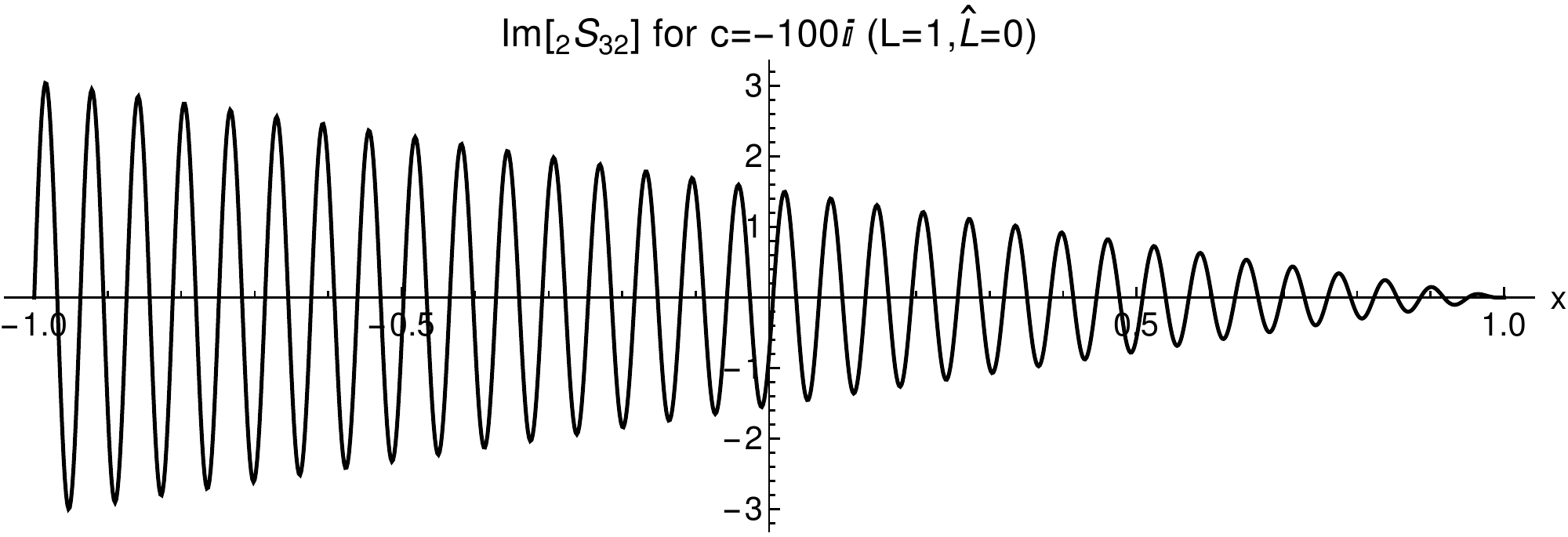}
    \caption{Eigenvector solution for $\swS{2}{32}{x}{-100i}=\swSanom{2}{22}{x}{-100i}$. Notice that this differs from the expected behavior described in Ref.~\cite{berti2005eigenvalues}. $\swS{2}{32}{x}{c}$ does not have a number of real zero-crossing equivalent to $L=1$ nor does the eigenvector go to zero at both endpoints.}
    \label{fig:2x2 anomalous vector}
\end{figure}

The behavior of the SWSFs, $\swS{s}{\ell m}{x}{-i|c|}$, of the anomalous solutions also differ from that predicted in Ref.~\cite{berti2005eigenvalues}, as is demonstrated in Fig.~\ref{fig:2x2 anomalous vector}.  In the caption of this figure, we introduce a change in notation where we will differentiate eigensolutions exhibiting asymptotic anomalous behavior by adding a hat(or carat).  We will discuss this new notation in more detail below.
Notice that Eq.~(\ref{eqn:sylm substitution}), which is used in the derivation of Eq.~(\ref{eqn:sAlm leading order}), demands that $\swS{s}{\ell m}{x}{-i|c|}\rightarrow0$ at the endpoints $x=\pm1$. It is interesting to note that this anomalous eigenvector solution $\swSanom{2}{32}{x}{-i|c|}$ does not go to zero at one endpoint. One would also predict three zero-crossings for $\text{Re}\big[\swS{2}{32}{x}{-i|c|}\big]$; one zero crossing in the inner region for $\swS[inner]{2}{32}{x}{-i|c|}$ and two more at $x=\pm2^{-\frac12}$ for $\swS[outer]{2}{32}{x}{-i|c|}$. Notice that neither of these expectations hold true for our numerical approximation of $\swSanom{2}{32}{x}{-i|c|}$. Another assumption used in the derivation of Eq.~(\ref{eqn:sAlm leading order}) is that the number of zero-crossings of $\text{Re}\big[\swS[inner]{s}{\ell m}{x}{-i|c|}\big]$ be constant for all $|c|$. The number of real zero-crossings for the eigenvector shown in Fig.~\ref{fig:2x2 anomalous vector} increases with $|c|$. It was found to be true for all anomalous eigenfunctions that the number of zero-crossings of $\text{Re}\big[\swSanom{s}{\ell m}{x}{-i|c|}\big]$ was not constant in $|c|$. In counterpoint, it was also found that some anomalous eigenvector solutions did go to zero at both of the endpoints.  However, these solution do exhibit the anomalous behavior of having an increasing number of zero-crossings as $|c|$ increases. All of these behaviors of the anomalous eigenfunctions will be explored in more detail in Sec.~\ref{sec:anomalous eigenfunctions}.  It seems likely that the assumptions used in the derivation of Eq.~(\ref{eqn:sAlm leading order}) did not allow previous works to predict the existence of the anomalous class of solutions for Eq.~(\ref{eqn:Angular Teukolsky Equation}).

\subsection{Normal Sequences}
\label{sec:Normal Sequences}

In order to generalize a power-series expansion for prolate $\scA{s}{\ell {m}}{c}$, we must explore the behavior of the anomalous and normal sequences separately. However, the presence of anomalous eigensolutions introduces some ambiguity in the labeling of sequences by $L$. This ambiguity is best explained through example. Consider the data set shown in Fig.~\ref{fig:2x2 anomalous small} for $m=2$ and $s=2$, which contains an anomalous sequence with $\ell=3$($L=1$). Based upon Eq.~(\ref{eqn:sAlm leading order}), one would anticipate a linear-order behavior of $2L+1$ for all normal sequences.

Using the $\scA{2}{\ell {2}}{c}$ sequences, we numerically determined that $\scA{2}{2 {2}}{c}=|c|+\mathcal{O}(|c|^{0})$ ($L=0$), agreeing with Eq.~(\ref{eqn:sAlm leading order}). However, all subsequent normal eigenvalue sequences ($L>1$) fit as if $L\rightarrow L-1$. For example, we numerically determined that $\scA{2}{4 {2}}{c}=3|c|+\mathcal{O}(|c|^{0})$ ($L=2$), in disagreement with Eq.~(\ref{eqn:sAlm leading order}). This trend holds true for all combinations of $m$ and $s$ for which there exists an anomalous eigensolution.

This shift in $L$ is due to an incompatibility in the labeling of $L$ in the spherical and asymptotic limits.  Recall that the labeling of $L$ used in our data sets is based upon the spherical limit eigensolutions of $\scA{s}{\ell {m}}{c}$; we merely carried over this label of $L$ as $|c|$ increased into the asymptotic regime. However, Eq.~(\ref{eqn:sAlm leading order}) uses an index of $L$ based on the parabolic cylinder functions, $D_{L}(x)$. The derivation of Eq.~(\ref{eqn:sAlm leading order}) guarantees a one-to-one relation between each normal solution and a corresponding $D_{L}(x)$. In the presence of anomalous sequences, the index $L$ we used in the spherical limit and the label used in Eq.~(\ref{eqn:sAlm leading order}) for the asymptotic prolate limit are not the same. To ensure that our labeling of $L$ is consistent with analytic predictions, we find it useful to define two parameters. We define $\sNlm{s}{\ell m}$ as the number of anomalous eigensolutions that exist for $m$ and $s$ with smaller values of $\ell$.\footnote{For example, if the first anomalous eigensolution occurred for $\ell=2$, then $\sNlm{s}{2m}=0$ and $\sNlm{s}{3m}=1$.} We also define $\bar{L}=L-\sNlm{s}{\ell m}$. It then becomes true that all of our normal asymptotic sequence data, regardless of the presence of anomalous sequences, obeys the leading-order asymptotic fit of
\begin{align}\label{eqn:sAlm star leading order}
	\scAnorm{s}{\ell m}{c} = (2\bar{L}+1)|c| + \mathcal{O}(|c|^{0}).
\end{align}
Here, and in Fig.~\ref{fig:eigenvector 2x2x10} below, we introduce another change in notation where we will differentiate eigensolutions exhibiting asymptotic normal behavior by adding an over-bar.
\begin{figure}
\includegraphics[width=\linewidth,clip]{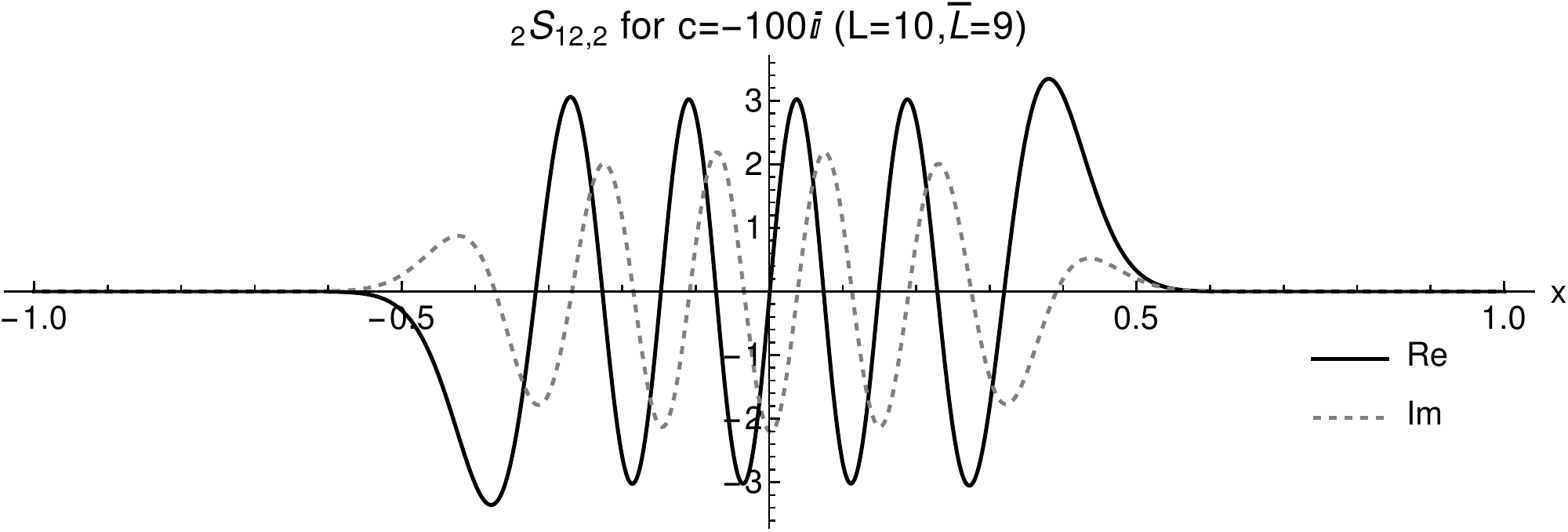}
    \caption{The real and imaginary components of $\swS{2}{12,2}{x}{-100i}=\swSnorm{2}{11,2}{x}{-100i}$. In the $c\to0$ limit the sequence of solutions corresponds to a value of $L=10$. As such, one would expect to see ten real zero-crossings in the inner region. Here we see nine zero crossings. We account for this by noting that $\bar{L}=L-\,_{2}N_{12,2}=9$, since there exists an anomalous sequence $\swSanom{2}{32}{x}{c}$.}
    \label{fig:eigenvector 2x2x10}
\end{figure}

The normal sequences of $\swS{s}{\ell m}{x}{c}$ also exhibit their expected behavior when using the label $\bar{L}$. For example, the sequence $\swS{2}{12, {2}}{x}{c}=\swSnorm{2}{11, {2}}{x}{c}$, shown in Fig~\ref{fig:eigenvector 2x2x10}, demonstrates that there are $\bar{L}=L-1=9$ real zero crossings in the inner region of the eigenvector.  As a result, we shall be using the index $\bar{L}$ in the numerical fits of our normal asymptotic prolate data.  Furthermore, we will compute the sequence labels $\ell$ for asymptotic normal sequences via Eq.~(\ref{eqn:L def}) using $\bar{L}$ instead of $L$.  Because $\bar{L}$ takes on consecutive values starting with $0$ regardless of the presence of anomalous sequences, relabeling $\ell$ for asymptotic normal eigensolutions removes any dependence on knowledge of associated anomalous sequences.

While in the scalar case of Eq.~(\ref{eqn:flammer solution}) we could expand the solution in terms of $|c|$ for both positive and negative imaginary values of $c$, the general case is more complicated.  For fitting, we will not expand in terms of $|c|$ but instead restrict ourselves to fitting sequences with negative imaginary values of $c$.  In this case, we can express our asymptotic expansion as
\begin{widetext}
\begin{align}\label{eqn:model fit}
	\scAnorm{s}{\ell m}{c} &= (2\bar{L}+1)ic-\frac14\left(2\bar{L}(\bar{L}+1)+3-4m^2\right) 
	+ \frac14C_0 \nonumber \\
	& \qquad - \frac{iB_1(\bar{L},m)-\frac1{16}C_1}{c} - \frac{B_2(\bar{L},m)-\frac1{64}C_2}{c^2} + \frac{iB_3(\bar{L},m)+\frac1{256}C_3}{c^3} + \sum_{n=4}^\infty\frac{C_n}{c^n}.
\end{align}
\end{widetext}
To find the values of $C_{0\to3}$, we fit the last 40 (largest values of $|c|$) data points from our numerically generated normal sequences for each value of $s$, $m$, and $\bar{L}$ using a greedy approach.

To obtain $C_0$, we use Eq.~(\ref{eqn:model fit}) with 5 unknown terms, $C_{0\to4}$ and construct an intermediate data set containing the fit values for $C_0$ for each value of $s$, $m$, and $\bar{L}$.  We then do linear fitting to find $C_0$ as a function of $s$, $m$, and $\bar{L}$.  In this case, based on the expansion for $\scA{0}{\ell m}{c}$\footnote{We note that for $s=0$, $\scA{0}{\ell m}{c}=\scAnorm{0}{\ell m}{c}$}, we expect that $C_0$ will have at most quadratic terms in these variables and must include $s$ in each term.  Including symmetry arguments, we can deduce that $C_0$ must include only terms involving $s$ and $s^2$.  From linear fitting using {\em Mathematica}, we find that 
\begin{align}\label{eqn:C0 value}
C_0 = 4s(s-1).  
\end{align}
The exact fit values depend on exactly what subset of data is used in the linear fitting.  The results we display below will all be obtained from limiting the fit to $s\le4$, $|m|\le4$, and $\bar{L}\le8$.  Including more data in the fit tends to slightly increase the uncertainty in each coefficient, but not significantly, and the resulting fit functions agree very well with all data sets.  The results from {\em Mathematica} include a best linear fit to each term along with a constant term which should be consistent with zero.  The results for fitting the real part of $C_0$ are displayed in Table~\ref{table:Real C0}.
\begin{table}
 \begin{tabular}{c| d{9} d{7}}
 &\multicolumn{1}{c}{Estimate} & \multicolumn{1}{c}{$\sigma$}  \\
\hline
 $1$ & -8.15\times10^{-9} & 5.8\times10^{-8} \\
 $s$ & -3.99999991 & 6.8\times10^{-8}  \\
 $s^2$ & 3.99999995 & 1.6\times10^{-8} \\
\end{tabular}
\caption{Linear fit results for the real part of $C_0$.  The first column lists each term in the linear function, the second column displays the linear fit value for the constant multiplying each term, and the third column displays the uncertainty in the fit.  We see that the constant term in the fit is consistent with 0 resulting in a fit of $C_0=4s(s-1)$.}
\label{table:Real C0}
\end{table}
The imaginary part of $C_0$ is consistent with $0$ as expected, and we see that the coefficients in front of the $s$ and $s^2$ terms are consistent with integer values.

Having determined the value for $C_0$, we repeat the full fitting process but we include the determined value of $C_0$ in our fitting function Eq.~(\ref{eqn:model fit}).  The unknowns in our fitting function are now $C_{1\to5}$, where now $C_5$ is included in the fit since $C_0$ has been determined.  We found that it is important to keep several terms in the expansion beyond the term we are trying to determine, in this case, $C_1$.  Tables~\ref{table:Real C1} and \ref{table:Imag C1} show the final results for the linear fits for the real and imaginary parts of $C_1$.
\begin{table}
 \begin{tabular}{c| d{9} d{7}}
 &\multicolumn{1}{c}{Estimate} & \multicolumn{1}{c}{$\sigma$}  \\
\hline
$1$ & 3.64\times10^{-6} & 1.7\times10^{-6}  \\
 $m s^2$ & 31.99999906 & 7.6\times10^{-8}
\end{tabular}
\caption{Linear fit results for the real part of $C_1$.  See the caption for Table~\ref{table:Real C0} for details.  We see that $\text{Re}[C_1]=32ms^2$.}
\label{table:Real C1}
\end{table}
\begin{table}
 \begin{tabular}{c| d{9} d{7}}
 &\multicolumn{1}{c}{Estimate} & \multicolumn{1}{c}{$\sigma$}  \\
\hline
 $1$ & -7.28\times10^{-7} & 1.8\times10^{-6}  \\
 $s^2$ & -16.00000162 & 3.2\times10^{-7}  \\
 $\bar{L}s^2$ & -31.99999932 & 5.9\times10^{-8}
\end{tabular}
\caption{Linear fit results for the imaginary part of $C_1$.  See the caption for Table~\ref{table:Real C0} for details.  We see that $\text{Im}[C_1]=-16(2\bar{L}+1)s^2$}
\label{table:Imag C1}
\end{table}
Again, we find that the possible non-vanishing coefficients are consistent with integer values and the combined result is
\begin{align}\label{eqn:C1 value}
	C_1 = -16i(2\bar{L}+1+2im)s^2.
\end{align}

The remaining two coefficients, $C_2$ and $C_3$ are found by a similar procedure.  For $C_2$, we replace $C_0$ and $C_1$ in Eq.~(\ref{eqn:model fit}) with the fit values given in Eqs.~(\ref{eqn:C0 value}) and (\ref{eqn:C1 value}) and fit all sequences to Eq.~(\ref{eqn:model fit}) keeping the remaining terms out to $C_6$.  The fits for $C_2$ are then used to find a functional form for $C_2$.  Inserting this into Eq.~(\ref{eqn:model fit}) and keeping terms out to $C_7$, we fit all sequences again and use the results to find a functional form for $C_3$.  With each successive order in the expansion, our ability to accurately fit for the coefficients diminishes, but we are confident in our results out to $C_3$.  Tables~\ref{table:Real C2}--\ref{table:Imag C3} display the results of the linear fits for $C_2$ and $C_3$.
\begin{table}
 \begin{tabular}{c| d{9} d{7}}
 &\multicolumn{1}{c}{Estimate} & \multicolumn{1}{c}{$\sigma$}  \\
\hline
 $1$ & -0.00009615 & 0.00011  \\
 $s^2$ & -95.99968481 & 0.000047 \\
 $\bar{L}s^2$ & -192.00006712 & 0.000011 \\
 s$^4$ & 63.99998780 & 2.6\times10^{-6} \\
 $m^2 s^2$ & 127.99998708 & 1.4\times10^{-6} \\
 $\bar{L}^2 s^2$ & -191.99999233 & 1.4\times10^{-6}
\end{tabular}
\caption{Linear fit results for the real part of $C_2$.  See the caption for Table~\ref{table:Real C0} for details.  We see that $\text{Re}[C_2]=-32\bigl(6\bar{L}(\bar{L}+1)-4m^2-2s^2+3\bigr)s^2$}
\label{table:Real C2}
\end{table}
\begin{table}
 \begin{tabular}{c| d{9} d{7}}
 &\multicolumn{1}{c}{Estimate} & \multicolumn{1}{c}{$\sigma$}  \\
\hline
 $1$ & -0.00030416 & 0.00013 \\
 $m s^2$ & -256.00008151 & 0.000011 \\
 $\bar{L}m s^2$ & -511.99996086 & 2.3\times10^{-6}
\end{tabular}
\caption{Linear fit results for the imaginary part of $C_2$.  See the caption for Table~\ref{table:Real C0} for details.  We see that $\text{Im}[C_2]=-256(2\bar{L}+1)ms^2$}
\label{table:Imag C2}
\end{table}
\begin{table}
 \begin{tabular}{c| d{9} d{7}}
 &\multicolumn{1}{c}{Estimate} & \multicolumn{1}{c}{$\sigma$}  \\
\hline
 $1$ & -0.01117050 & 0.0058 \\
 $m s^2$ & -2431.99572667 & 0.0013 \\
 $\bar{L}m s^2$ & -5376.00498626 & 0.00038 \\
 $\bar{L}^2 m s^2$ & -5375.99915279 & 0.000046 \\
 $m^3 s^2$ & 511.99997465 & 0.000055 \\
 $m s^4$ & 1023.99994750 & 0.000072
\end{tabular}
\caption{Linear fit results for the real part of $C_3$.  See the caption for Table~\ref{table:Real C0} for details.  We see that $\text{Re}[C_3]=-256\bigl(21\bar{L}(\bar{L}+1)-2m^2-4s^2+\frac{19}2\bigr)ms^2$}
\label{table:Real C3}
\end{table}
\begin{table}
 \begin{tabular}{c| d{9} d{7}}
 &\multicolumn{1}{c}{Estimate} & \multicolumn{1}{c}{$\sigma$}  \\
\hline
 $1$ & 0.00355322 & 0.014 \\
 $s^2$ & 816.02922552 & 0.0084 \\
 $\bar{L}s^2$ & 2223.98136420 & 0.0035 \\
 $\bar{L}^2 s^2$ & 1776.00111392 & 0.00095 \\
 $\bar{L}^3 s^2$ & 1183.99999057 & 0.000078 \\
 $m^2 s^2$ & -1920.00184852 & 0.00031 \\
 $\bar{L}m^2 s^2$ & -3839.99902142 & 0.000065 \\
 $s^4$ & -1024.00117837 & 0.00052 \\
 $\bar{L}s^4$ & -2047.99938243 & 0.00010
\end{tabular}
\caption{Linear fit results for the imaginary part of $C_3$.  See the caption for Table~\ref{table:Real C0} for details.  We see that $\text{Im}[C_3]=16(2\bar{L}+1)\bigl(37\bar{L}(\bar{L}+1)+51\bigr)s^2-32(2\bar{L}+1)(60m^2s^2+32s^4)$}
\label{table:Imag C3}
\end{table}
In both cases, we find that the possible non-vanishing coefficients are consistent with integer values and the combined results are
\begin{align}
\label{eqn:C2 value}
	C_2 &= -32\Bigl[6\bar{L}(\bar{L}+1)-4m^2-2s^2+3 \\
	& \hspace{1.75in}+ 8i(2\bar{L}+1)m\Bigr]s^2  \nonumber \\
\label{eqn:C3 value}
	C_3 &= 16i\Bigl[(2\bar{L}+1)\bigl(37\bar{L}(\bar{L}+1)+51-120m^2-64s^2\bigr) \nonumber\\
	& \hspace{0.5in} + 16i\Bigl(21\bar{L}(\bar{L}+1)-2m^2-4s^2+\frac{19}2\Bigr)m\Bigr]s^2
\end{align}

In full, we can write the asymptotic expansion of $\scAnorm{s}{\ell m}{c}$ for normal sequences in the prolate case where $c=-i|c|$ (negative imaginary values) as
\begin{widetext}
\begin{align}\label{eqn:prolate normal solution}
	\scAnorm{s}{\ell m}{c} = i c (2 \bar{L}+1)
& -\frac14 \left[2 \bar{L} (\bar{L}+1) -4 m^2 -4 s(s-1)+3\right] \nonumber \\
\mbox{} &+\frac{i}{16 c}\left[(2 \bar{L}+1) \left(\bar{L} (\bar{L}+1)-8 m^2-16 s^2+3\right)-32i m s^2 \right] \nonumber \\
\mbox{} & +\frac1{64 c^2}\biggl[5
\left(\bar{L}(\bar{L}+1)(\bar{L}(\bar{L}+1)+7)+3\right)-48 \left(2 \bar{L} (\bar{L}+1)+1\right) m^2 \nonumber\\
\mbox{} &\hspace{70pt}-32 \left(6 \bar{L} (\bar{L}+1)-4 m^2-2 s^2+3\right)s^2-256 i (2 \bar{L}+1)m s^2 \biggr] \nonumber \\
\mbox{} &-\frac{i}{256 c^3}\biggl[
\biggl(\frac14(2\bar{L}+1)\left(\bar{L}(\bar{L}+1)(33\bar{L}(\bar{L}+1)+415)+453\right) \nonumber\\
\mbox{} &\hspace{100pt}-8 (2 \bar{L}+1) \left(37 \bar{L}(\bar{L}+1)+51\right) \left(m^2+2 s^2\right) \nonumber \\
\mbox{} &\hspace{160pt}+32 (2 \bar{L}+1) \left(m^4+60 m^2 s^2+32 s^4\right)\biggr) \nonumber \\
\mbox{} &\hspace{130pt}-256i\left(21 \bar{L} (\bar{L}+1)-2 m^2-4 s^2+\frac{19}2\right)m s^2\biggr]
+ \mathcal{O}(c^{-4})
\end{align}
\end{widetext}
As with the oblate expansion in Eq.~(\ref{eqn:oblate solution all}), the prolate asymptotic expansion for normal sequences given in Eq.~(\ref{eqn:prolate normal solution}) cannot be expressed in a way that is valid for both positive and negative imaginary values of $c$.  The form of $C_0$, along with the fact that all subsequent coefficients include only even powers of $s$ guarantees that Eq.~(\ref{eq:swSF_sA_ident}) is satisfied.  Together, Eqs.~(\ref{eq:swSF_mcA_ident}) and (\ref{eq:swSF_cA_ident}) demand, when $c$ is imaginary, that the $\text{Re}[\scAnorm{s}{\ell m}{c}]$ must include only even powers of $m$ and the $\text{Im}[\scAnorm{s}{\ell m}{c}]$ must include only odd powers of $m$.  Equation~(\ref{eqn:prolate normal solution}) satisfies these conditions.  Furthermore, most of the individual terms in Eq.~(\ref{eqn:prolate normal solution}) could be written in a way that satisfies Eqs.~(\ref{eq:swSF_mcA_ident}) and (\ref{eq:swSF_cA_ident}) individually by using $|c|$ for all of the terms which are real when evaluated for imaginary values of $c$.  For the imaginary terms, those that involve odd powers of $c$ obey Eqs.~(\ref{eq:swSF_mcA_ident}) and (\ref{eq:swSF_cA_ident}).  But, any imaginary term at even powers of $c$ will violate Eqs.~(\ref{eq:swSF_mcA_ident}) and (\ref{eq:swSF_cA_ident}) individually, and we see no way to re-express such terms in a way that preserves these symmetries.

\begin{figure}
\begin{tabular}{cc}
\includegraphics[width=0.5\linewidth,clip]{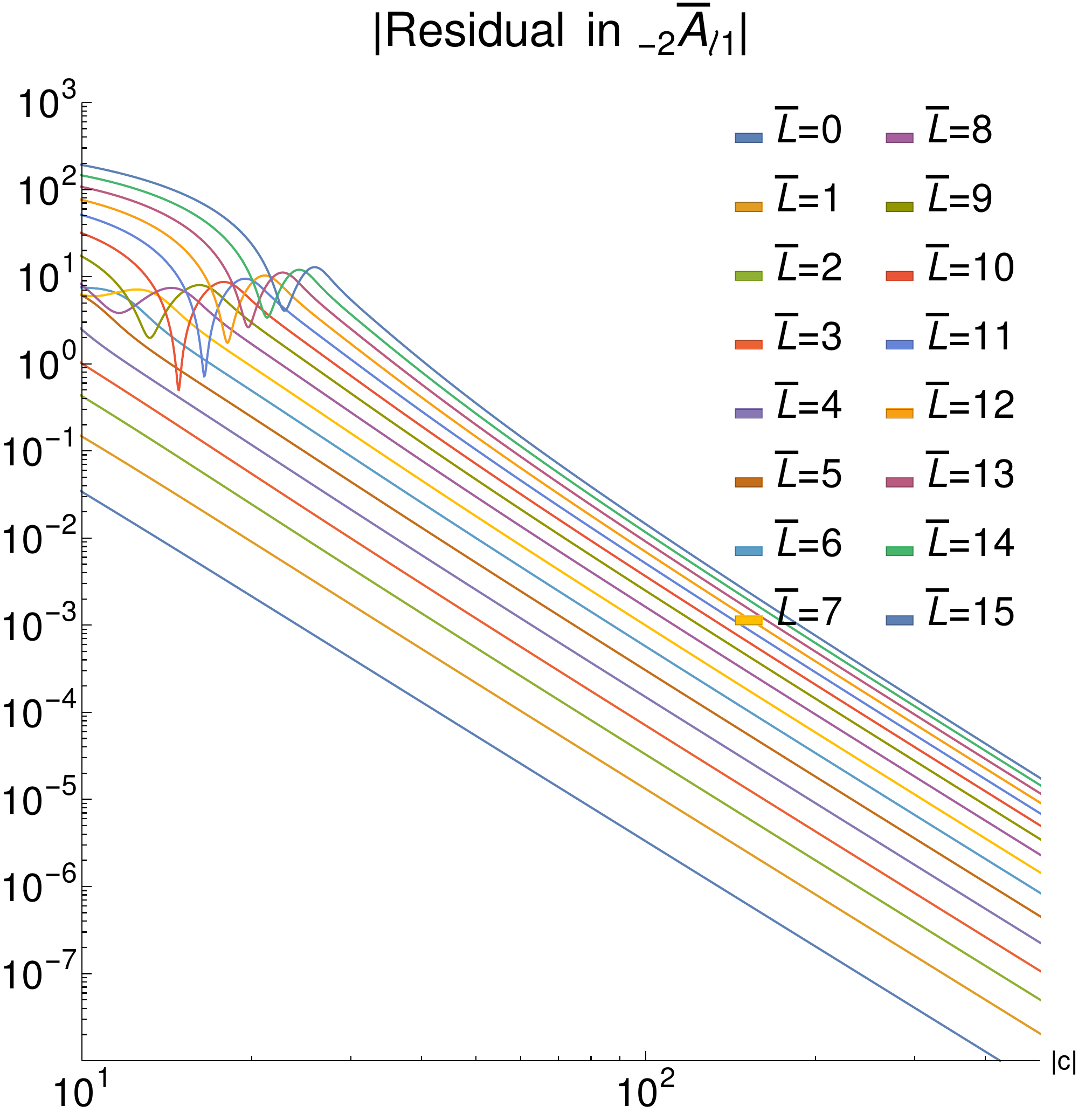} &
\includegraphics[width=0.5\linewidth,clip]{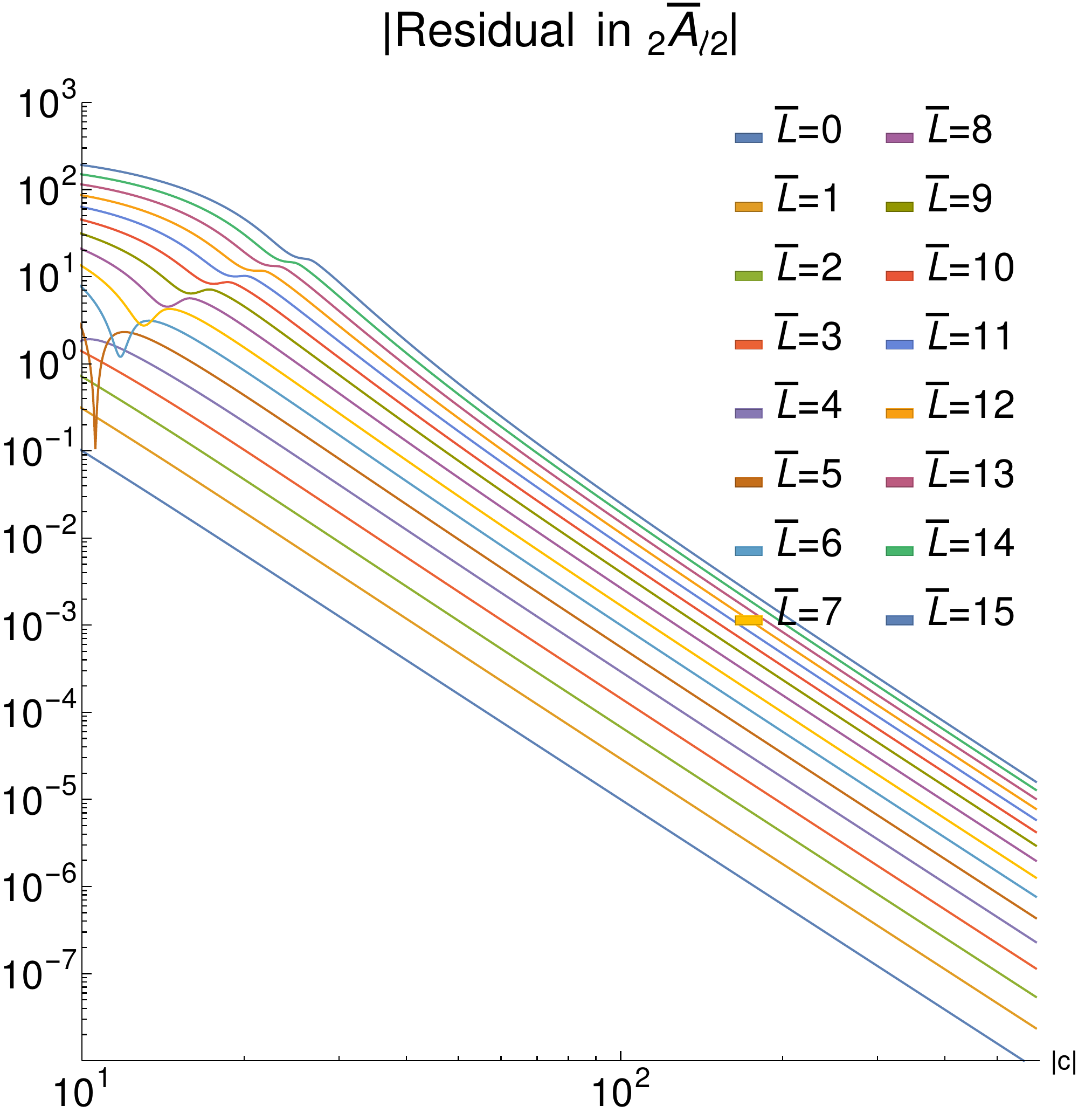} \\
\includegraphics[width=0.5\linewidth,clip]{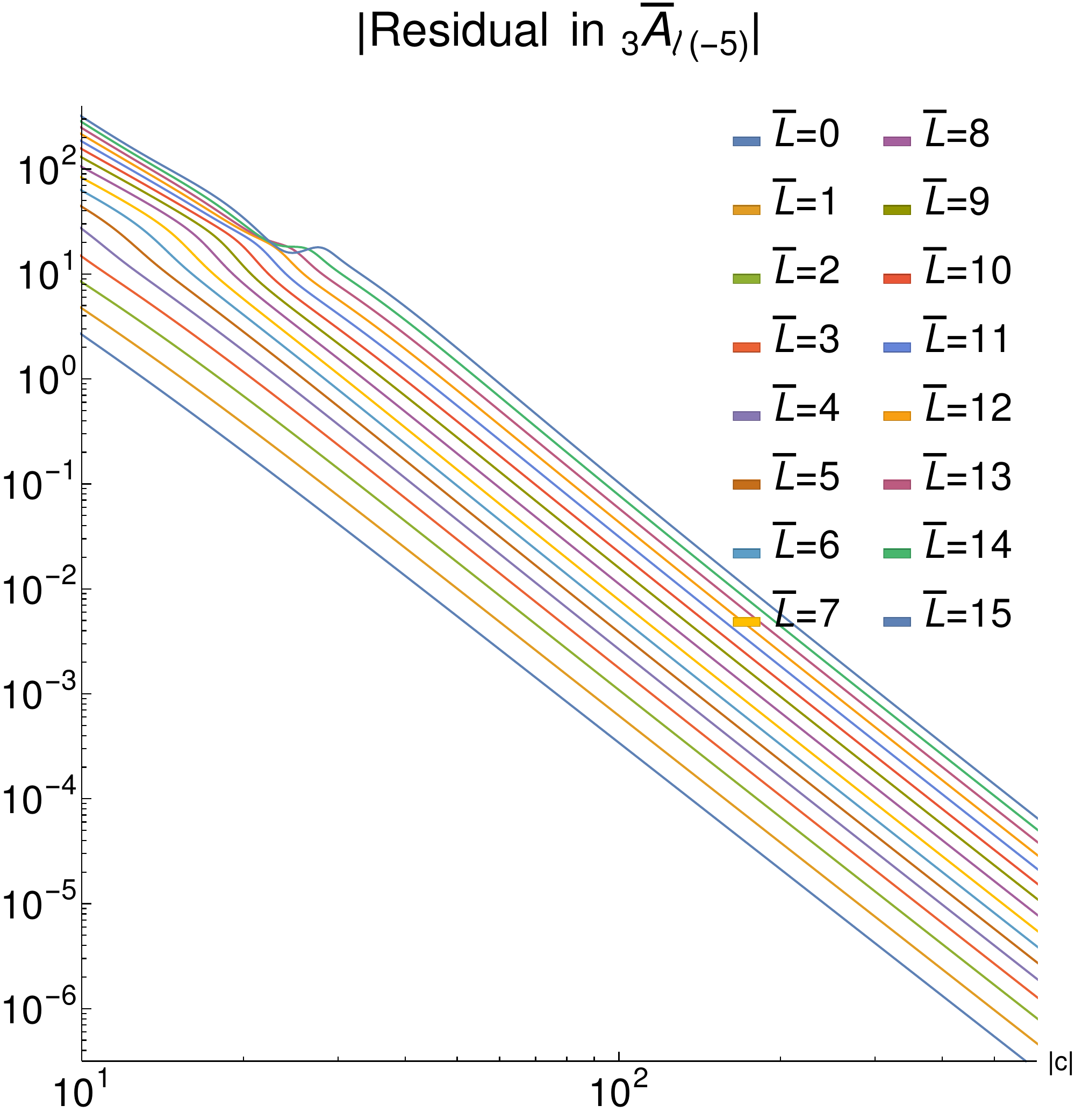} &
\includegraphics[width=0.5\linewidth,clip]{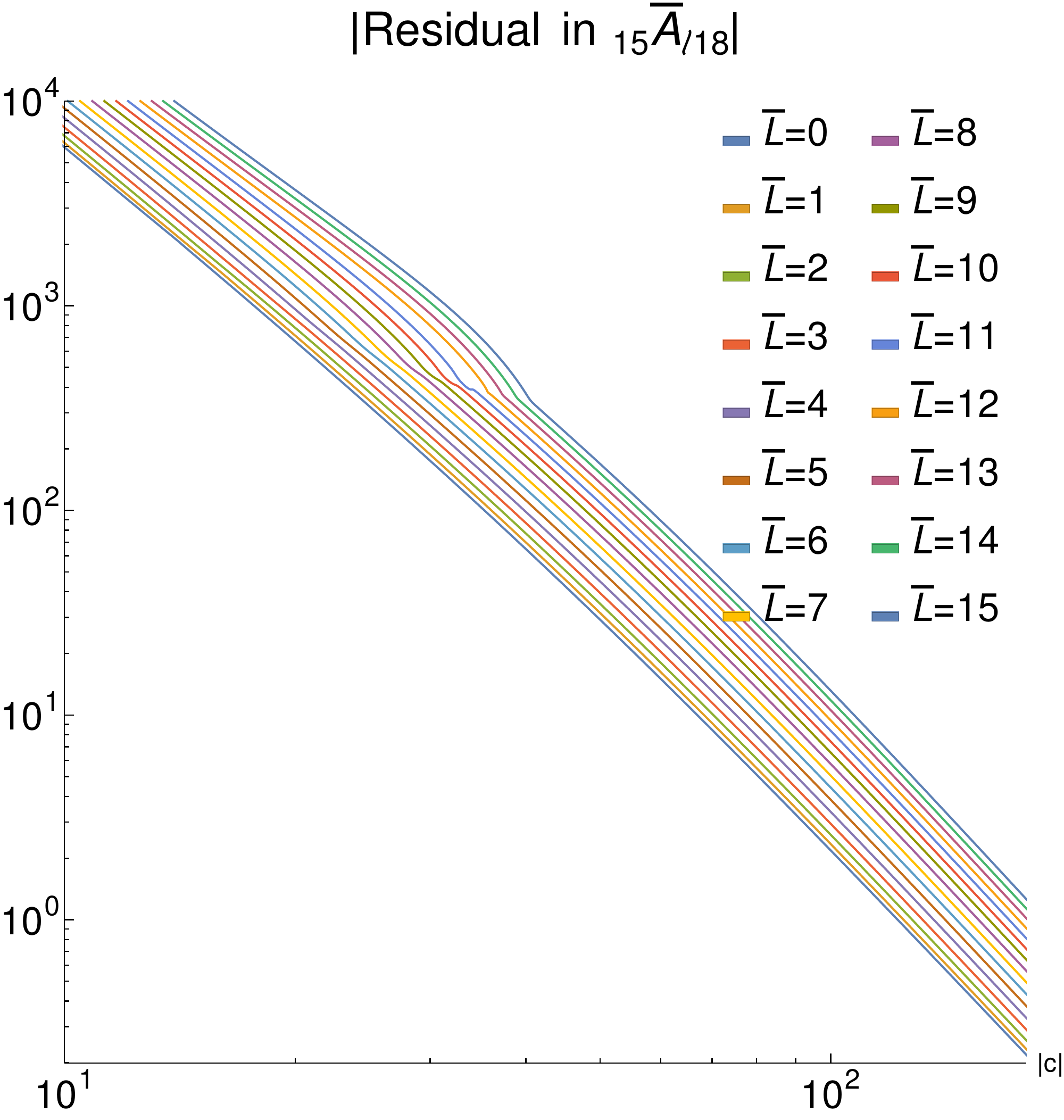}
\end{tabular}
\caption{\label{fig:loglognormresiduals} Log-log plots of the magnitude of the difference between the normal asymptotic fitting function of Eq.~(\ref{eqn:prolate normal solution}) and corresponding numerically computed values of $\scA{s}{\ell m}{-i|c|}$ versus $|c|$.  The $(m,s)$ pairs of $(1,-2)$, $(2,2)$, $(-5,3)$, and $(18,15)$ are each displayed in a separate plot showing the first $16$ normal sequences.  For large values of $|c|$ the slope of each line is approximately $-4$.}
\end{figure}

Finally, to demonstrate the fidelity of the the asymptotic fitting function given in Eq.~(\ref{eqn:prolate normal solution}), we show in Fig.~\ref{fig:loglognormresiduals} plots of the residuals obtained by subtracting Eq.~(\ref{eqn:prolate normal solution}) from the corresponding numerical data.
Four examples are displayed corresponding to the $(m,s)$ pairs of $(1,-2)$, $(2,2)$, $(-5,3)$, and $(15,18)$.  In each case, the absolute values of the residual for the first $16$ normal sequences versus $|c|$ are shown in a log-log plot.  In the asymptotic regime at large values of $|c|$, we find the slope of each line is very close to $-4$.  This is consistent with the fact that the fitting function in Eq.~(\ref{eqn:prolate normal solution}) is only defined through $\mathcal{O}(c^{-3})$ and demonstrates that the fitting function is correct through this order for a wide range of values for $s$, $m$, and $\bar{L}$.

\subsection{Anomalous Sequences}

The remainder of this section is devoted to exploring the behavior of the anomalous eigensolutions. Our investigation of the anomalous eigenvalue sequences was primarily driven by two goals.  The first was to try to determine a method to predict which sequences would exhibit anomalous behavior after transitioning to the asymptotic regime. The second goal was to try to find a power-series expansion in $c$ for these anomalous sequences, similar to Eq.~(\ref{eqn:prolate normal solution}).  We will show below, the behavior of the anomalous sequences is sufficiently complex that we have not been able to fully achieve either goal.  We hope that the results we have uncovered will inspire future analytical work that may further illuminate these very unusual sequences.

\begin{figure}
\includegraphics[width=\linewidth,clip]{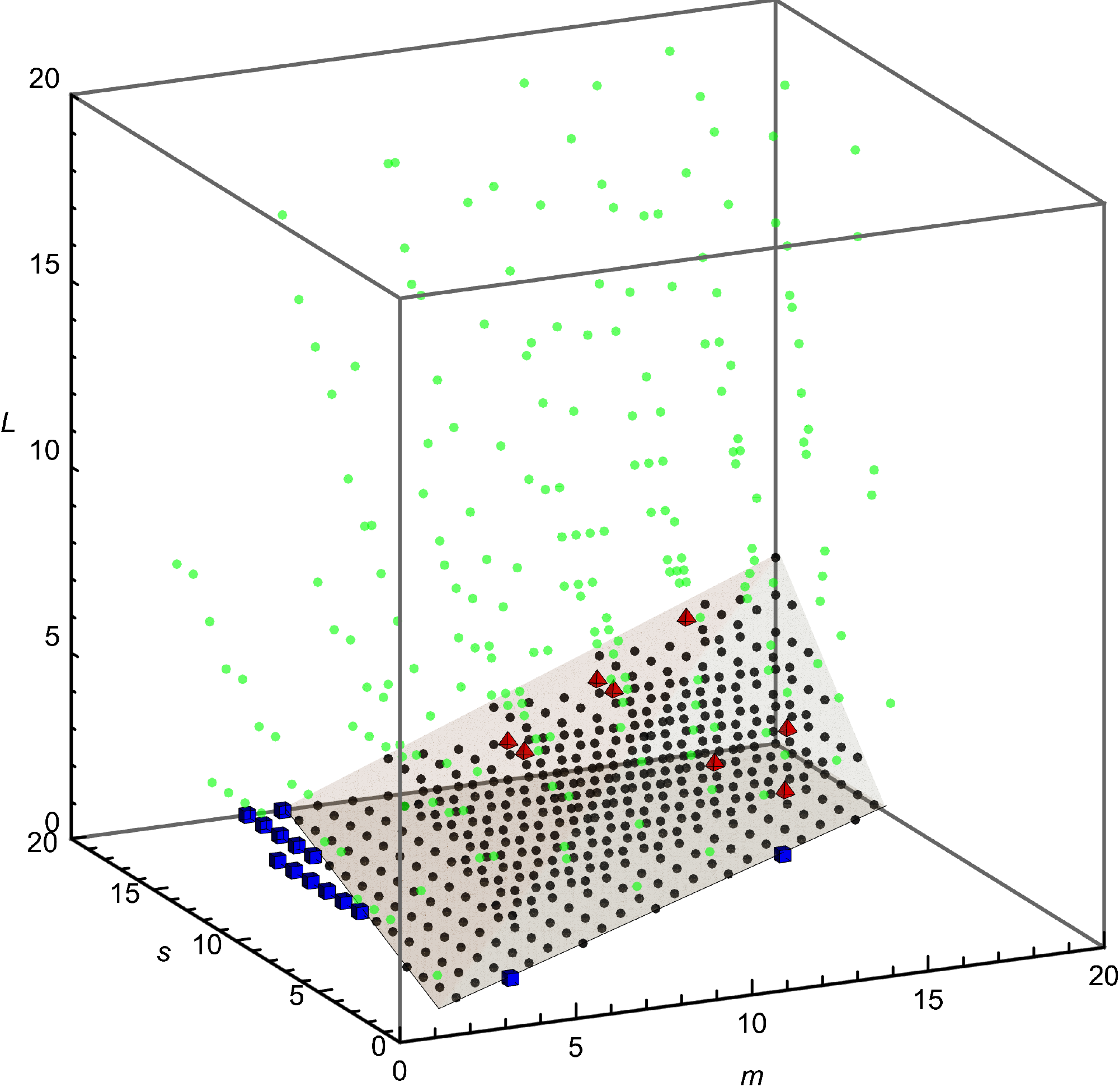}
    \caption{Combinations of $m$, $s$, and $L$ which yield anomalous sequences. The black dots and blue cubes designate sequences which have $L={}_sN_{\ell m}$ and are contiguous as described in the text.  The gray region displays the densely filled region $\mathcal{D}$ as defined by $L\ge0$, $14L\le11|s|-7|m|-10$, and $25L\le9|m|-2|s|-15$.  The red tetrahedra denote the 8 points within $\mathcal{D}$ that are not anomalous.  Finally, the light green dots outside of $\mathcal{D}$ denote the remaining non-contiguous anomalous sequences.  See Sec.~\ref{sec:anomalous data overview} for additional details.}
    \label{fig:anomalous sequence scatter plot}
\end{figure}

\subsubsection{Overview of anomalous data sets}\label{sec:anomalous data overview}
As described in Sec.~\ref{sec:Classes of Prolate Eigensolutions}, we examined each of our prolate sequences to determine which exhibited general anomalous behavior as defined by Eq.~(\ref{eqn:anomalous leading order}).  Out of all 15,376 prolate sequences we constructed, 775 of the sequences were anomalous.  Appendix~\ref{sec:anomalous table appendix} contains tables listing details for all of the 103 anomalous sequences from the high-resolution data for $0\le m\le10$ and $0\le s\le10$.  Figure~\ref{fig:anomalous sequence scatter plot} displays a scatter plot in $\{m,s,L\}$ showing sequences which are anomalous.  Recall that for the labels $\{m,s,L\}$, while $m$ and $s$ are fixed parameters in Eq.~(\ref{eqn:Angular Teukolsky Equation}), the $L$ label is simply a convenient choice with no unique definition.  In the limit that $c\to0$, $L$ simply labels the infinite set of solutions in order of increasing value of $\scA{s}{\ell{m}}{0}$ [see Eq.~(\ref{eqn:spherical sAlm})] where $\ell$ and $L$ are related by Eq.~(\ref{eqn:L def}).  Most of the anomalous sequences we have obtained with $m,s\ge0$ are represented within the plot, but a small number with large values of $L$ are omitted.

A careful inspection of Fig.~\ref{fig:anomalous sequence scatter plot} shows that there exists a region of $\{m,s,L\}$ space which is densely filled with anomalous sequences.  In this dense region, for given values of $m$ and $s$, the first few contiguous values of $L$ are anomalous implying that $L={}_sN_{\ell m}$.  For example, for $m=9$ and $s=10$, we have found $4$ anomalous sequences occurring at $L=\{0,1,3,78\}$ (see Table~\ref{table:anomalous data 2}).  For this case, the first $2$ sequences are anomalous so that ${}_{10}N_{10,9}=0$ and ${}_{10}N_{11,9}=1$.  In general, for $L=0$, a bounded area $\mathcal{A}$ defined by $11|s|\ge 7|m|+9$ and $|s|\le10|m|-24$ contains all sequences which are anomalous with $L=0$.  We caution that these limits are empirically obtained and are only known to apply for $|m|\le20$ and $|s|\le20$.  For $L\ge0$, we define a  region $\mathcal{D}$ which is bounded by $2$ additional planes: $14L\le11|s|-7|m|-10$ and $25L\le9|m|-2|s|-15$.  This regions is displayed in light gray in Fig.~\ref{fig:anomalous sequence scatter plot} and contains nearly all of the anomalous sequences with contiguous values of $L$.  If we consider only positive values of $m$ and $s$ no larger than $20$, then $\mathcal{D}$ contains $435$ points.  Unfortunately, $\mathcal{D}$ does not perfectly define the set of points which are contiguously anomalous.  There are $441$ anomalous points for which $L={}_sN_{\ell m}$.  These are displayed in Fig.~\ref{fig:anomalous sequence scatter plot} as $427$ black dots within $\mathcal{D}$, and $14$ blue cubes which lie outside $\mathcal{D}$.  Note that these $14$ anomalous sequences  are captured by the additional condition specified above specifically for $L=0$.  In addition, there are $8$ values of $\{m,s,L\}$ that lie within $\mathcal{D}$ but do not correspond to anomalous sequences.  These are displayed as the $8$ red tetrahedra in Fig.~\ref{fig:anomalous sequence scatter plot}.  These $8$ points are listed in Table~\ref{table:missing anomalous}.  The light green dots in Fig.~\ref{fig:anomalous sequence scatter plot} represent most of the remaining $243$ anomalous sequences for which ${}_sN_{\ell m}\ne L$.  
\begin{figure}[h]
\includegraphics[width=\linewidth,clip]{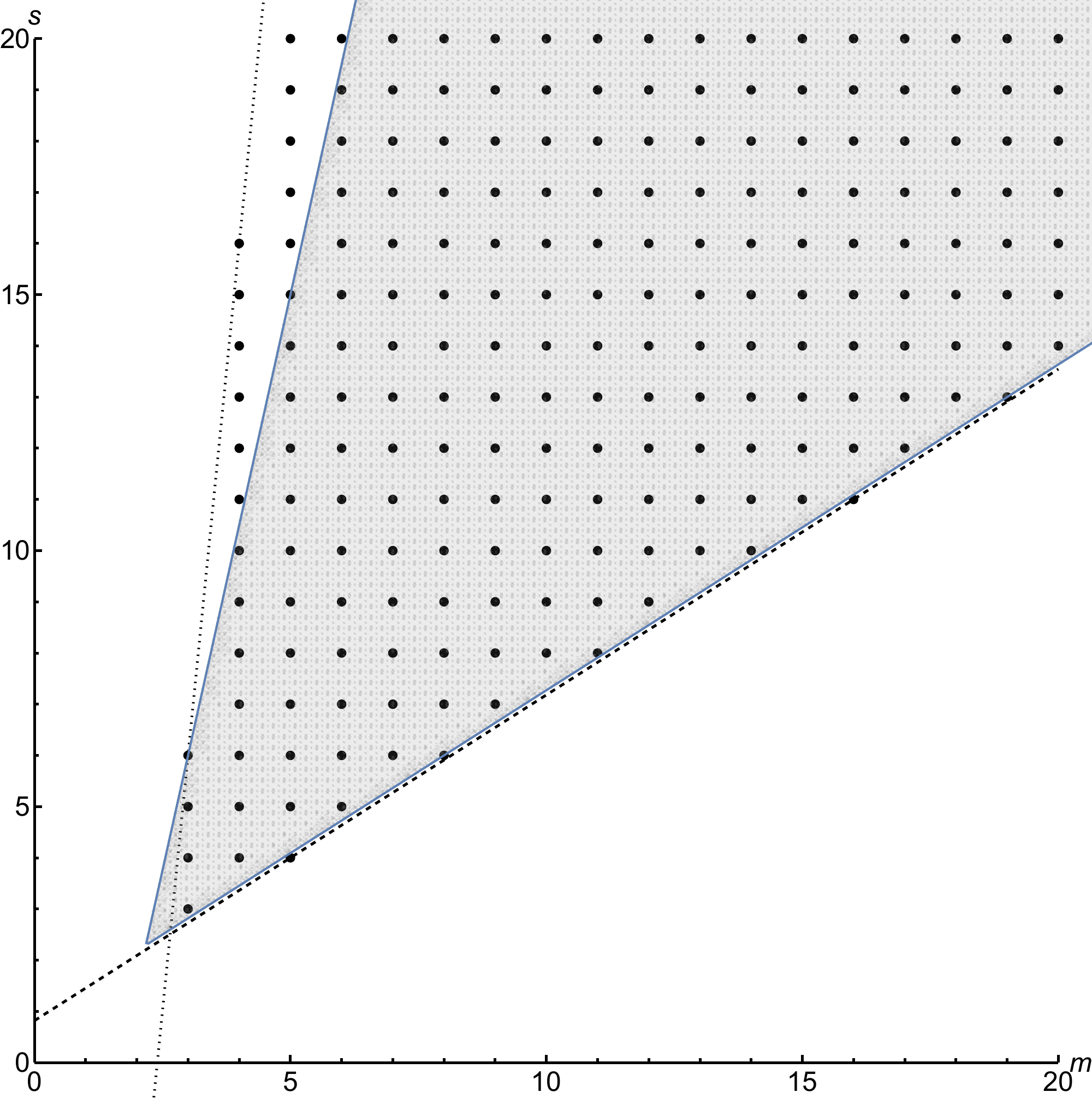}
    \caption{The black dots represent anomalous sequences for which $L={}_sN_{\ell m}=0$.  The gray region illustrates the intersection of region $\mathcal{D}$ (see Fig.~\ref{fig:anomalous sequence scatter plot}) with $L=0$.  The dashed and dotted black lines represent the boundaries of an area defined respectively by $0\le 11|s|-7|m|-9$ and $0\le10|m|-|s|-24$ within which all points are anomalous with $L={}_sN_{\ell m}=0$.  This area is defined as $\mathcal{A}$ in the text.  The points outside of the gray region are the sequences denoted by blue squares in Fig.~\ref{fig:anomalous sequence scatter plot}.}
    \label{fig:anomalous L0 slice}
\end{figure} 
Figure~\ref{fig:anomalous L0 slice} includes the boundaries of the area $\mathcal{A}$ in the $L=0$ plane of Fig.~\ref{fig:anomalous sequence scatter plot}.  The shaded region in this figure, which lies mostly within $\mathcal{A}$, is the intersection of the region $\mathcal{D}$ with the $L=0$ plane.  It is advantageous to omit the small number of anomalous sequences from region $\mathcal{D}$ in the $L=0$ plane because expanding the region to include them dramatically increases the number of points within $\mathcal{D}$ which are not anomalous.  Together, the area $\mathcal{A}$, region $\mathcal{D}$, and the points in Table~\ref{table:missing anomalous} provide a complete description for which of the first few contiguous sequences are  anomalous for given values of $m$ and $s$.  Unfortunately, the locations of the remaining non-contiguous anomalous sequences which occur at $L\ne {}_sN_{\ell m}$ are not so easy to predict.
\begin{figure}[h]
\includegraphics[width=\linewidth,clip]{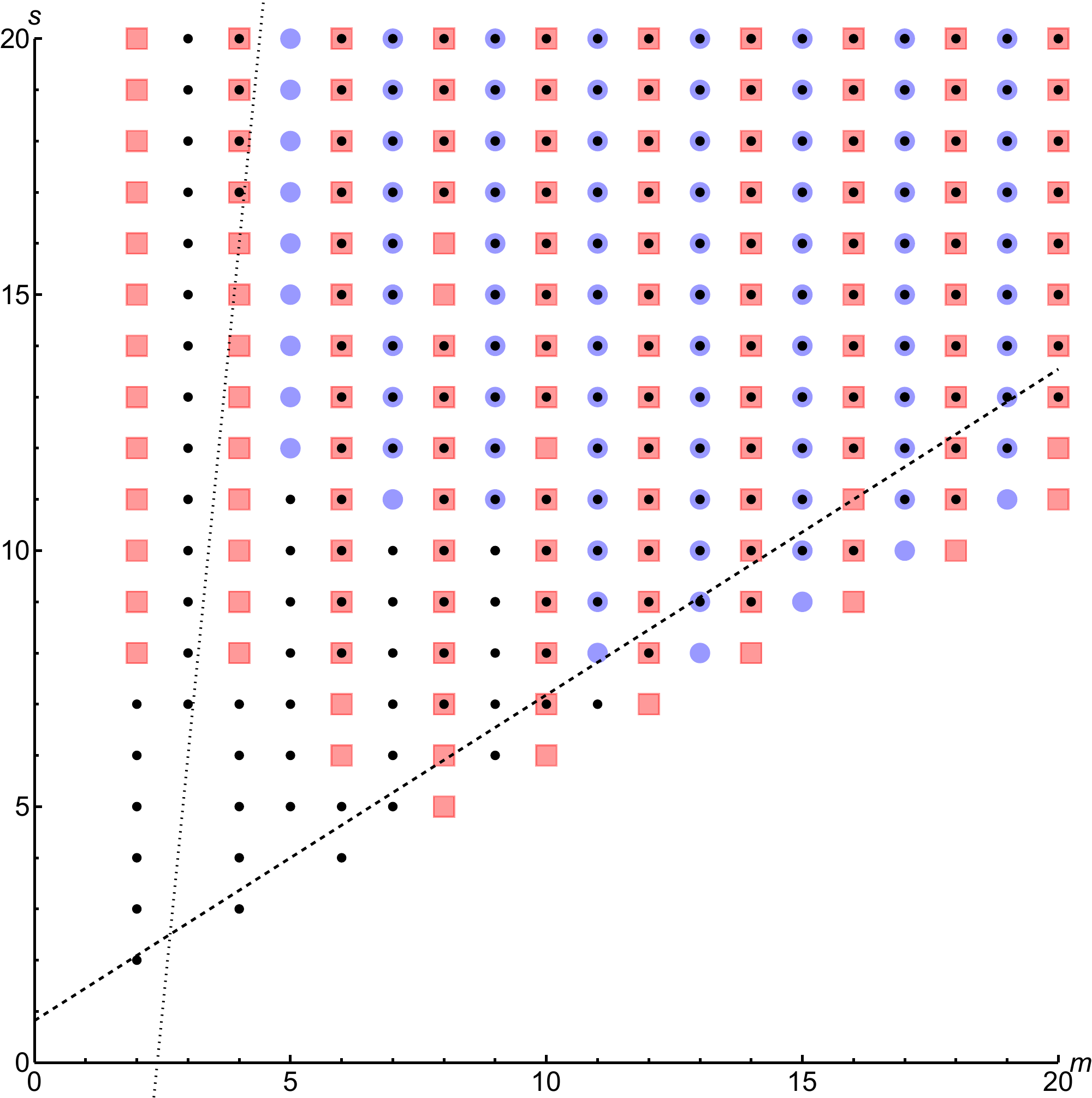}
    \caption{The black dots, light red squares, and light blue circles represent values of $|m|$ and $|s|$ at which at least one anomalous sequence with $L\ne{}_sN_{\ell m}$ could exist.  The black dots represent locations where at least one such anomalous sequence has been found within our limited data set. The light red squares represent locations where Type-1 anomalous sequences could exist, but are not present in our limited data set.  The light blue circles represent locations where Type-3 anomalous sequences should exist, but are not present in our limited data set.  The dashed and dotted black lines represent the boundaries of the area $\mathcal{A}$ defined in the text (see also the caption for Fig.~\ref{fig:anomalous L0 slice})
    }
    \label{fig:all non-contiguous anom}
\end{figure} 
As shown by the black dots in Fig.~\ref{fig:all non-contiguous anom} they appear at values of $L>0$ for values of $m$ and $s$ that can be outside of the area $\mathcal{A}$ defined for $L=0$.  The black dots in Fig.~\ref{fig:all non-contiguous anom} may not be complete since we have only explored a limited range of $L$.  The light red and blue symbols in Fig.~\ref{fig:all non-contiguous anom} represent locations we could expect additional non-contiguous anomalous sequences to exist, but which are not present in our limited data set.  Why we expect anomalous sequences in these locations will be discussed in Sec.~\ref{sec:predict anomalous}.

\begin{table}
 \begin{tabular}{cc|c}
$|m|$ & $|s|$ & $L$ \\
\hline
11 & 16 & 2 \\
11 & 17 & 2 \\
14 & 17 & 3 \\
14 & 18 & 3 \\
15 & 13 & 2 \\
17 & 13 & 1 \\
17 & 19 & 4 \\
18 & 15 & 2
\end{tabular}
\caption{Points with the region $\mathcal{D}$ and with $|m|\le20$ and $|s|\le20$ which do not correspond to anomalous sequences.}
\label{table:missing anomalous}
\end{table}

\subsubsection{Asymptotic form for the anomalous eigenvalues}\label{sec:anomalous form}
We initially recognized the presence of the anomalous asymptotic sequences of eigenvalues by the fact that they grow as $|c|^2$ rather than linearly in $|c|$.  This behavior is similar to that in Eq.~(\ref{eqn:oblate solution all}) for the oblate sequences where $c$ is real.  We find in fact that, with 2 small redefinitions, Eq.~(\ref{eqn:oblate solution all}) fits all of the anomalous sequences well up to $\mathcal{O}(c^{-1})$ if we simply insert a purely imaginary value of $c$.  Interestingly, for some anomalous sequences such as the $s=m=4$ sequence with $L=0$, Eq.~(\ref{eqn:oblate solution all}) fits well up $\mathcal{O}(c^{-5})$ which is the first undefined term in the analytic expansion.  On the other hand, the $s=m=4$ sequence with $L=28$ disagrees with the analytic expansion at $\mathcal{O}(c^{-1})$ and the $s=m=3$ anomalous sequence with $L=0$ disagrees with the analytic expansion at $\mathcal{O}(c^{-3})$.

In order for Eq.~(\ref{eqn:oblate solution all}) to apply to the anomalous asymptotic prolate sequences, we must let $L\to \hat{L}=\,_{s}N_{\ell m}$ in a similar way in which we redefined $L\to \bar{L}$ for the normal asymptotic prolate sequences.  We must further modify $_{s}q_{\ell m}$ for the cases when $m<0$ by defining $_{s}q_{\ell m}\to-_{s}q_{\ell(-m)}$ for $m<0$.  Finally, we have only found anomalous sequences for the cases when $\ell < \max\left({_s\ell_m},{_{-s}\ell_m}\right)$ in Eq.~(\ref{eqn:sqlm all}).  This means that we can replace the definition of $_{s}q_{\ell m}$ with a greatly simplified version
\begin{align}\label{eqn:anomalous sqlm}
        _{s}\hat{q}_{\ell m}=\left\{\begin{array}{lc}
            \sign(m)(2\hat{L}+\bigl||m| - |s|\bigr| - |s| + 1)&\quad m\ne0\\
            2\hat{L}+1&\quad m=0
            \end{array}\right..
\end{align}
We now define the base fitting function, in the asymptotic regime, for the anomalous prolate eigenvalue sequences as Eq.~(\ref{eqn:oblate solution all}) with $_{s}q_{\ell m}\to{}_{s}\hat{q}_{\ell m}$, and denote it as $\scAanom{s}{\ell m}{c}$.  We also find that, because $_{s}\hat{q}_{\ell m}=-{}_{s}\hat{q}_{\ell(-m)}$, our base fitting function satisfies all of the fundamental symmetries outlined in Eqs.~(\ref{eq:swSF_sA_ident}), (\ref{eq:swSF_mcA_ident}), and (\ref{eq:swSF_cA_ident}), and thus is valid for both positive and negative imaginary values of $c$.  
Following the approach for the asymptotic normal eigensolutions, we redefine all asymptotic anomalous eigensolutions in terms of the hatted quantities $\scAanom{s}{\ell m}{c}$, $\swSanom{s}{\ell m}{c}$, ${}_{s}\hat{q}_{\ell m}$, and $\hat{L}$; and the values of $\ell$ for all hatted quantities is computed using Eq.~(\ref{eqn:L def}) with $L$ replaced by $\hat{L}$.

\begin{figure}[h]
\begin{tabular}{cc}
\includegraphics[width=0.5\linewidth,clip]{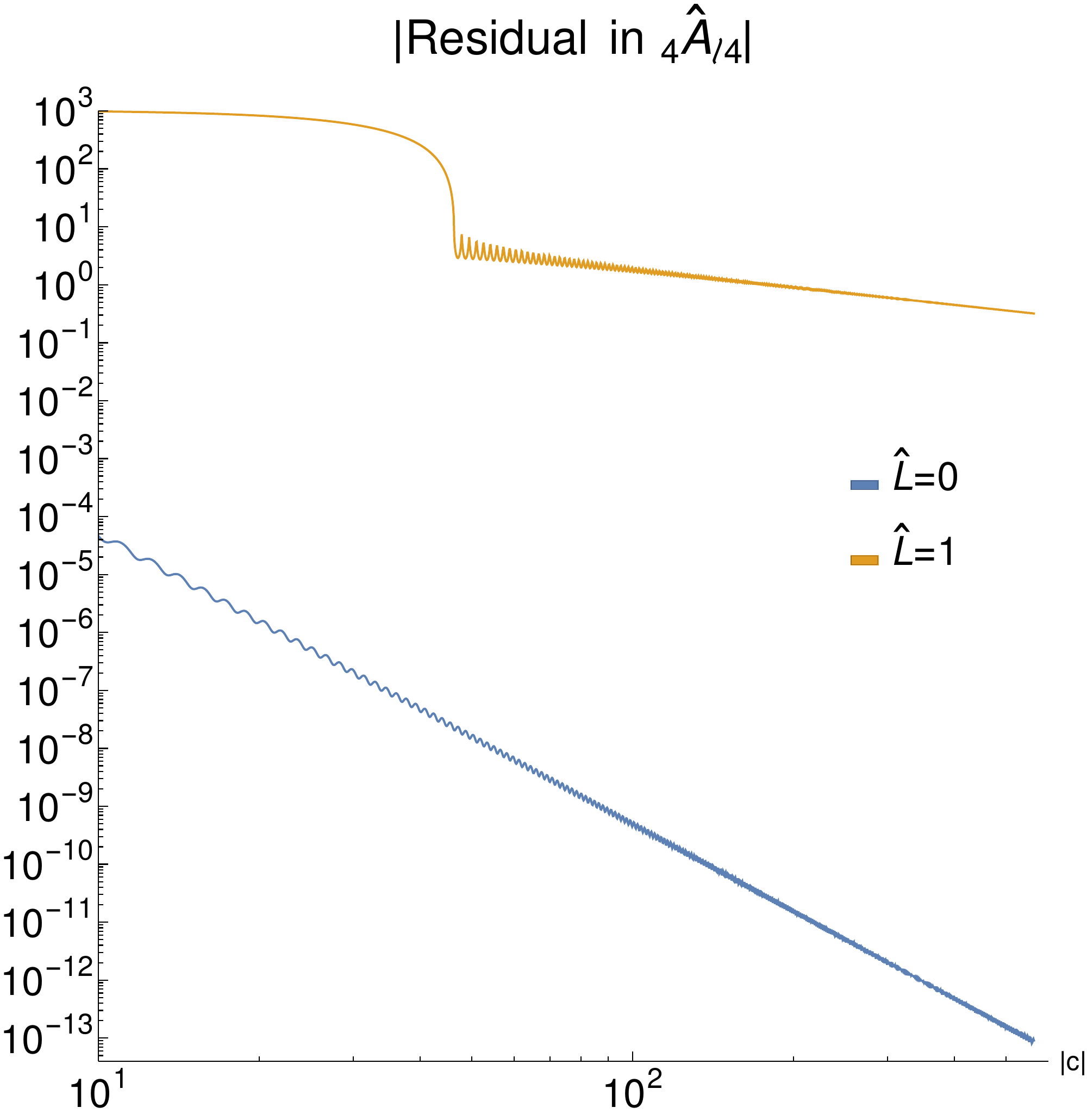} &
\includegraphics[width=0.5\linewidth,clip]{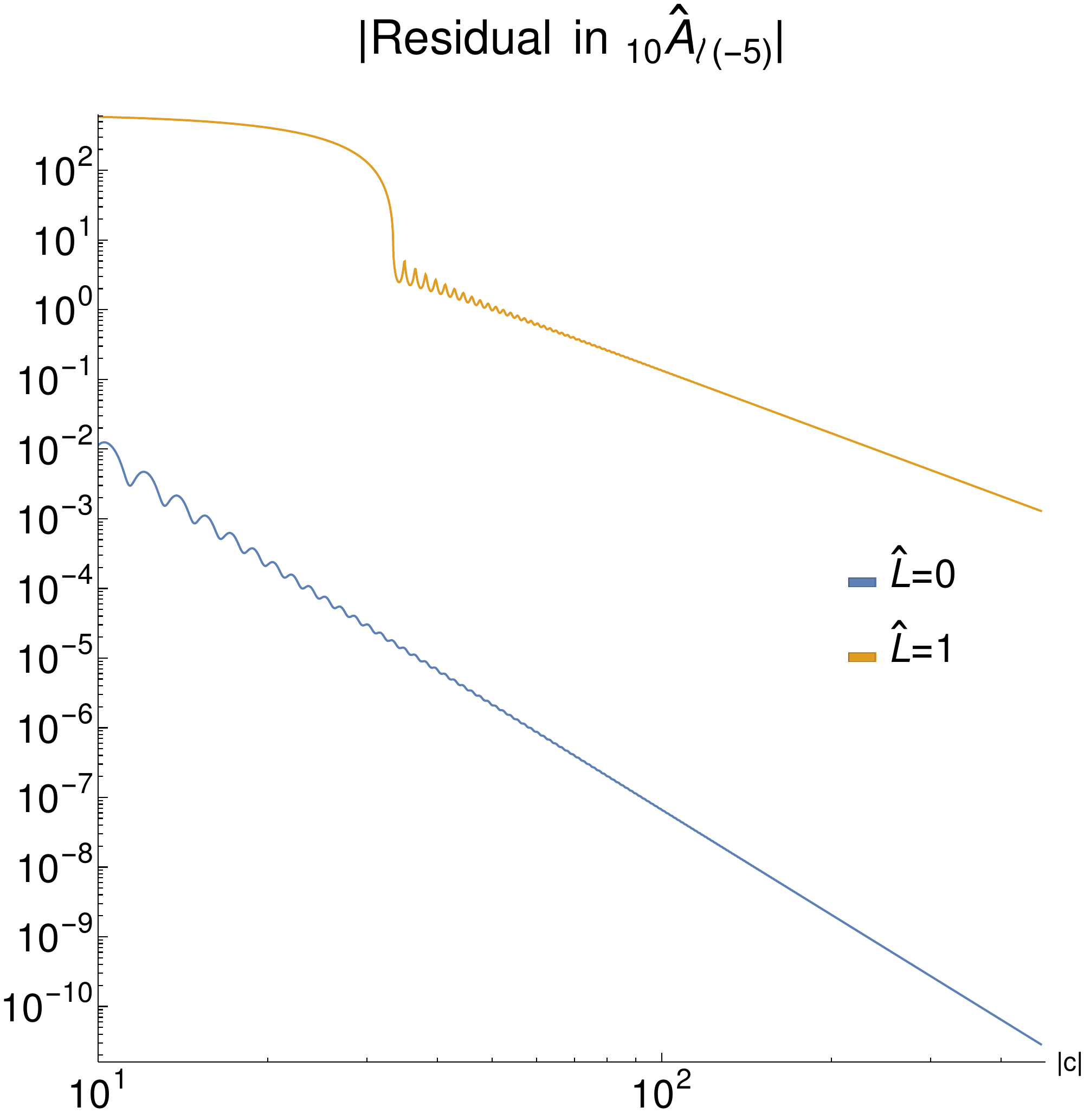} \\
\includegraphics[width=0.5\linewidth,clip]{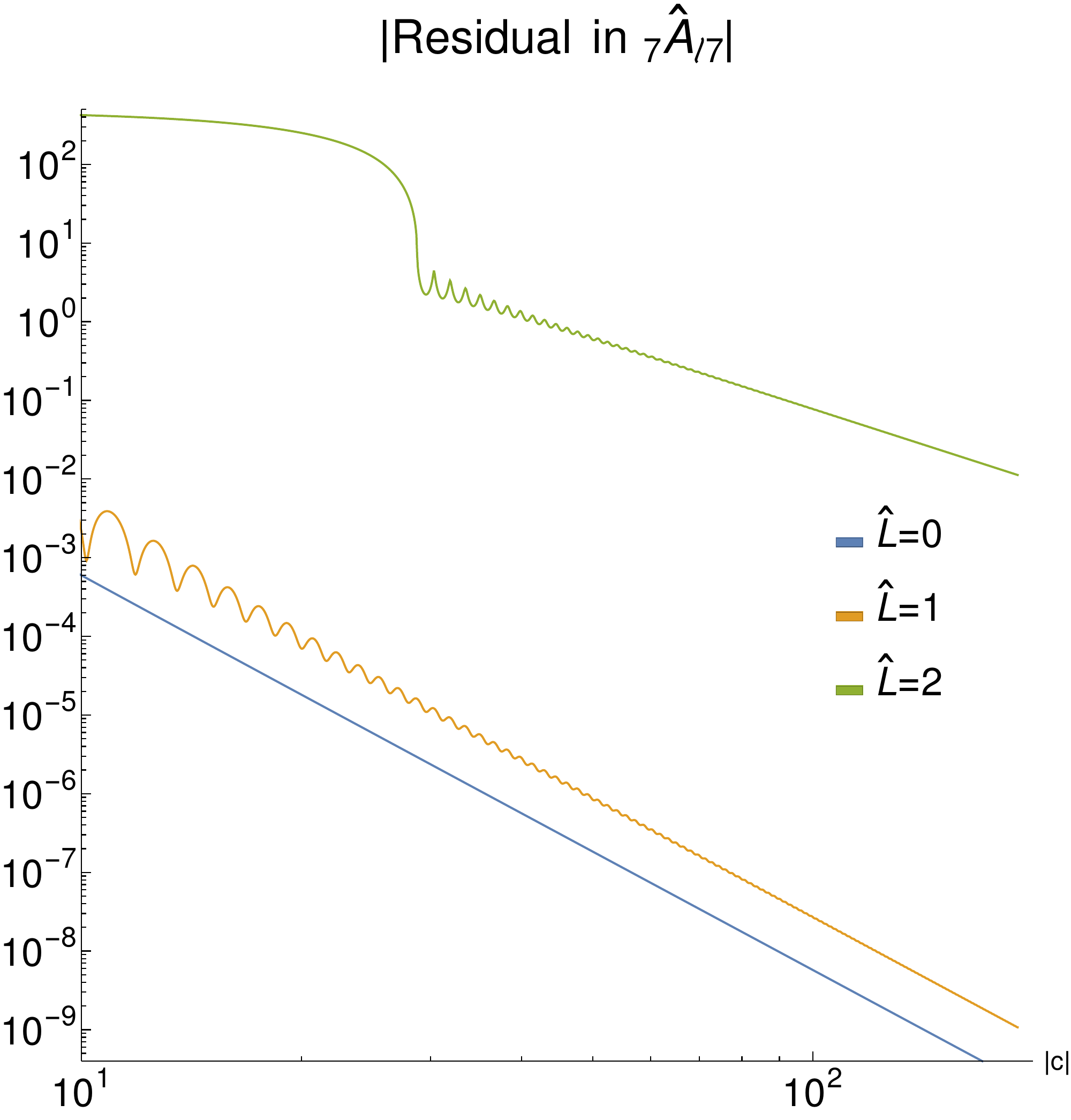} &
\includegraphics[width=0.5\linewidth,clip]{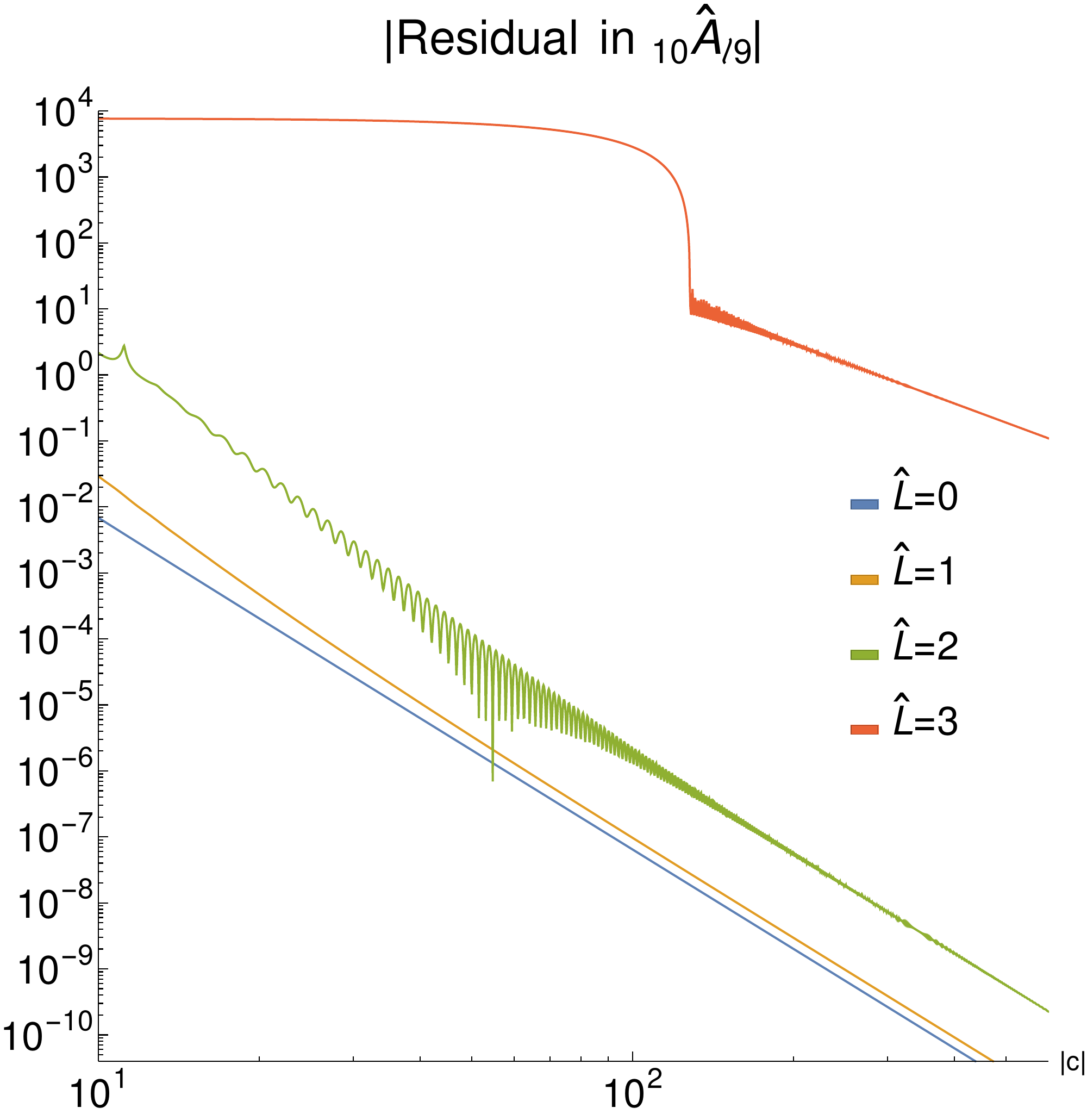}
\end{tabular}
\caption{\label{fig:logloganomresiduals} Log-log plots of the magnitude of the difference between the anomalous asymptotic base fitting function of Eq.~(\ref{eqn:oblate solution all}) with $_{s}q_{\ell m}$ replaced by $_{s}\hat{q}_{\ell m}$ as given in Eq.~(\ref{eqn:anomalous sqlm}) and corresponding numerically computed values of $\scAanom{s}{\ell m}{-i|c|}$ versus $|c|$.  The $(m,s)$ pairs of $(4,4)$, $(-5,10)$, $(7,7)$, and $(9,10)$ are each displayed in a separate plot showing all known anomalous sequences.  The asymptotic slope for the residual of $\scAanom{4}{5,4}{-i|c|}$ is $-1$.  The asymptotic slopes for $\scAanom{10}{11(-5)}{-i|c|}$, $\scAanom{7}{9,7}{-i|c|}$, and $\scAanom{10}{13,9}{-i|c|}$ are $-3$.  The asymptotic slopes for all remaining anomalous sequences are $-5$.}
\end{figure}

Figure~\ref{fig:logloganomresiduals} illustrates the degree to which the base fitting function for the anomalous sequences, $\scAanom{s}{\ell m}{c}$, agrees with the numerical data in the asymptotic regime.  Here we plot the magnitude of the residual obtained by taking the difference between the anomalous base fitting function and the corresponding numerical values for $\scAanom{s}{\ell m}{c}$.  Each plot in the figure shows all of the anomalous sequences for a particular $(m,s)$ pair.
In the upper left plot we show the residuals for the 2 anomalous sequences with $m=s=4$.  As mentioned above, the sequence with $L=\hat{L}=0$ has a slope of $-5$ in this plot showing that the entire fitting function through $\mathcal{O}(c^{-4})$ agrees with the numerical data.  However, the sequence with $L=28$ ($\hat{L}=1$) has an asymptotic slope of $-1$.  This shows that the at $\mathcal{O}(c^{-1})$ the fitting function does not fully account for the behavior seen in the numerical data.  If we look separately at the real and imaginary parts of the residual we see that there is error in both components.  This is interesting since, for purely imaginary $c$, Eq.~(\ref{eqn:oblate solution}) suggests that the contributions at each order in $c$ should be either real or imaginary.  We find this to be true in both the base fitting function and the numerical data until we reach the order at which the base fitting function fails to work.  At this order, the numerical data now have both real and imaginary parts.  
\begin{figure}[h]
\begin{tabular}{cc}
\includegraphics[width=0.5\linewidth,clip]{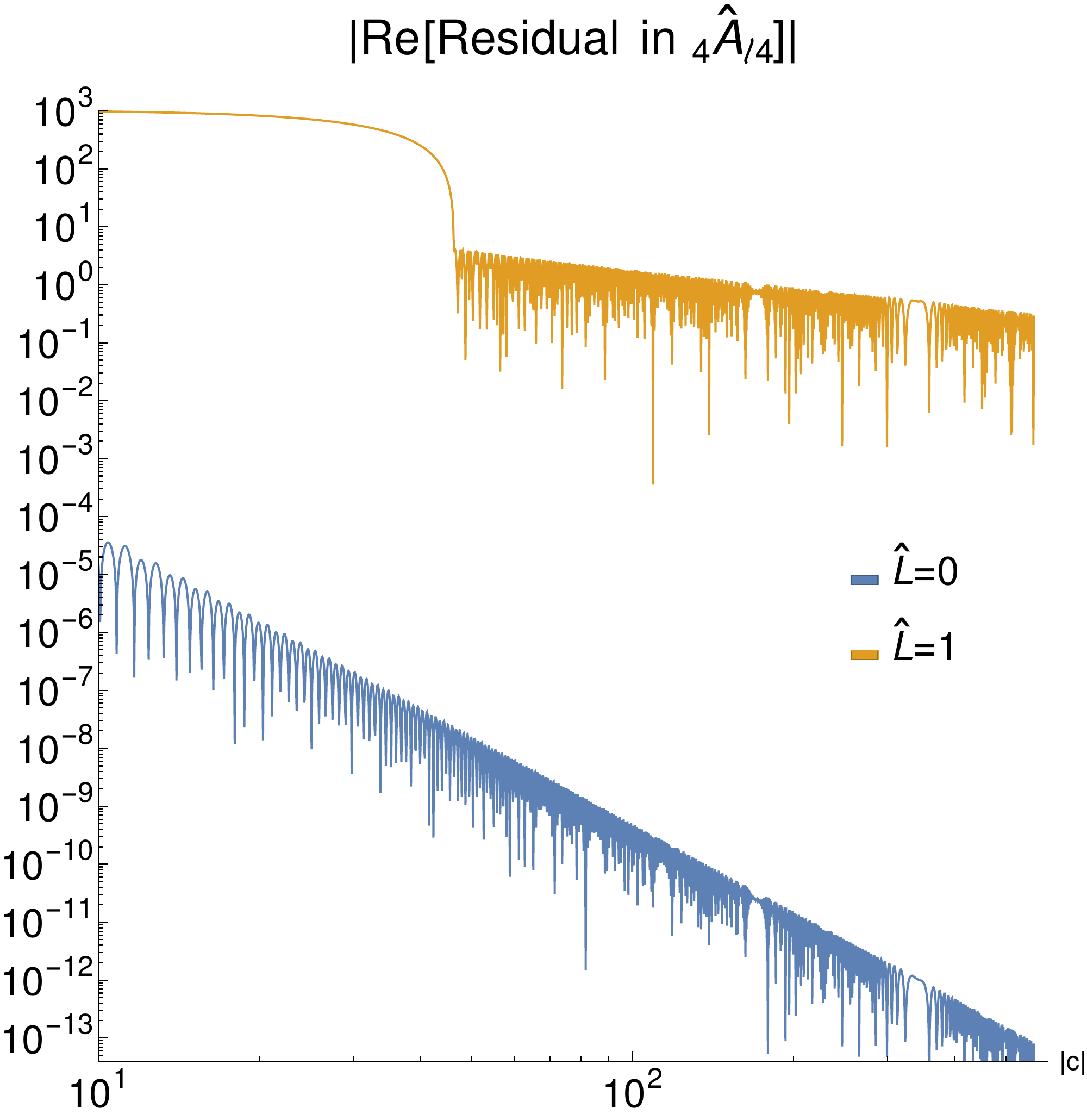} &
\includegraphics[width=0.5\linewidth,clip]{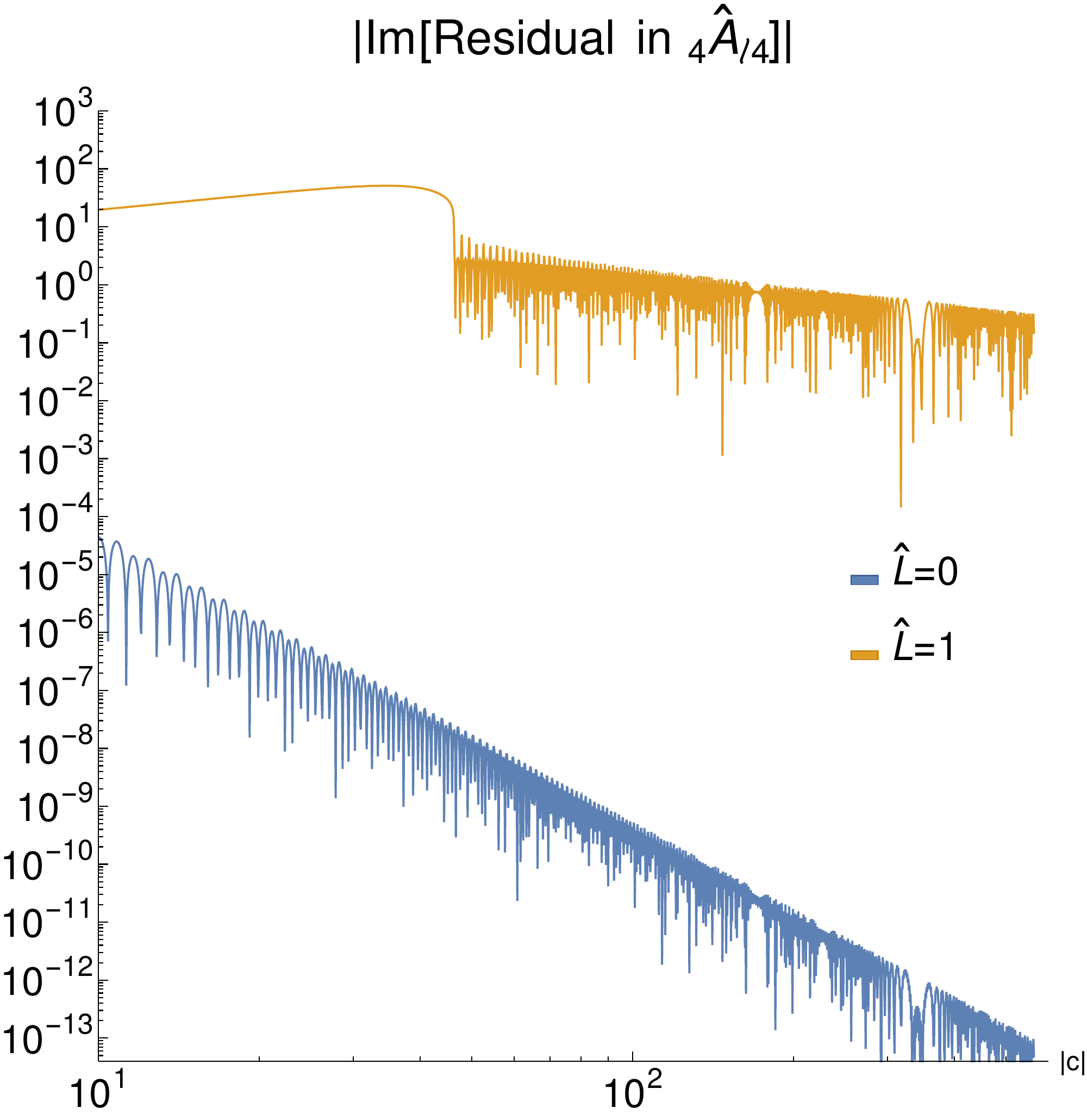} 
\end{tabular}
\caption{\label{fig:loglogReImanomresiduals} Log-log plots of the magnitudes of the real and imaginary parts of the difference between the anomalous asymptotic base fitting function of Eq.~(\ref{eqn:oblate solution all}) with $_{s}q_{\ell m}$ replaced by $_{s}\hat{q}_{\ell m}$ as given in Eq.~(\ref{eqn:anomalous sqlm}) and corresponding numerically computed values of $\scAanom{s}{\ell m}{-i|c|}$ versus $|c|$ for the case of $m=s=4$.  The asymptotic slopes of both the real and imaginary parts of the residuals of $\scAanom{4}{5,4}{-i|c|}$ are $-1$, while the asymptotic slopes of both the real and imaginary parts for $\scAanom{4}{44}{-i|c|}$ are $-5$.}
\end{figure}
For example, Fig.~\ref{fig:loglogReImanomresiduals} shows separately the real and imaginary parts of the residuals for the two anomalous sequences for $m=s=4$.
Equation~(\ref{eqn:oblate solution all}) says that the term at $\mathcal{O}(c^{-1})$ should be purely imaginary.  For the anomalous sequences with $m=s=4$, the numerical solution for the $L=\hat{L}=0$ sequences is purely imaginary at $\mathcal{O}(c^{-1})$, while the numerical data for the $L=28$ ($\hat{L}=1$) sequence includes a non-vanishing real part at $\mathcal{O}(c^{-1})$.  Similarly, at $\mathcal{O}(c^{-5})$, Eq.~(\ref{eqn:oblate solution all}) suggests that the fitting function should again be purely imaginary.  But, the residual for both the real and imaginary parts of the $L=\hat{L}=0$ sequence are non-vanishing at this order.  In all cases, we find that the slope of the magnitude of the numerical residual is either $-1$, $-3$, or $-5$.

Examining the behavior of the residual between the numerical data and the anomalous base fitting function for all of the anomalous sequences reveals a clear pattern.  In Tables~\ref{table:anomalous data 1} and \ref{table:anomalous data 2} in Appendix~\ref{sec:anomalous table appendix}, the column labeled by $n$ designates the slope of the magnitude of the residual in a log-log plot and thus the order in $c$ at which the numerical data deviates from the behavior of the anomalous base fitting function.  We find this behavior to be governed by the leading order behavior of the imaginary part of the separation constant $\scAanom{s}{\ell m}{c}$.

For the oblate case with purely imaginary values of $c$, the leading order behavior in the imaginary part of $\scAanom{s}{\ell m}{c}$ [see Eq.~(\ref{eqn:oblate solution})] is at linear order in $c$.  The coefficient of this term is $2{}_s\hat{q}_{\ell m}$.  Tables~\ref{table:anomalous data 1} and \ref{table:anomalous data 2} include a column labeled by $2{}_s\hat{q}_{\ell|m|}+1$ which lists the value of this combination\footnote{We note that $-2|{}_s\hat{q}_{\ell m}|+1$ will also agree with the slope of the residual for anomalous prolate solutions.  However, the value of the combination given in the text can also server as discriminant for whether a given sequence can be anomalous.}.  We note that when $n\ge-3$ it agrees with $2{}_s\hat{q}_{\ell|m|}+1$.  Whenever $n=-5$, we also find that $2{}_s\hat{q}_{\ell|m|}+1 \le -5$.  Because the base fitting function only extends to $\mathcal{O}(c^{-4})$ we cannot expect to find $n<-5$.  We conjecture that $2{}_s\hat{q}_{\ell|m|}+1$ should correctly give the order in $c$ at which the anomalous base fitting function deviates from the true asymptotic behavior of the anomalous sequences.

Further evidence supporting this conjecture can be found by examining the real and imaginary parts of the residual.  Tables~\ref{table:anomalous data 1} and \ref{table:anomalous data 2} include a column labeled by $\{\text{Re}[n],\text{Im}[n]\}$ where $\text{Re}[n]$ and $\text{Im}[n]$ designates the slope of the real and imaginary parts of the residual in a log-log plot and thus the order in $c$ at which the real and imaginary parts of the numerical data deviate from the behavior of the anomalous base fitting function.  When $n=2{}_s\hat{q}_{\ell|m|}+1\ge-5$, we find that both $\text{Re}[n]=n$ and $\text{Im}[n]=n$.  This further illustrates the point made previously that when anomalous asymptotic sequences deviate from the base fitting function at order $n$, the true fitting function should gain a term at $\mathcal{O}(c^n)$ that is in general complex.  Looking at all of the sequences for which $2{}_s\hat{q}_{\ell|m|}+1 \le -7$ we find that $\text{Re}[n]=-6$.  This suggests that, for these sequences, if the base fitting function were known at $\mathcal{O}(c^{-5})$, this term would agree with the numerical data.  There is one exception seen in Table~\ref{table:anomalous data 2} in the case with $m=9$, $s=10$, $\hat{L}=2$ for which $\text{Re}[n]\approx-7$.  This case could represent a special case where the term  at $\mathcal{O}(c^{-6})$ is very small, or could simply be due to the difficulty in extracting the power-law falloff of the envelope of a highly oscillatory function.

Because a true fitting function for the prolate anomalous separation constant would necessarily be different for different values of $2{}_s\hat{q}_{\ell|m|}+1$, we now define a Type for each anomalous sequence based on the value of $2{}_s\hat{q}_{\ell|m|}+1$ for that sequence.  So, all sequences with $2{}_s\hat{q}_{\ell|m|}+1=-1$ are designated as Type-1 anomalous sequences.  All sequences with $2{}_s\hat{q}_{\ell|m|}+1=-3$ are designated as Type-3 anomalous sequences, and so on.

\subsubsection{Predicting the existence of anomalous sequences}
\label{sec:predict anomalous}
For given values of $m$ and $s$, we can predict with reasonable, but not absolute confidence, whether or not anomalous sequences will exist, and also how many anomalous sequences should exist.  Anomalous sequences should exist whenever $2{}_s\hat{q}_{\ell|m|}+1<0$.  The first such anomalous sequence will be labeled $\hat{L}=0$ and so will have a value of $\ell=\max(|m|,|s|)$ and the smallest (most negative) value of $2{}_s\hat{q}_{\ell|m|}+1$.  The next anomalous sequence, if it exists, will have $\hat{L}=1$ and $2{}_s\hat{q}_{\ell|m|}+1$ will increases by 4.  Additional anomalous sequences should exist with $\hat{L}$ increasing until the set of anomalous sequences terminates with $2{}_s\hat{q}_{\ell|m|}+1=-3$ or $-1$.

Figure~\ref{fig:all non-contiguous anom} helps to illustrate, in part, why we do not claim to be able to predict with certainty the existence of anomalous sequences for given $m$ and $s$.  The red squares in Fig.~\ref{fig:all non-contiguous anom} represent the locations where Type-1 anomalous sequences could exist, but have not been found within our set of numerical sequences.  Each of these missing sequences correspond to the largest value of $\hat{L}$ for each given $m$ and $s$ where they may exist, and more importantly may correspond to a very large value of $L$.  As discussed in Sec.~\ref{sec:prolate data sets}, for most values of $m$ and $s$, we only constructed sequences up to $L\sim15$.  In only a few cases did we construct sequences out to $L\sim150$.  Because of this, these missing sequences may simply be due to the fact that we did not extend our search to large enough $L$ to find them.  The blue circles in Fig.~\ref{fig:all non-contiguous anom} represent the locations where Type-3 anomalous sequences could exist, but have not been found within our set of numerical sequences.  Because the anomalous type decreases by 4 as we increase $\hat{L}$ for given $m$ and $s$, the Type-3 anomalous sequences correspond to the largest value of $\hat{L}$ for each given $m$ and $s$ where blue circles are displayed.  And, as with the missing Type-1 sequences, these missing sequences may simply be due to the fact that we did not extend our search to large enough $L$ to find them.

\begin{figure}[h]
\includegraphics[width=\linewidth,clip]{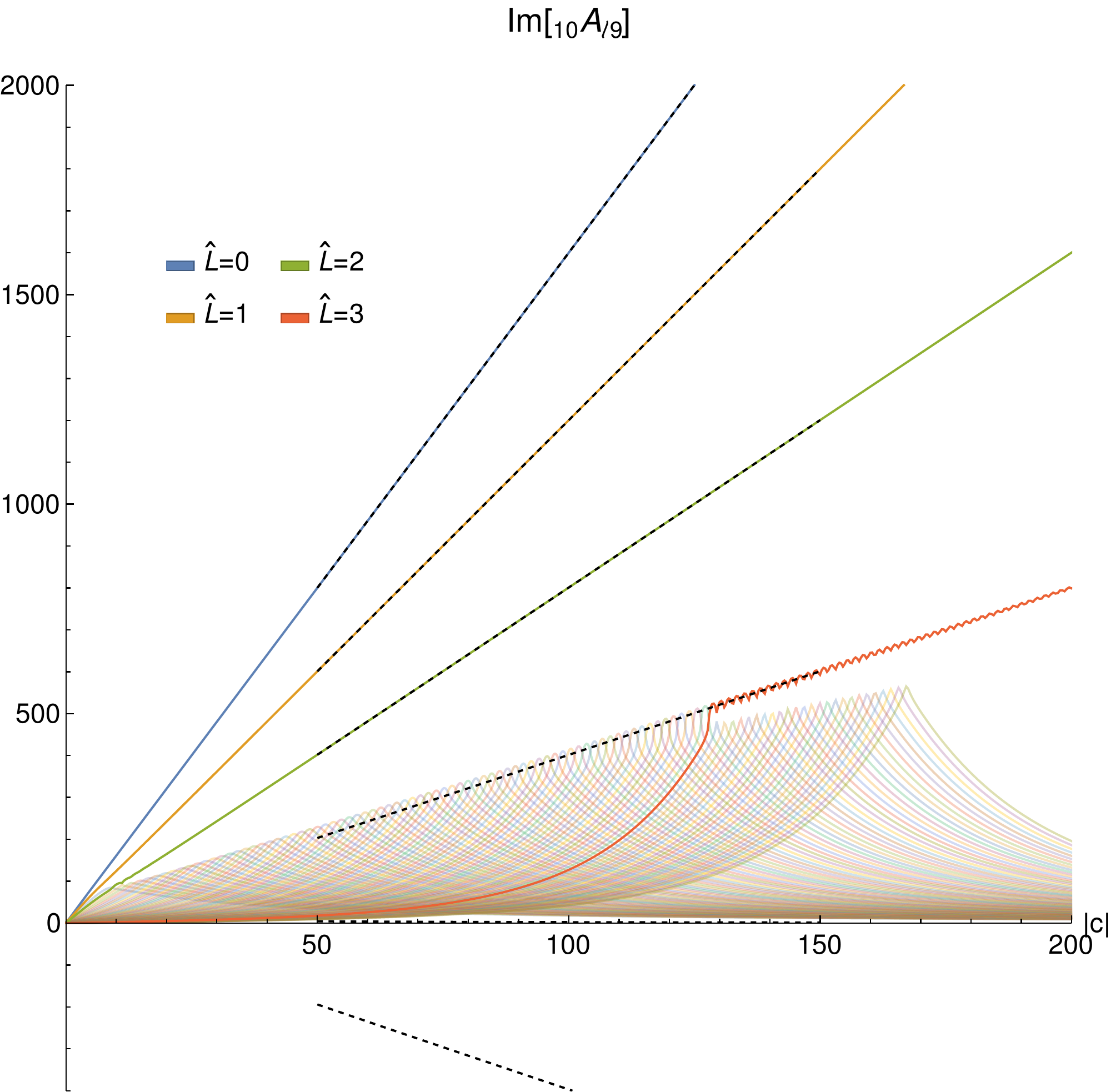}
\caption{\label{fig:ImA9x10anom} Imaginary part of $\scA{10}{\ell9}{-i|c|}$.  The labeled curves are the 4 anomalous sequences.  The faint curves are the first 100 normal sequences.  The dashed lines show the base anomalous asymptotic fitting function for the first 6 values of $\hat{L}$ plotted over the limited range of $50\le|c|\le150$.  The last 2, with $2{}_s\hat{q}_{\ell|m|}+1=1$ and 5, do not correspond to valid anomalous sequences.}
\end{figure}

Figure~\ref{fig:ImA9x10anom} presents an example for $m=9$ and $s=10$ in which the Type-3 anomalous sequence with $\hat{L}=3$ was not found in the original data set, but was found at $L=78$ in a data set that includes many additional values of $L$.
Figure~\ref{fig:ImA9x10anom} also illustrates why we should not be surprised that anomalous sequences do not exist for $2{}_s\hat{q}_{\ell|m|}+1>-1$.  Anomalous sequences arise when two neighboring eigenvalues become nearly degenerate.  As we have seen, they do not actually become degenerate, but the real part of the eigenvalues deflect while the imaginary parts cross as illustrated in  Fig.~\ref{fig:2x2 anomalous small}. The crossing of the imaginary parts does not occur in the neighborhood of the sequences with $2{}_s\hat{q}_{\ell|m|}+1>-1$.  In Fig.~\ref{fig:ImA9x10anom}, the anomalous sequence that could correspond to $\hat{L}=4$ has $2{}_s\hat{q}_{\ell|m|}+1=1$ and the asymptotic slope of the imaginary part of the eigenvalue is zero.  Potential anomalous sequences with even larger values of $2{}_s\hat{q}_{\ell|m|}+1$ yield asymptotic slopes for the imaginary parts of the eigenvalues which are negative.

There is an additional level of uncertainty as to the existence of at least some Type-1 sequences.
\begin{figure}[h]
\includegraphics[width=\linewidth,clip]{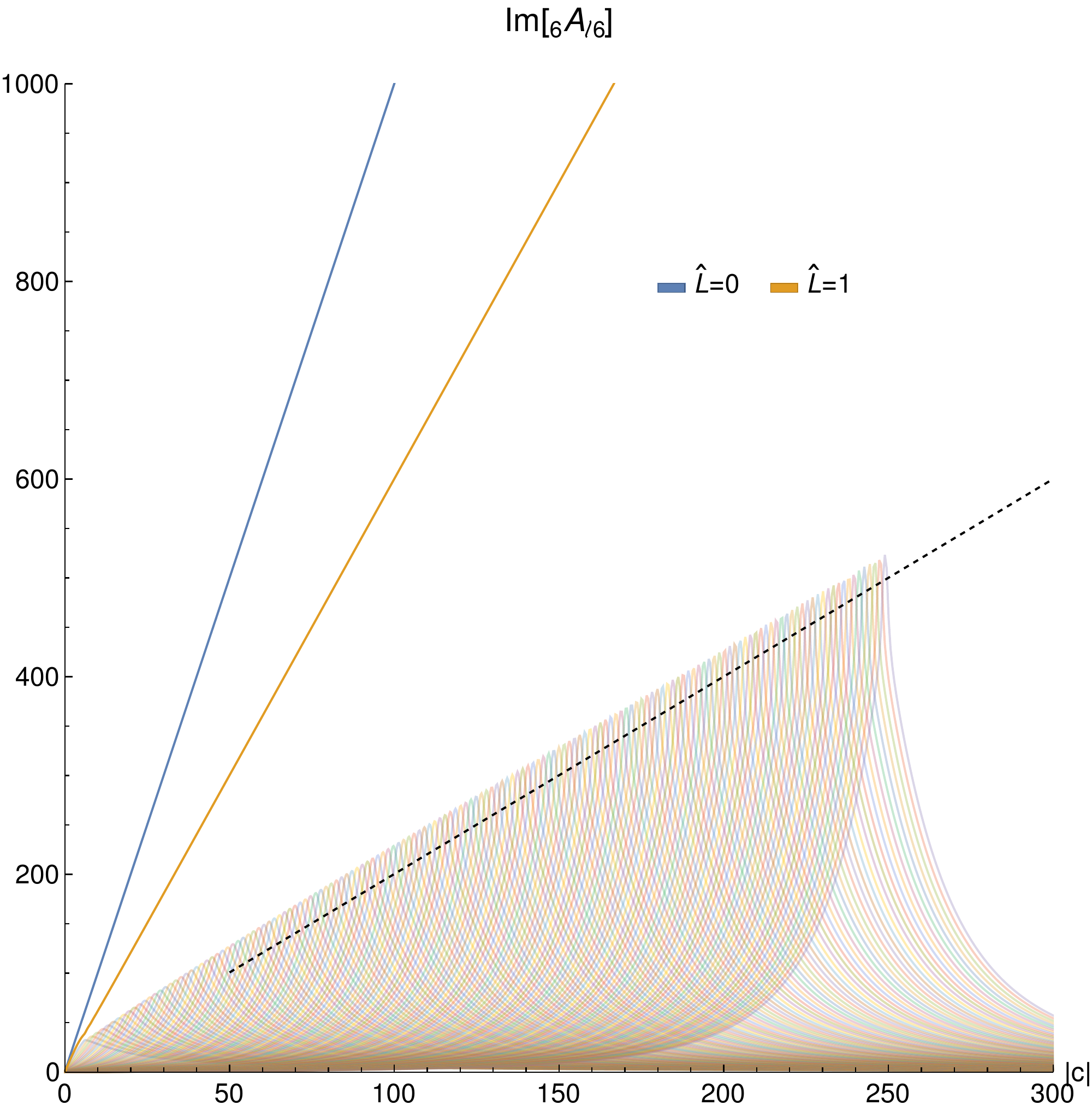}
\caption{\label{fig:ImA6x6anom} Imaginary part of $\scA{6}{\ell6}{-i|c|}$.  The labeled curves are the 2 anomalous sequences.  The faint curves are the first 155 normal sequences.  The dashed line shows the base anomalous asymptotic fitting function for $\hat{L}=2$ which corresponds to $2{}_s\hat{q}_{\ell|m|}+1=-1$.  This line could correspond to an anomalous sequence, but there is no evidence for this for $L<157$.}
\end{figure}
Figure~\ref{fig:ImA6x6anom} is similar to Fig.~\ref{fig:ImA9x10anom} but presents the case of $m=s=6$.  In this example, a Type-1 anomalous sequence could be present, but has not been found.  When the Type-1 sequence was not found in the original data set, we searched again keeping the first 157 sequences but again found no Type-1 anomalous sequence.  In Fig.~\ref{fig:ImA6x6anom}, the dashed line represents the asymptotic behavior of the imaginary part of the separation constant for a Type-1 sequences.  Interestingly, we see that the slope of this line seems to be the same as the sequence of peaks in the first 155 normal sequences shown as faint curves in the figure.  A general feature is that an anomalous sequence seems to transition to asymptotic anomalous behavior near the value of $|c|$ where the imaginary part of the base anomalous fitting function would exceed the peak of the imaginary part of the full normal sequence.  We see in Fig.~\ref{fig:ImA9x10anom} that this is not a precise condition.  But, clearly the slope of the line connecting the peaks of the imaginary parts of the normal sequences has a smaller slope than the slope of the Type-3 anomalous sequence, and the transition to asymptotic anomalous behavior occurs near the point where the dashed line representing the Type-3 asymptotic behavior intersects the line connecting the peaks of the normal sequences.  In Fig.~\ref{fig:ImA6x6anom}, it is clear that the dashed line is nowhere near to crossing the line connecting the peaks of the normal sequences.

The discussion above may provide insight into why the transition to anomalous behavior occurs, but it does not provide a way to predict whether or not a certain set of sequences will contain any specific anomalous sequence.  The peaks in the imaginary part of the normal sequences is a feature of the transition to asymptotic normal behavior and cannot be predicted simply from the normal prolate asymptotic fitting function of Eq.~({\ref{eqn:prolate normal solution}}).  Finding these peaks requires constructing the full numerical sequences and would also directly find the anomalous sequences if they existed within the set of $L$ values being tracked.  Unfortunately, the time required to accurately track sequences increases rapidly with the number of sequences being tracked.  

\begin{figure}[h]
\includegraphics[width=\linewidth,clip]{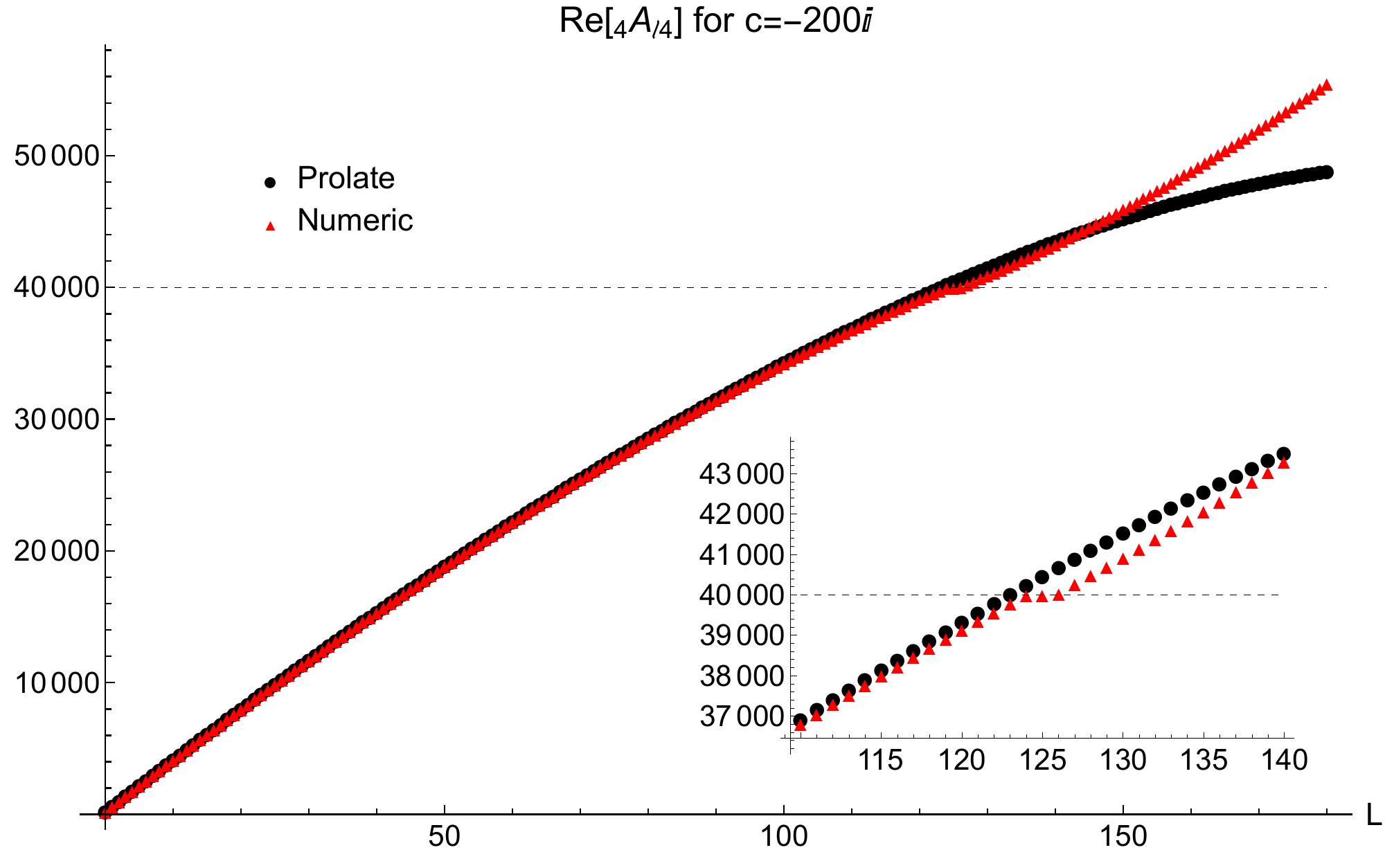}
\includegraphics[width=\linewidth,clip]{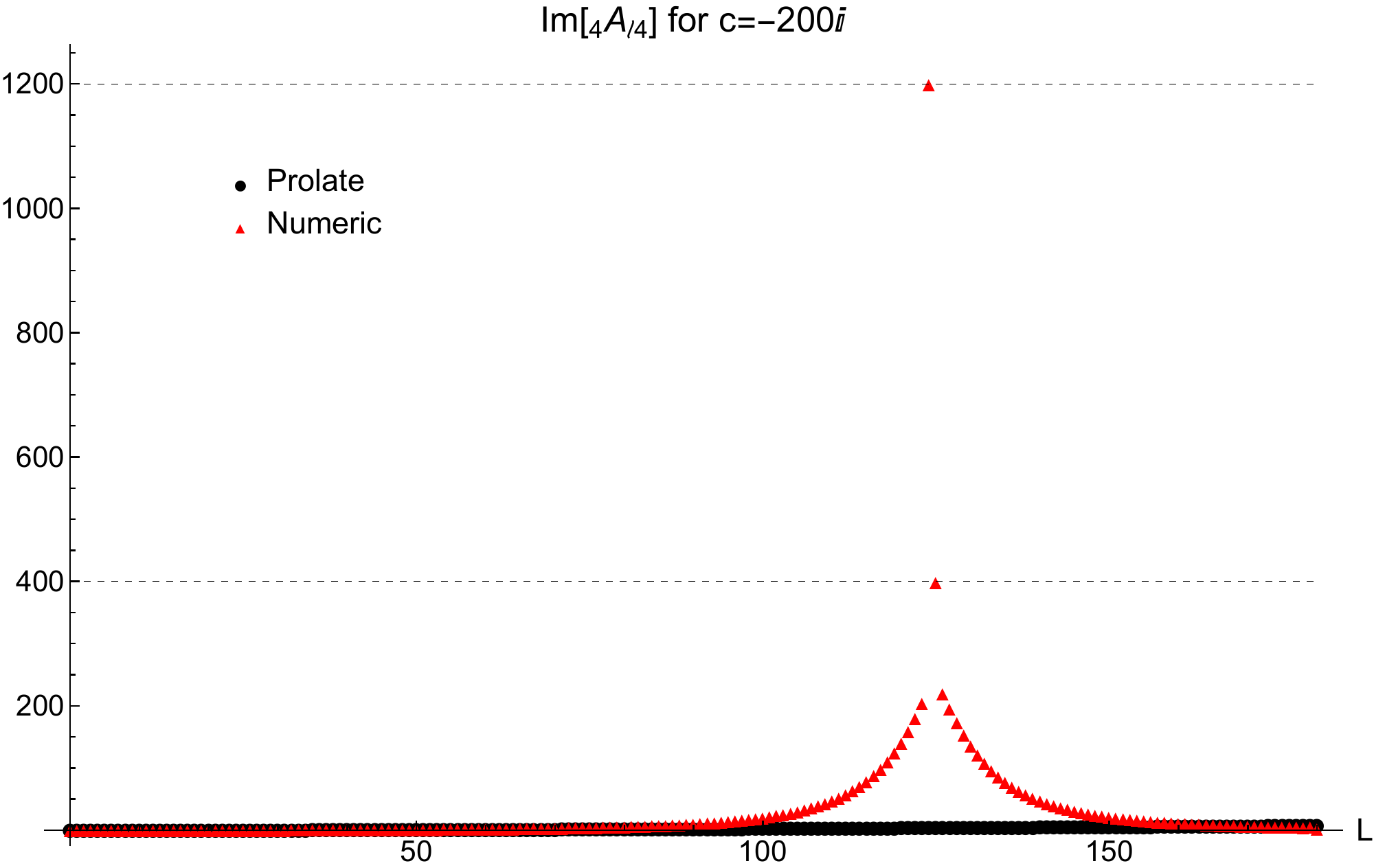}
    \caption{The real and imaginary parts of $\scA{4}{\ell4}{c}$ at $c=-200i$ for the first 180 eigenvalues when sorted by the value of the real part.  In these plots, $L$ represents the sorted position at $c=-200i$.  The black circles are computed using the asymptotic prolate fit in Eq.~(\ref{eqn:prolate normal solution}) with $\bar{L}=L$, and the red triangles are the numerically computed values.  The base asymptotic anomalous fit for $\hat{L}=0$ and $1$ are displayed as dashed lines since their position in the list of eigenvalues cannot be predicted.}
    \label{fig:4x4-200-Eigenvalues}
\end{figure}

An alternative approach for determining the existence of anomalous sequences is illustrated in Fig.~\ref{fig:4x4-200-Eigenvalues}.
Instead of constructing sequences of solutions as a function of $|c|$, we simply explore the behavior of a set of eigenvalues for given $m$ and $s$ to see if we can determine if any of the solutions are anomalous.  In Fig.~\ref{fig:4x4-200-Eigenvalues}, we consider $m=s=4$ for the case of $c=-200i$.  The upper plot in the figure displays $\text{Re}[\scA{4}{\ell4}{c}]$ verses $L$ with the red triangles representing the numerical solutions, and the black circles the corresponding values based on the prolate asymptotic fitting function Eq.~(\ref{eqn:prolate normal solution}).  The numerical eigenvalues are sorted based on their real part.  The horizontal dashed lines represent the values of the real part of the base asymptotic anomalous fitting function for each value of $\hat{L}$ which has $2{}_s\hat{q}_{\ell|m|}+1<0$.  Since we cannot assign a value of $L$ to each $\hat{L}$, these values are plotted as horizontal lines.  And, because the real parts of the asymptotic eigenvalues quickly become nearly degenerate, the individual lines appear as one line in this plot.  The lower plot in the figure shows similar information for the imaginary part of the eigenvalues.  The main difference is that the imaginary parts of the asymptotic anomalous fitting function for each value of $\hat{L}$ are no longer nearly degenerate.

In the plot for $\text{Im}[\scA{4}{\ell4}{c}]$ in Fig.~\ref{fig:4x4-200-Eigenvalues}, we can clearly see that the 2 anomalous sequences are present.  These are the two expected sequences with $\hat{L}=0$ ($2{}_s\hat{q}_{\ell|m|}+1=-5$) and $\hat{L}=1$ ($2{}_s\hat{q}_{\ell|m|}+1=-1$).
\begin{figure}[h]
\includegraphics[width=\linewidth,clip]{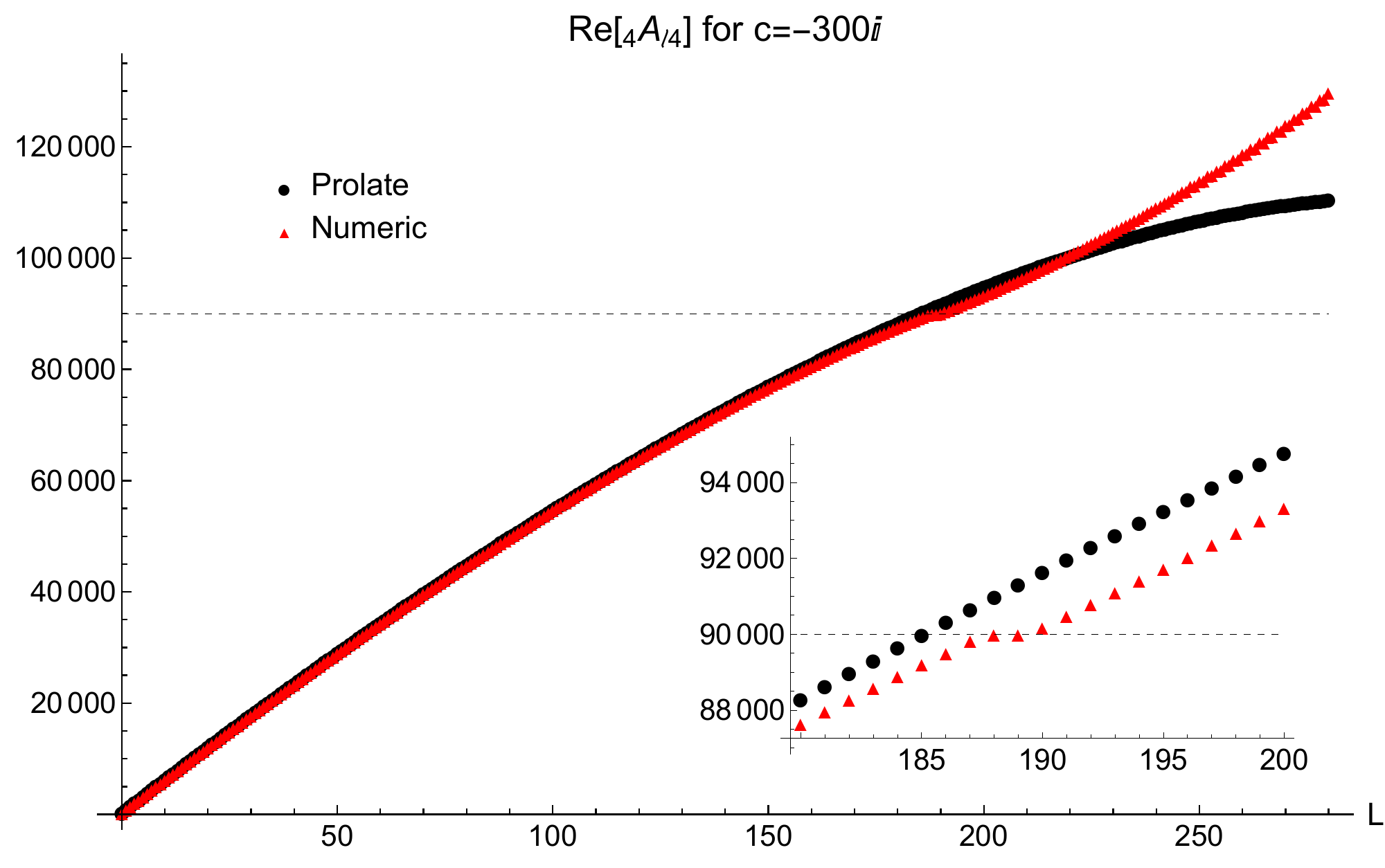}
\includegraphics[width=\linewidth,clip]{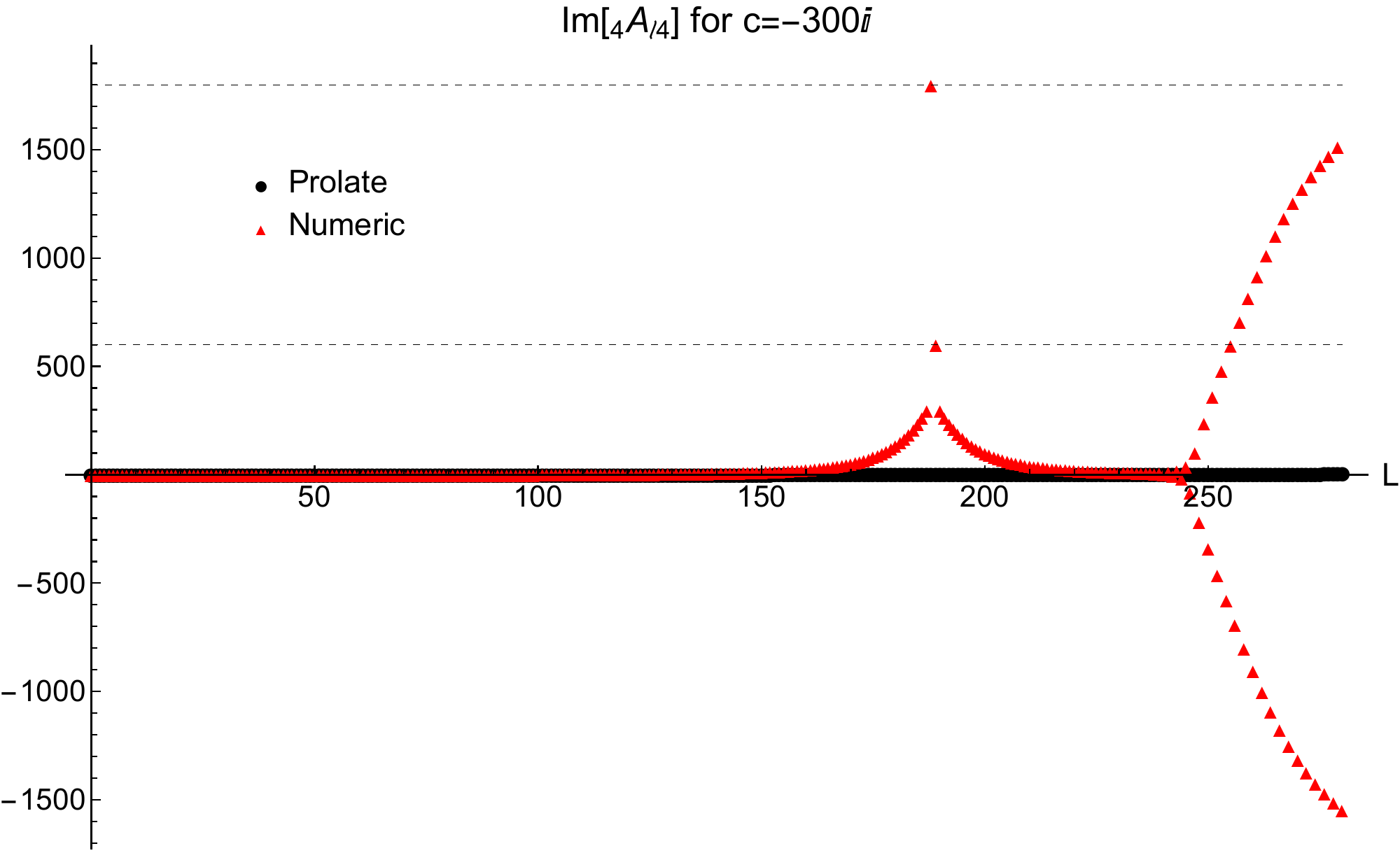}
    \caption{The real and imaginary parts of $\scA{4}{\ell4}{c}$ at $c=-300i$ for the first 280 eigenvalues when sorted by the value of the real part.  See Fig.~\ref{fig:4x4-200-Eigenvalues} for additional information.  In this figure, eigenvalues are plotted beyond $L=250$ where numerical errors are significant as can easily be seen in the plot of the imaginary part.}
    \label{fig:4x4-300-Eigenvalues}
\end{figure}
Because we have constructed full sequences for $m=s=4$, we know that the $\hat{L}=0$ anomalous eigenvalue is part of the $L=0$ sequences, but it appears at the $124\text{th}$ position in the sorted list of eigenvalues when $c=-200i$.  Similarly, the $\hat{L}=1$ anomalous eigenvalue is part of the $L=28$ sequence, but it appears at the $125\text{th}$ position.  It is a general feature of the anomalous eigenvalues that their position rapidly shifts in the sorted list of eigenvalues as $|c|$ increases because the leading order behavior grows as $|c|^2$.  Figure~\ref{fig:4x4-300-Eigenvalues} is similar to Fig.~\ref{fig:4x4-200-Eigenvalues} except that we are further into the asymptotic regime with $c=-300i$.  Now, the two anomalous eigenvalues occur at the $188\text{th}$ and $189\text{th}$ positions.  In Fig.~\ref{fig:4x4-300-Eigenvalues}, we also show some eigenvalue solutions which are not spectrally converged.  Given the size of the matrix used to numerically construct these eigenvalues, those eigenvalues starting just below the $250\text{th}$ sorted location are not accurately determined which can easily be seen  in the plot for $\text{Im}[\scA{4}{\ell4}{c}]$.

Figure~\ref{fig:ImA6x6anom} presented the example of $m=s=6$ for which the expected $\hat{L}=2$ anomalous sequence was not found within $L<157$.  Using this alternative method, we can explore larger values of $L$ to see if the $\hat{L}=2$ anomalous sequence is present without the need to construct an extremely large set of full sequences.
\begin{figure}[h]
\includegraphics[width=\linewidth,clip]{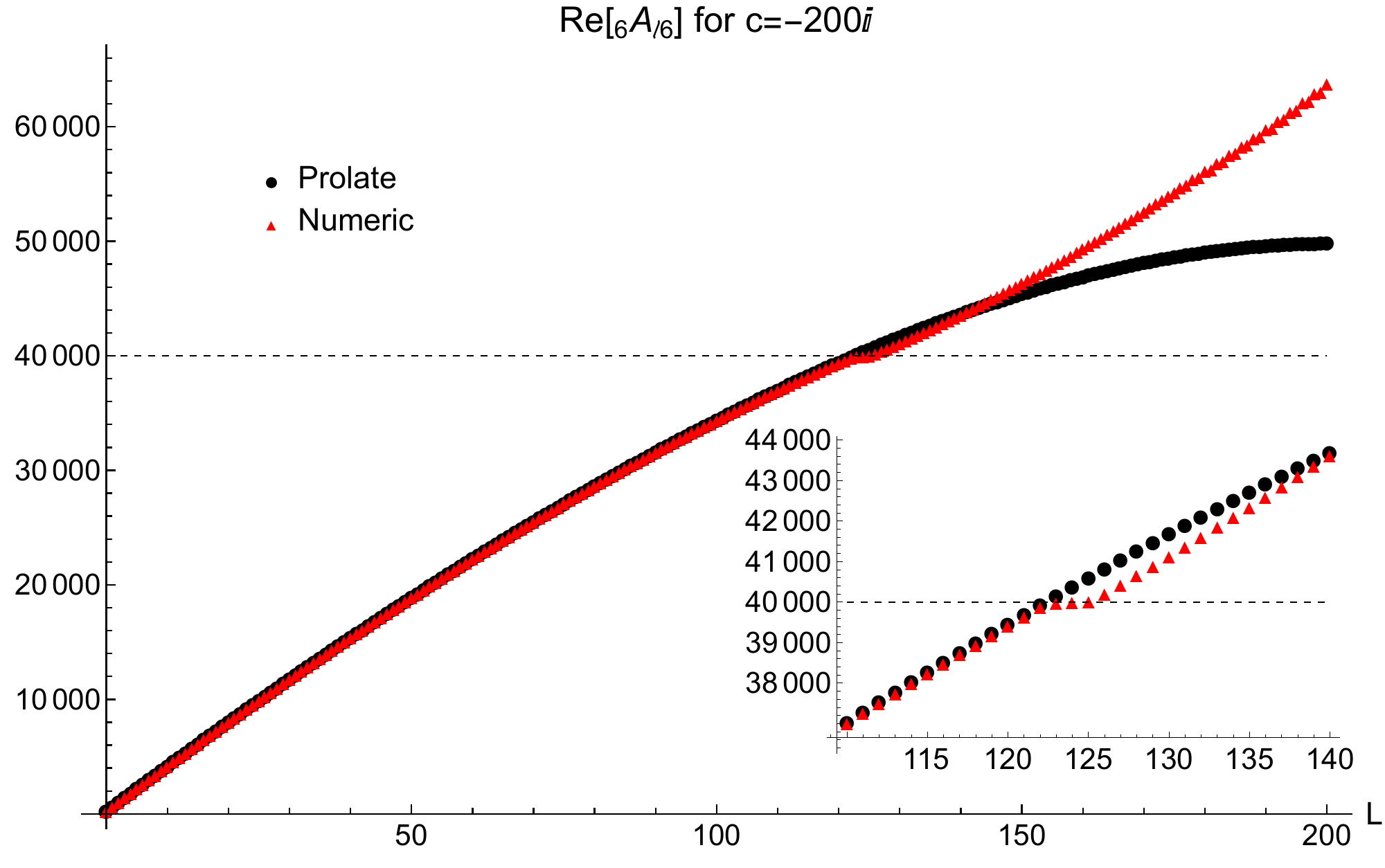}
\includegraphics[width=\linewidth,clip]{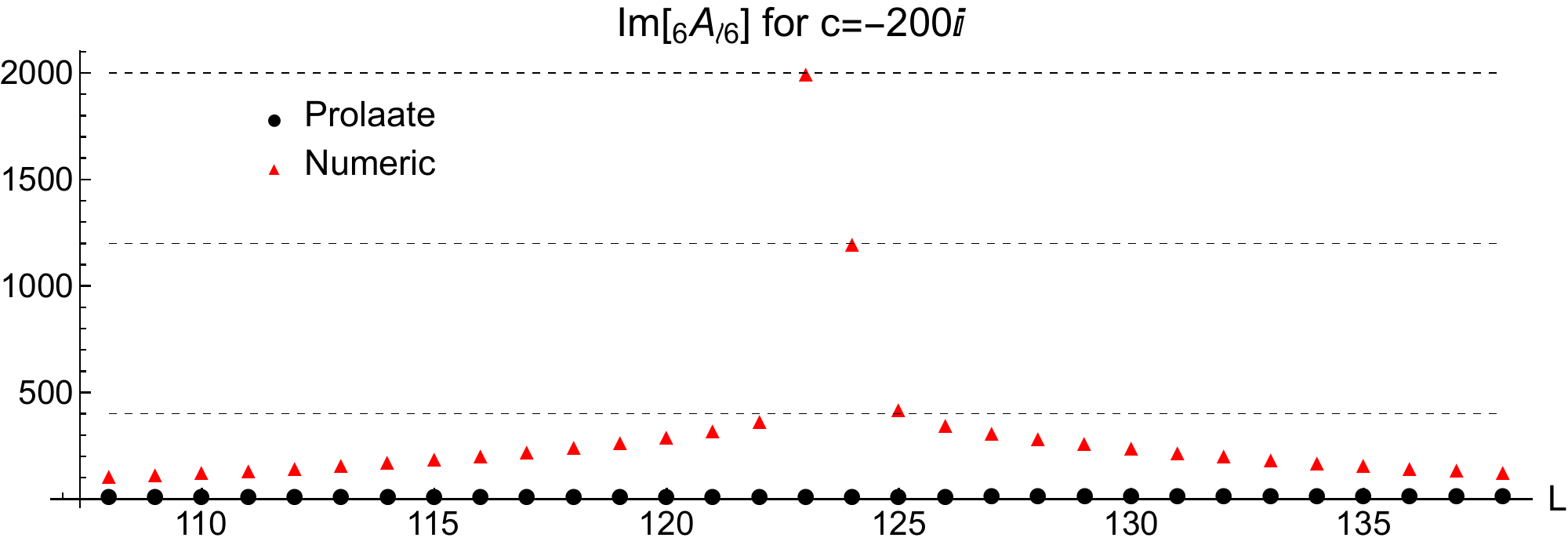}
\includegraphics[width=\linewidth,clip]{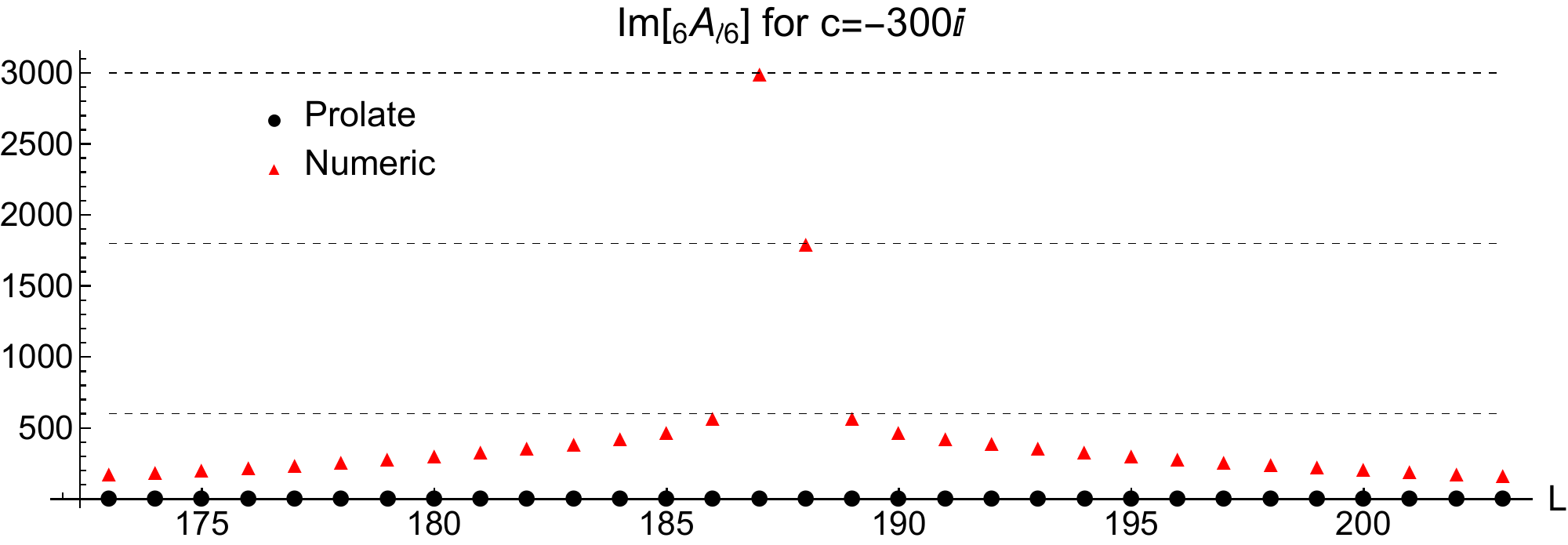}
\includegraphics[width=\linewidth,clip]{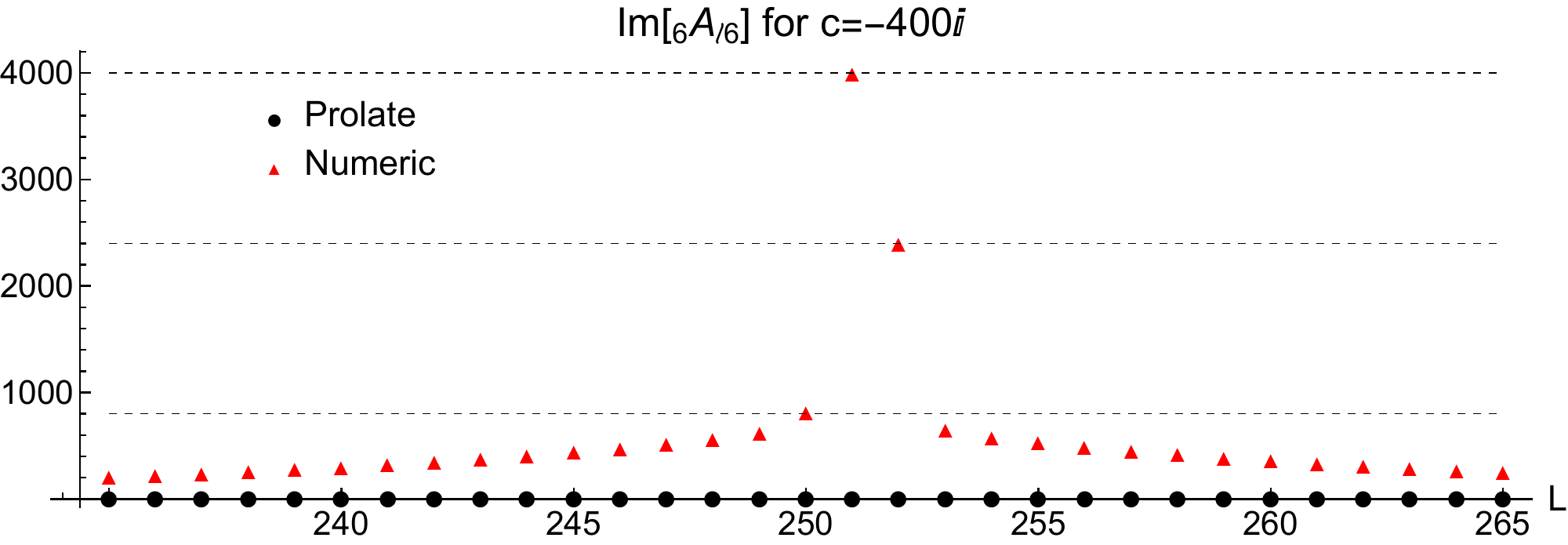}
\includegraphics[width=\linewidth,clip]{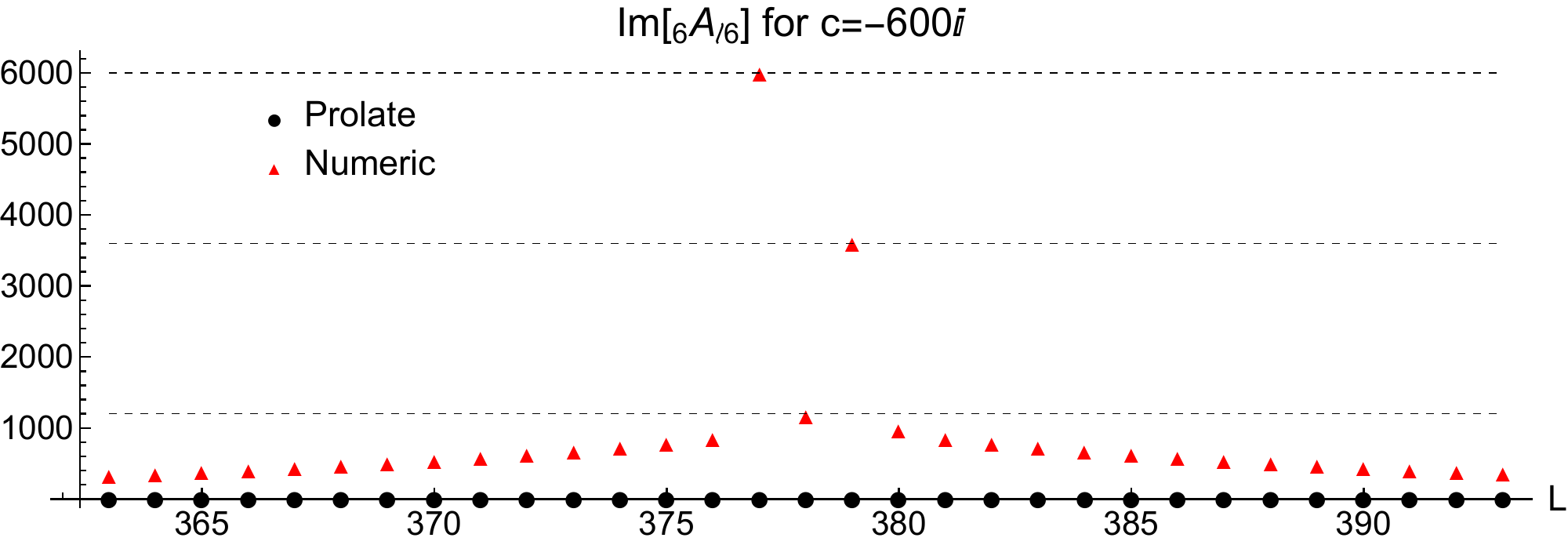}
    \caption{The real and imaginary parts of $\scA{6}{\ell6}{c}$.  The real part is plotted for $c=-200i$ while the imaginary part is plotted for $c=-200i$, $-300i$, $-400i$, and $-600i$.  See Fig.~\ref{fig:4x4-200-Eigenvalues} for additional information.  The base asymptotic anomalous fit for $\hat{L}=0$, $1$ and $2$ are displayed as dashed lines.  However, the numerical solutions only indicate anomalous sequences for $\hat{L}=0$ and $1$.}
    \label{fig:6x6-600-Eigenvalues}
\end{figure}
Figure~\ref{fig:6x6-600-Eigenvalues} explores the eigenvalues for $\scA{6}{\ell6}{c}$ at $c=-200i$, $-300i$, $-400i$, and $-600i$.  At each value of $c$, we clearly see the anomalous eigenvalues associated with $\hat{L}=0$ and $1$.  In the plot for $\text{Im}[\scA{6}{\ell6}{c}]$ at $c=-200i$, it appears that there is agreement between the numerical eigenvalue at $L=125$, and the $\hat{L}=2$ anomalous eigenvalue, but this is just coincidence as can be verified by Fig.~\ref{fig:ImA6x6anom}.  Moreover, as we let $|c|$ increase, we see the location of the numerical eigenvalue closest to the $\hat{L}=2$ anomalous eigenvalue shifts position relative the $\hat{L}=0$ and $1$ eigenvalues.  At $c=-600i$, we see that there is still no clear evidence for an $\hat{L}=2$ anomalous sequences out as far as $L=378$.

\subsubsection{Behavior of the anomalous eigenfunctions}\label{sec:anomalous eigenfunctions}

Together, Figs.~\ref{fig:2x2 anomalous small} and \ref{fig:6x6-600-Eigenvalues} clearly illustrate the way in which the locations of anomalous sequences, within the ordered list of eigensolutions, move through the normal sequences as $|c|$ increases.  Let us consider the behavior of the $\hat{L}=0$ eigenfunction for the case of $m=s=2$.
\begin{figure}\vspace{10pt}
\includegraphics[width=\linewidth,clip]{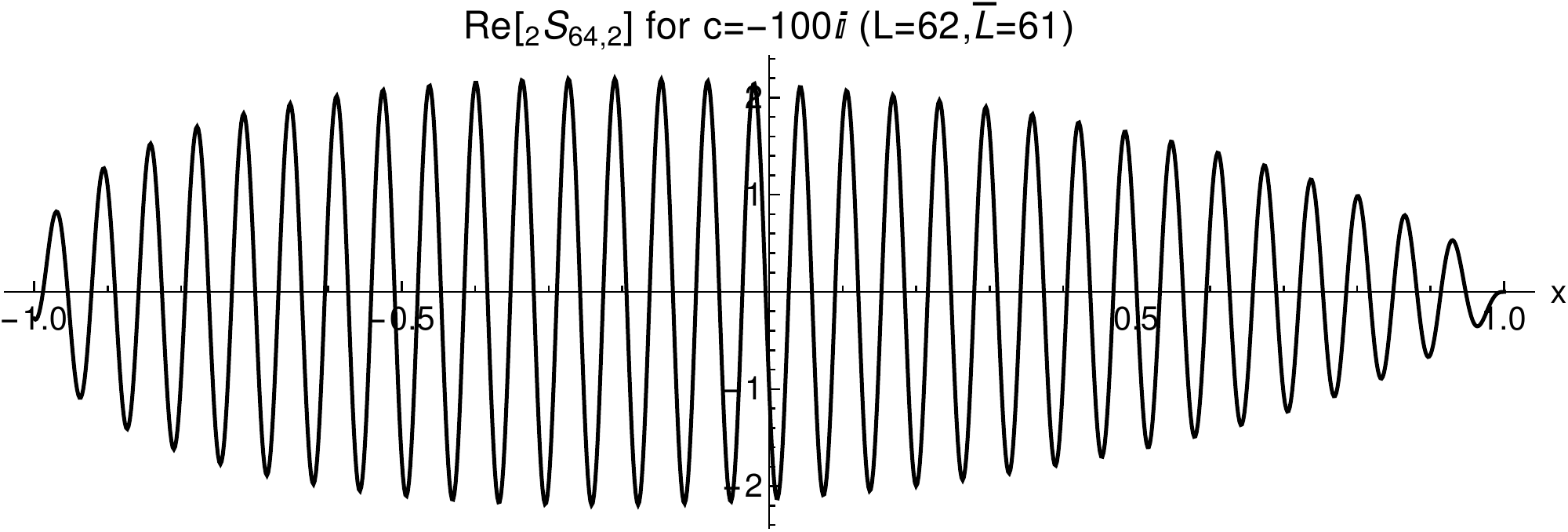}
\includegraphics[width=\linewidth,clip]{2x2x1Real}
\includegraphics[width=\linewidth,clip]{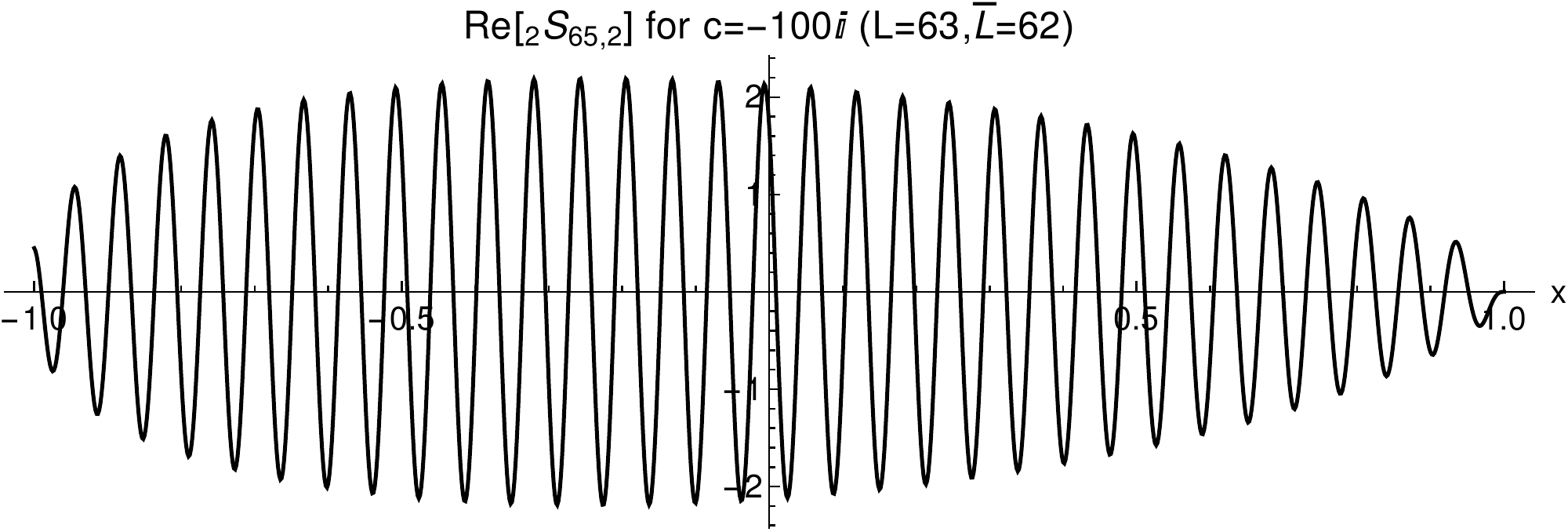}
    \caption{The real part of the $3$ consecutive eigenvector solutions $\swS{2}{64,2}{x}{c}$, $\swS{2}{32}{x}{c}$, and $\swS{2}{65,2}{x}{c}$ at $c=-100i$. When eigenvalues are sorted by the real part of $\scA{s}{\ell m}{c}$, these 3 eigenvectors correspond to $62^{\text{nd}}$, $63^{\text{rd}}$, and $64^{\text{th}}$ eigenvalues.  $\swS{2}{32}{x}{c}=\swSanom{2}{22}{x}{c}$ corresponds to an anomalous sequence with $L=1(\hat{L}=0)$, while $\swS{2}{64,2}{x}{c}=\swSnorm{2}{63,2}{x}{c}$ and $\swS{2}{65,2}{x}{c}=\swSnorm{2}{64,2}{x}{c}$ correspond to normal sequences with $L=62(\bar{L}=61)$ and $63(\bar{L}=62)$. At this large value of $L$, the normal sequences are not yet in the asymptotic regime.}
    \label{fig:2x2 anomalous compare 100}
\end{figure}
Figure~\ref{fig:2x2 anomalous compare 100} plots the real part of the eigenfunctions corresponding to the $62\text{nd}$, $63\text{rd}$, and $64\text{th}$ sorted eigenvalues when $c=-100i$.  The eigenfunction in the top plot corresponds to a normal sequence with $\bar{L}=61$ and $L=62$.  The eigenfunction in the bottom plot corresponds to the next normal sequence with $\bar{L}=62$ and $L=63$.  Between them is the eigenfunction corresponding to the lone anomalous sequence for $m=s=2$ having $\hat{L}=0$ and $L=1$.  The imaginary part of the anomalous eigenfunction can also be seen in Fig.~\ref{fig:2x2 anomalous vector}.  It is worth noting that the normal eigenfunctions with $L=62$ and $L=63$ do not yet display the expected behavior for solutions that are in the asymptotic regime.  The transition to asymptotic behavior for a normal sequence is just beginning at the value of $c$ where an anomalous eigenvalue passes a given normal eigenvalue.  We will discuss this point in more detail in Sec.~\ref{sec:conclusion}.
\begin{figure}\vspace{10pt}
\includegraphics[width=\linewidth,clip]{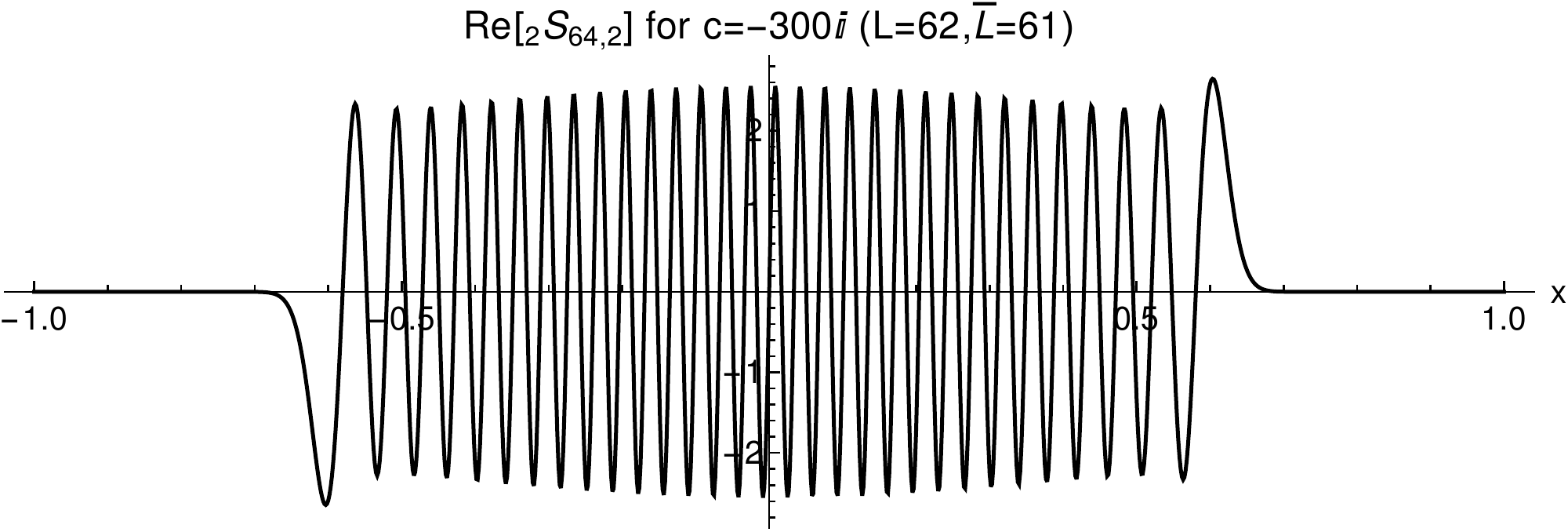}
\includegraphics[width=\linewidth,clip]{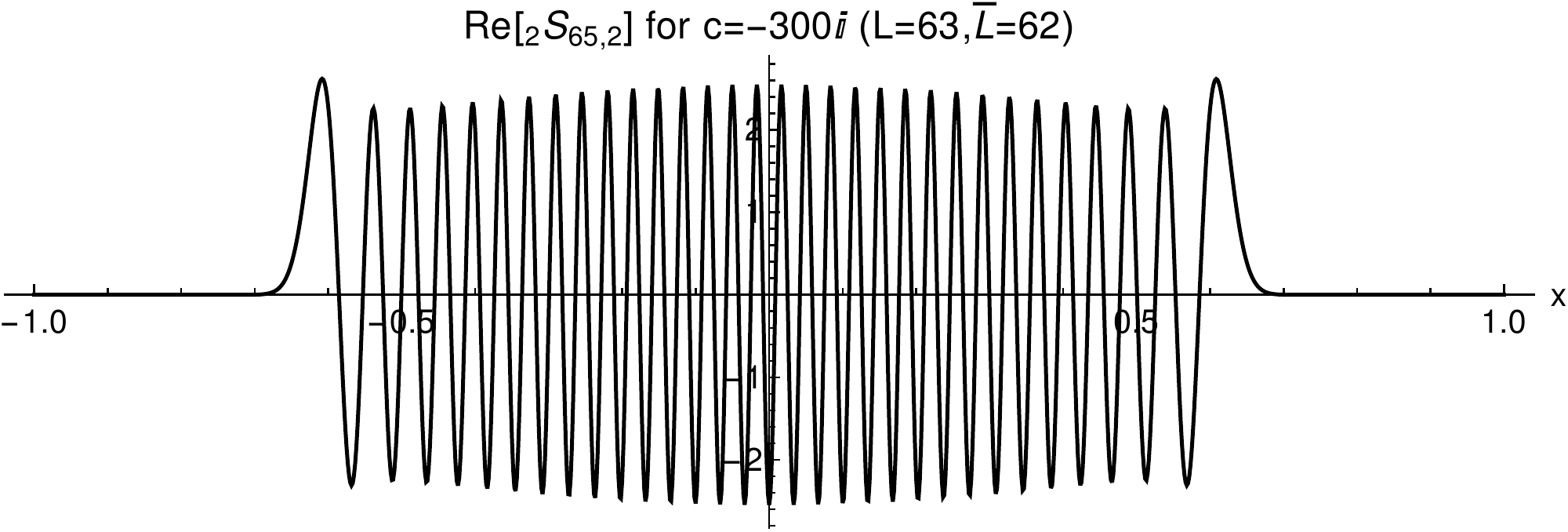}
\includegraphics[width=\linewidth,clip]{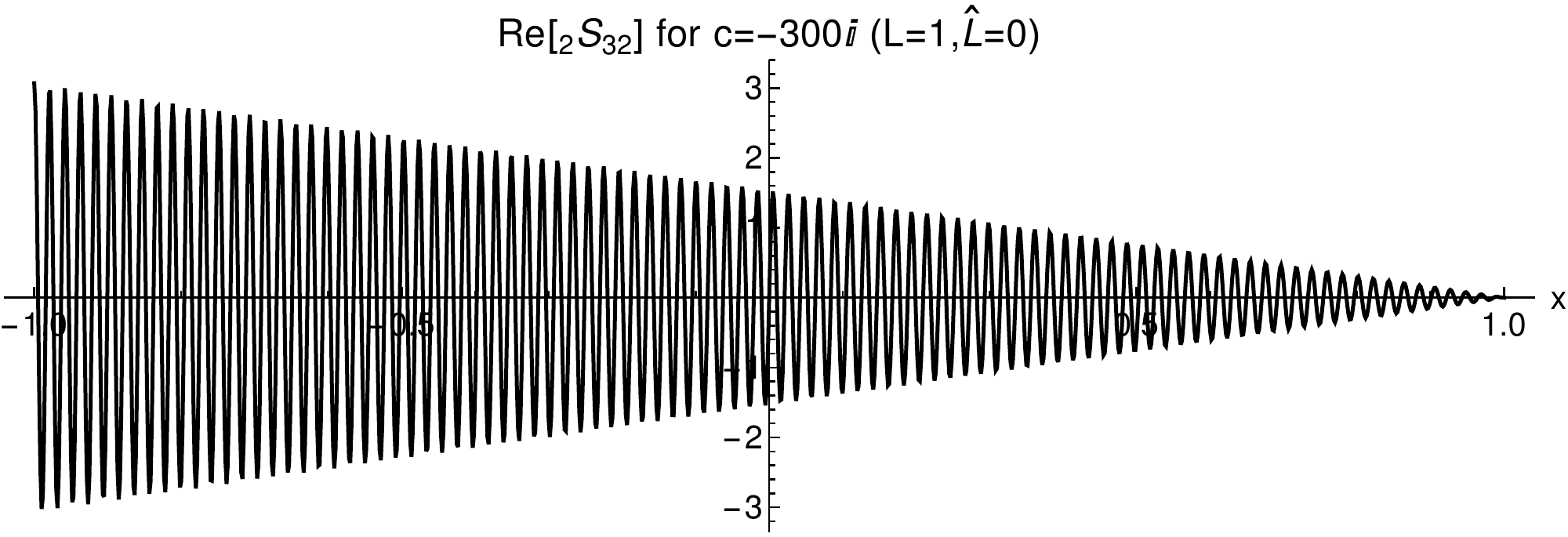}
    \caption{The real part of the $3$ eigenvector solutions $\swS{2}{64,2}{x}{c}$, $\swS{2}{65,2}{x}{c}$, and $\swS{2}{32}{x}{c}$ at $c=-300i$. These are eigenvectors along the same 3 sequences as in Fig.~\ref{fig:2x2 anomalous compare 100} but further into the asymptotic regime.  $\swS{2}{32}{x}{c}=\swSanom{2}{22}{x}{c}$ corresponds to an anomalous sequence with $L=1(\hat{L}=0)$, but it now corresponds to the $190^{\text{th}}$ eigenvalue. The normal sequences $\swS{2}{64,2}{x}{c}=\swSnorm{2}{63,2}{x}{c}$ and $\swS{2}{65,2}{x}{c}=\swSnorm{2}{64,2}{x}{c}$ correspond to the $62^{\text{nd}}$ and $63^{\text{rd}}$ eigenvalues and are now in the asymptotic regime.}
    \label{fig:2x2 anomalous compare 300}
\end{figure}
Figure~\ref{fig:2x2 anomalous compare 300} displays the real part of the same eigenfunctions as in Fig.~\ref{fig:2x2 anomalous compare 100}, but for $c=-300i$ instead of $c=-100i$.  The eigenfunctions for the two normal sequences corresponding to $L=62$ and $L=63$ are now clearly showing the behavior expected of an eigensolution in the asymptotic regime.  The anomalous eigensolution corresponding to $L=1$ has now shifted to being the $190\text{th}$ eigenvalue and it is clear that the number of real zero crossings has increased dramatically.  At $c=-100i$, the real part of the anomalous eigenfunction has $63$ real zero crossings.  At $c=-300i$, the real part of the anomalous eigenfunction has $190$ real zero crossings.  We find a clear pattern that, as we increase $|c|$ and move along an anomalous sequence, the associated anomalous eigenfunction gains an additional real zero crossing each time its eigenvalue moves past a normal eigenvalue.

The transition to anomalous asymptotic behavior for an anomalous sequence is just beginning at the value of $c$ along the sequence where the real part of the anomalous eigenvalue first deflects away from a neighboring normal eigenvalue and then crosses subsequent normal eigenvalues as can be seen in Fig.~\ref{fig:2x2 anomalous small}.  As seen in Figs.~\ref{fig:2x2 anomalous compare 100} and \ref{fig:2x2 anomalous compare 300}, the asymptotic form of an anomalous eigenfunction takes on an approximately constant envelope modulating some number of oscillations where the number of oscillations increases as we move to larger values of $|c|$.  However, the shape of the envelope depends on the specific values of $m$, $s$, and $\hat{L}$.
\begin{figure}\vspace{10pt}
\begin{tabular}{cc}
\includegraphics[width=0.5\linewidth,clip]{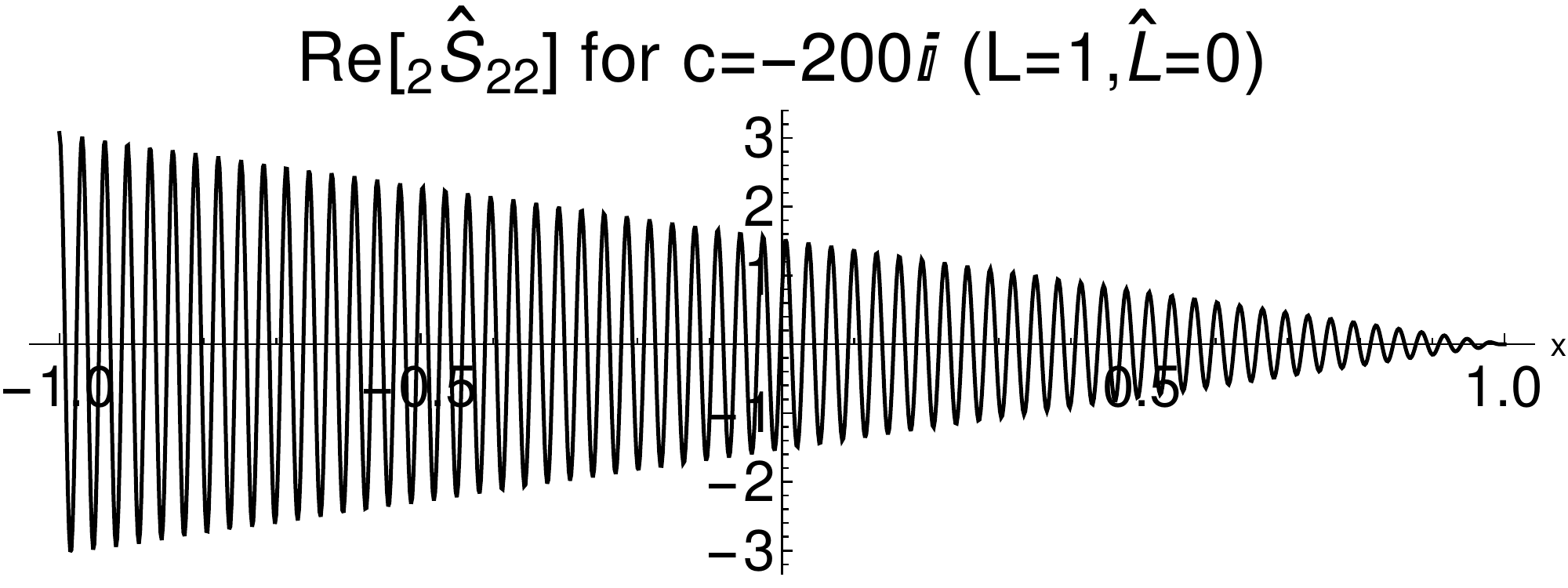} &
\includegraphics[width=0.5\linewidth,clip]{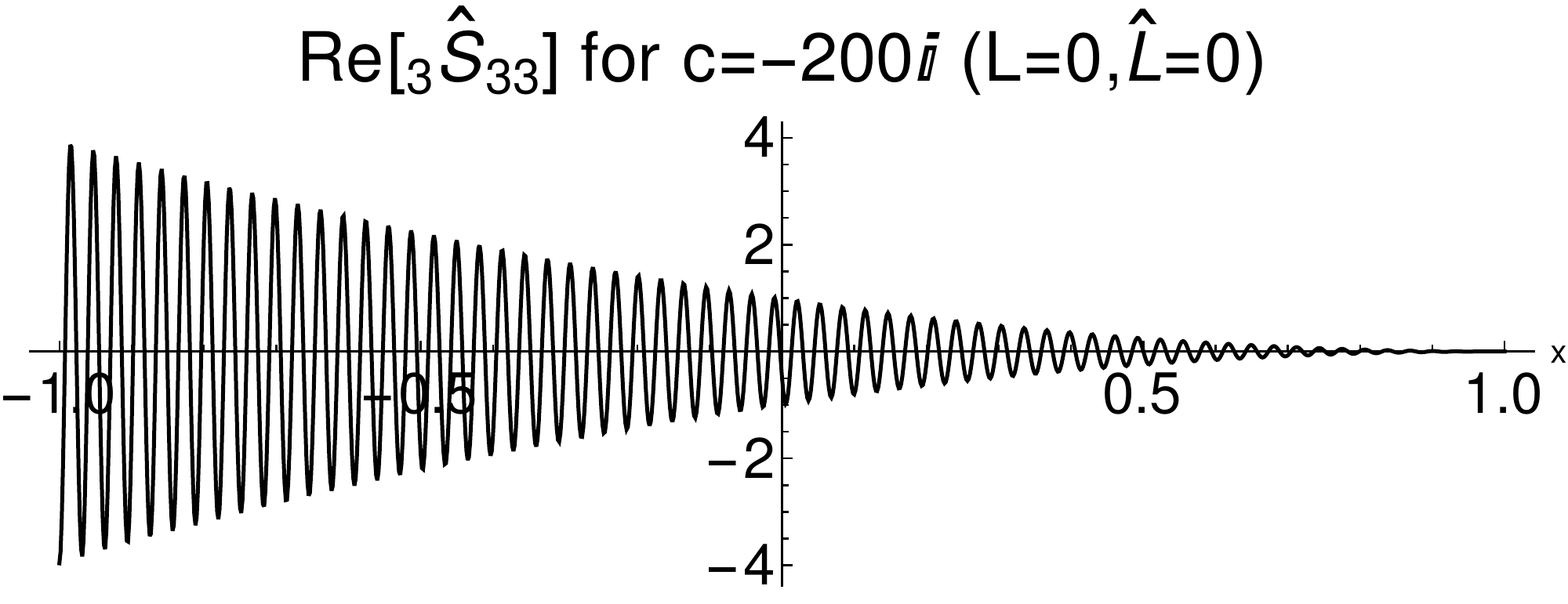} \\
\includegraphics[width=0.5\linewidth,clip]{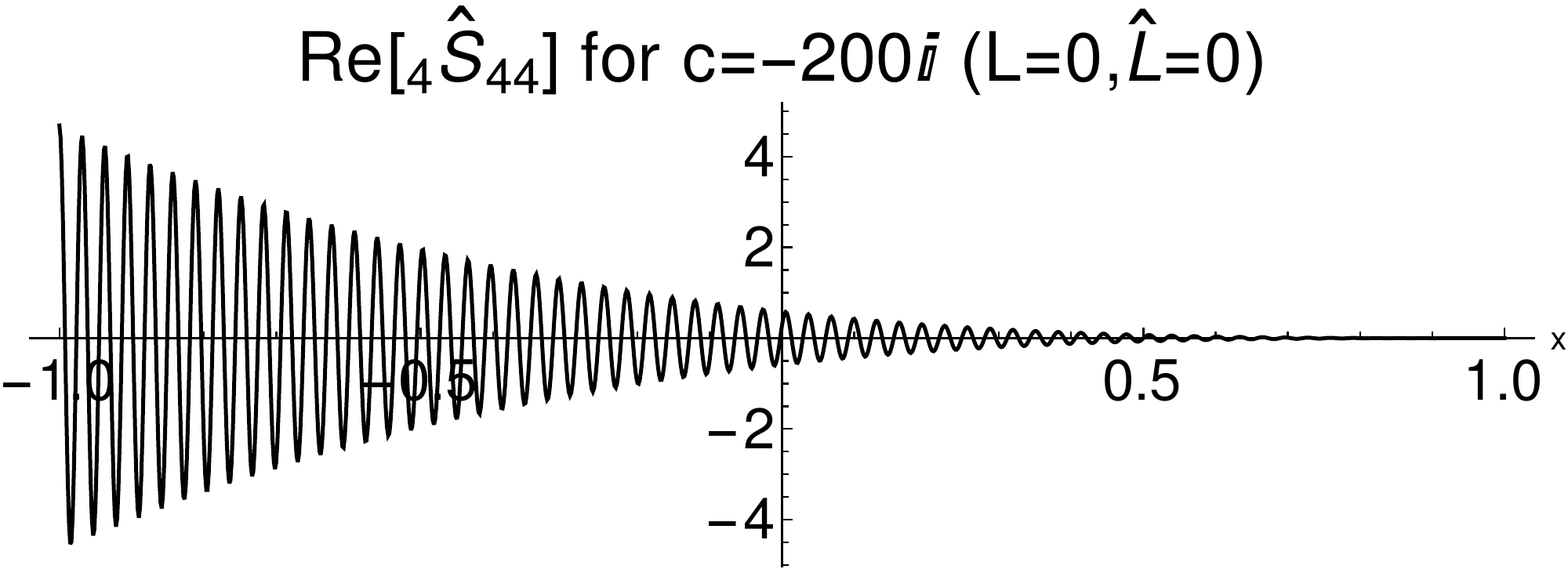} &
\includegraphics[width=0.5\linewidth,clip]{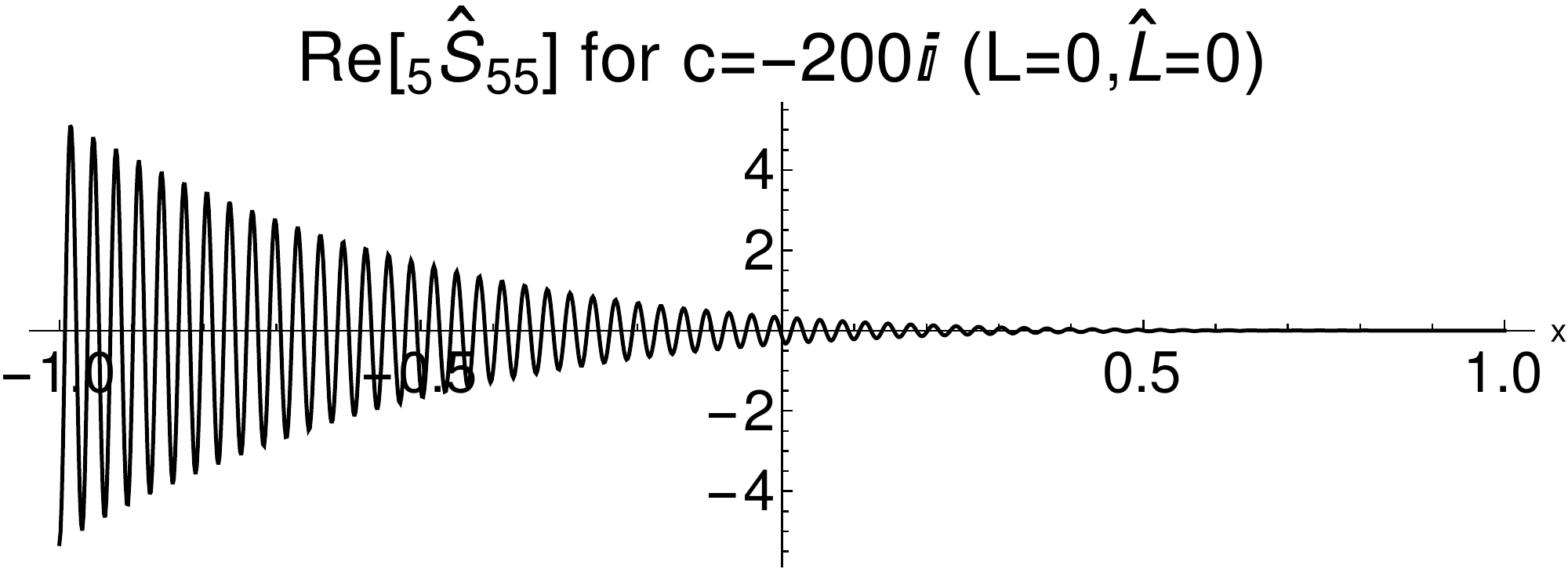} \\
\includegraphics[width=0.5\linewidth,clip]{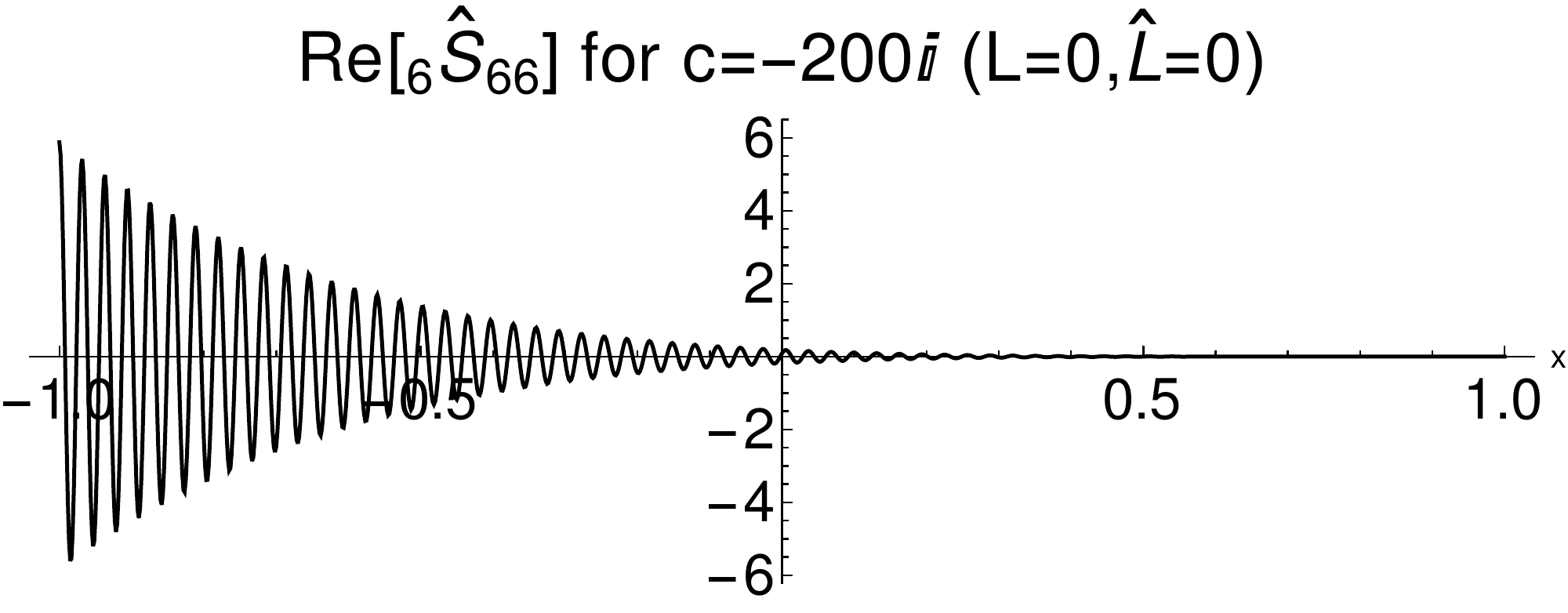} &
\includegraphics[width=0.5\linewidth,clip]{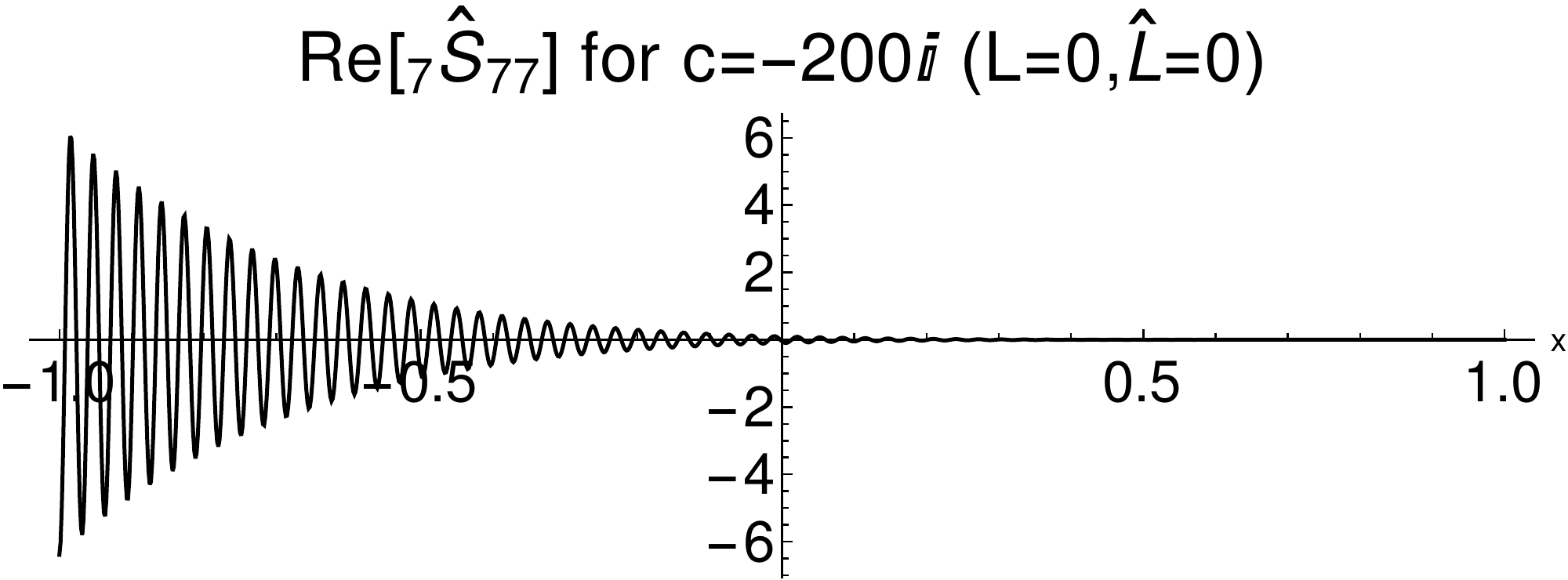}
\end{tabular}
    \caption{The real part of the $\hat{L}=0$ eigenvector solutions  $\swSanom{s}{\ell m}{x}{c}$ with $s=m=2\to7$ and $c=-200i$.  The anomalous type of each increases by $2$ from Type-$1$ for $m=s=2$ to Type-$11$ for $m=s=7$.}
    \label{fig:Lstar0seqmeigenfunctions}
\end{figure}
Figure~\ref{fig:Lstar0seqmeigenfunctions} shows a representative set of anomalous eigenfunctions for the case when $m=s=2\to7$ and $\hat{L}=0$.
\begin{figure}\vspace{10pt}
\begin{tabular}{cc}
\includegraphics[width=0.5\linewidth,clip]{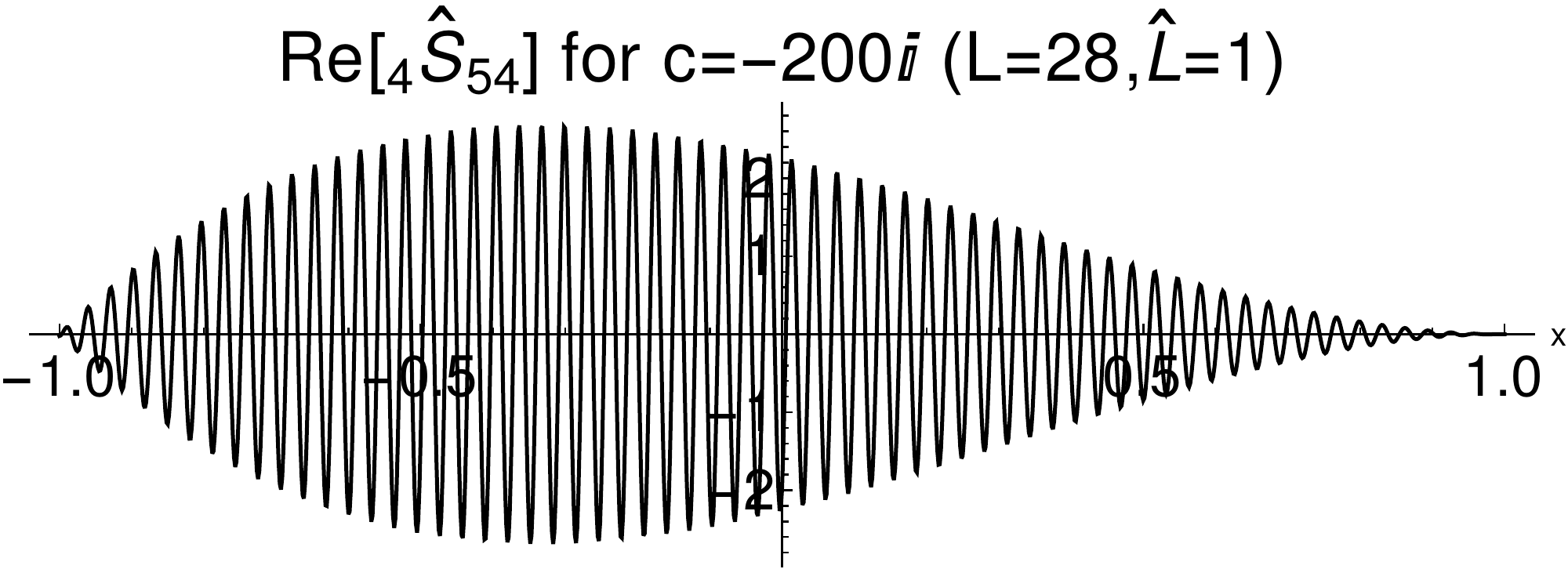} &
\includegraphics[width=0.5\linewidth,clip]{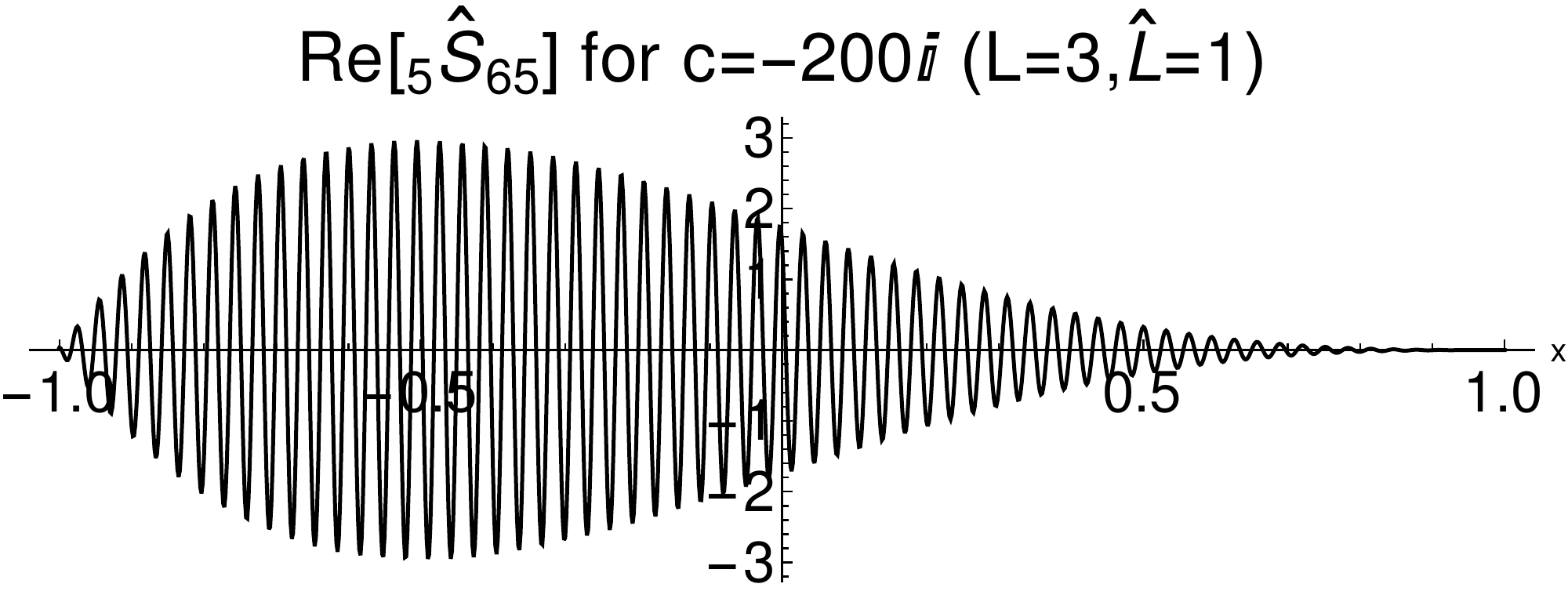} \\
\includegraphics[width=0.5\linewidth,clip]{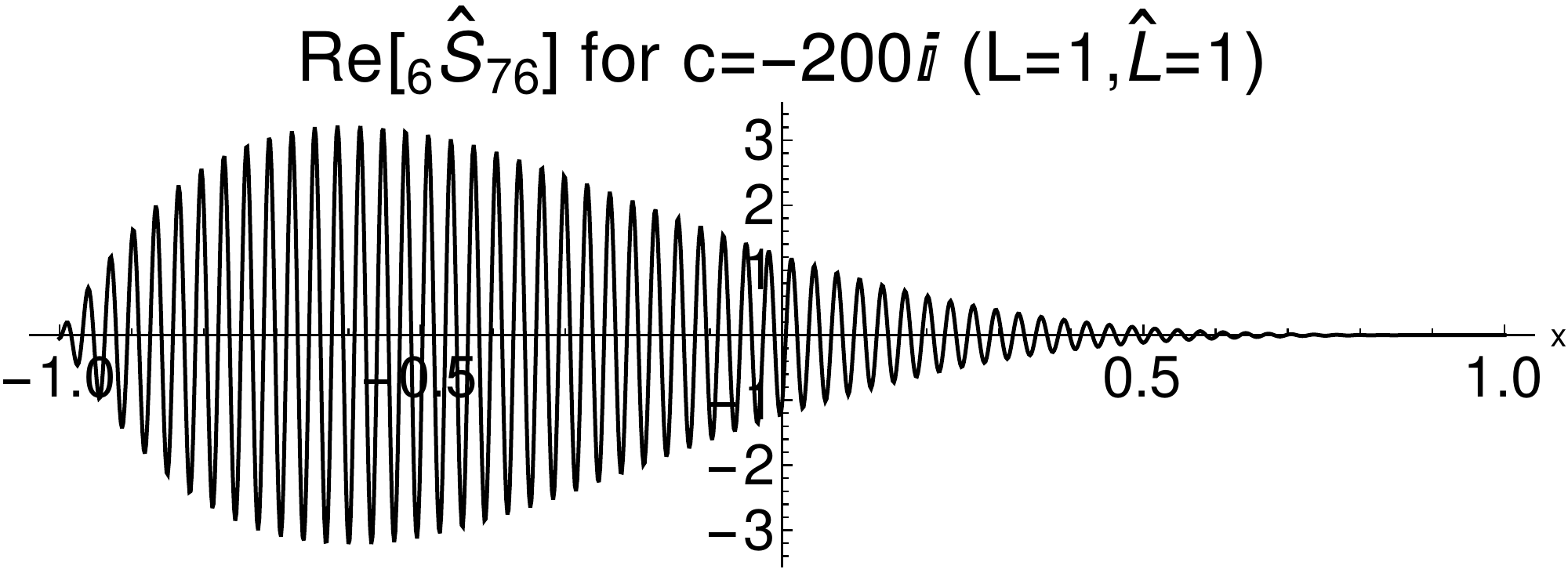} &
\includegraphics[width=0.5\linewidth,clip]{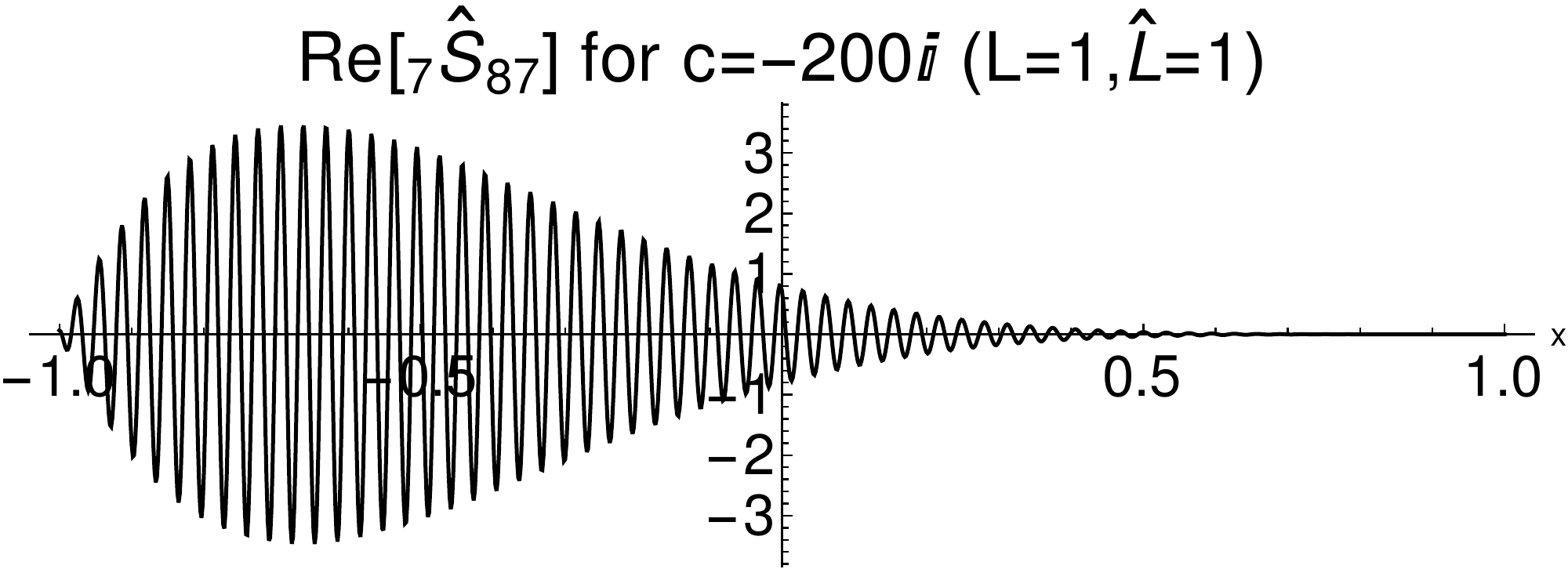}
\end{tabular}
    \caption{The real part of the $\hat{L}=1$ eigenvector solutions  $\swSanom{s}{\ell m}{x}{c}$ with $s=m=4\to7$ and $c=-200i$.  The anomalous type of each increases by $2$ from Type-$1$ for $m=s=4$ to Type-$7$ for $m=s=7$.}
    \label{fig:Lstar1seqmeigenfunctions}
\end{figure}
Figure~\ref{fig:Lstar1seqmeigenfunctions} shows a representative set of anomalous eigenfunctions for the case when $m=s=4\to7$ and $\hat{L}=1$.
\begin{figure}\vspace{10pt}
\includegraphics[width=\linewidth,clip]{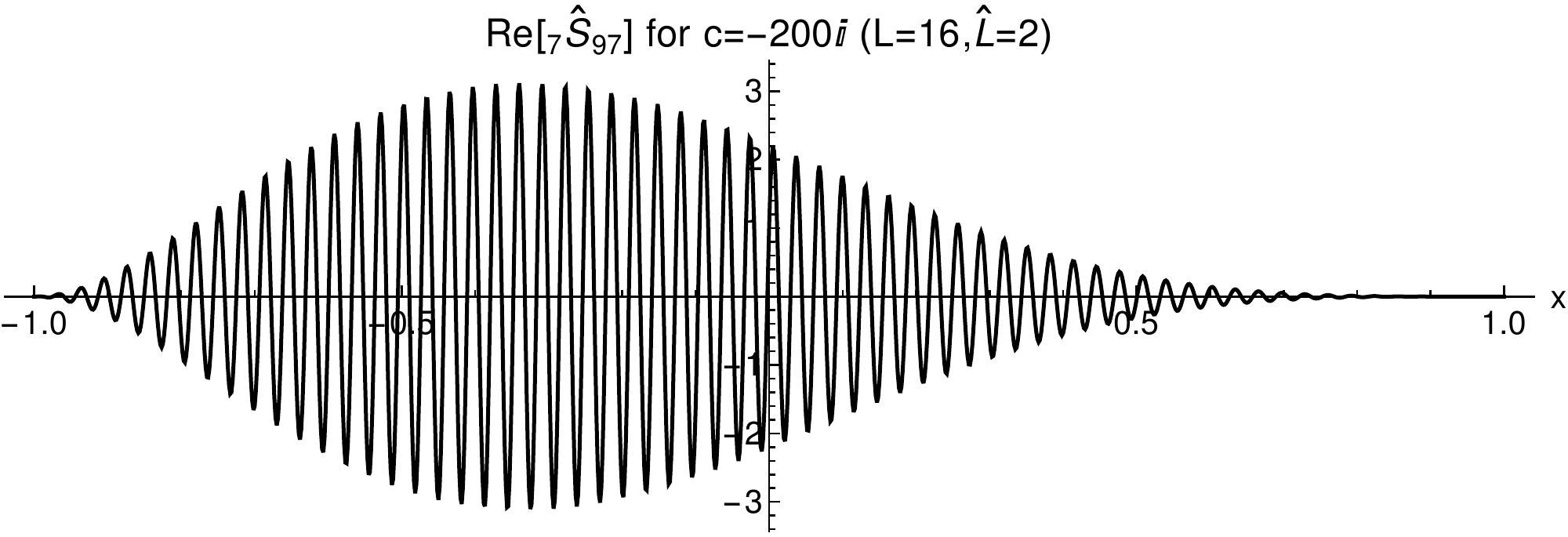}
    \caption{The real part of the $\hat{L}=2$ eigenvector solution  $\swSanom{7}{97}{x}{c}$ at $c=-200i$.  This solution is Type-$3$ anomalous.  A Type-$1$ anomalous solution might exist as $\swSanom{6}{86}{x}{c}$, but was not found after searching as far out as $L=378$.}
    \label{fig:Lstar2seqmeigenfunctions}
\end{figure}
Figure~\ref{fig:Lstar2seqmeigenfunctions} shows a representative  anomalous eigenfunction for the case when $m=s=7$ and $\hat{L}=2$.
\begin{figure}\vspace{10pt}
\begin{tabular}{cc}
\includegraphics[width=0.5\linewidth,clip]{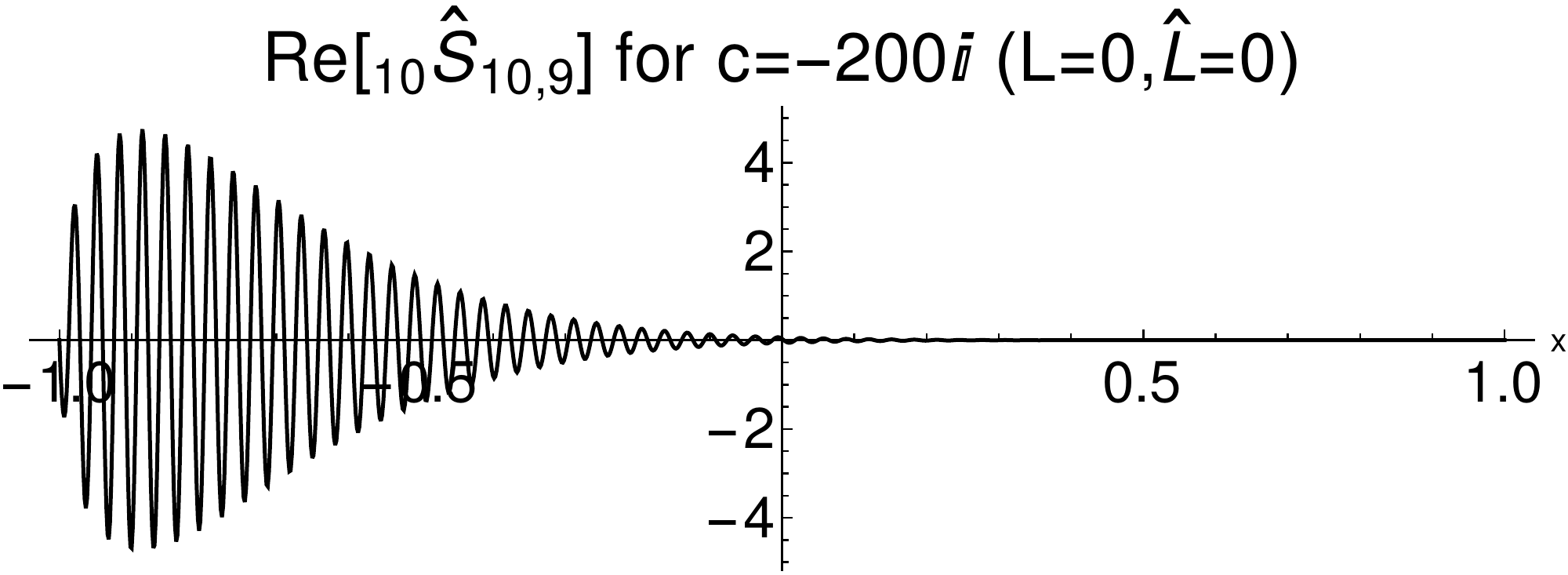} &
\includegraphics[width=0.5\linewidth,clip]{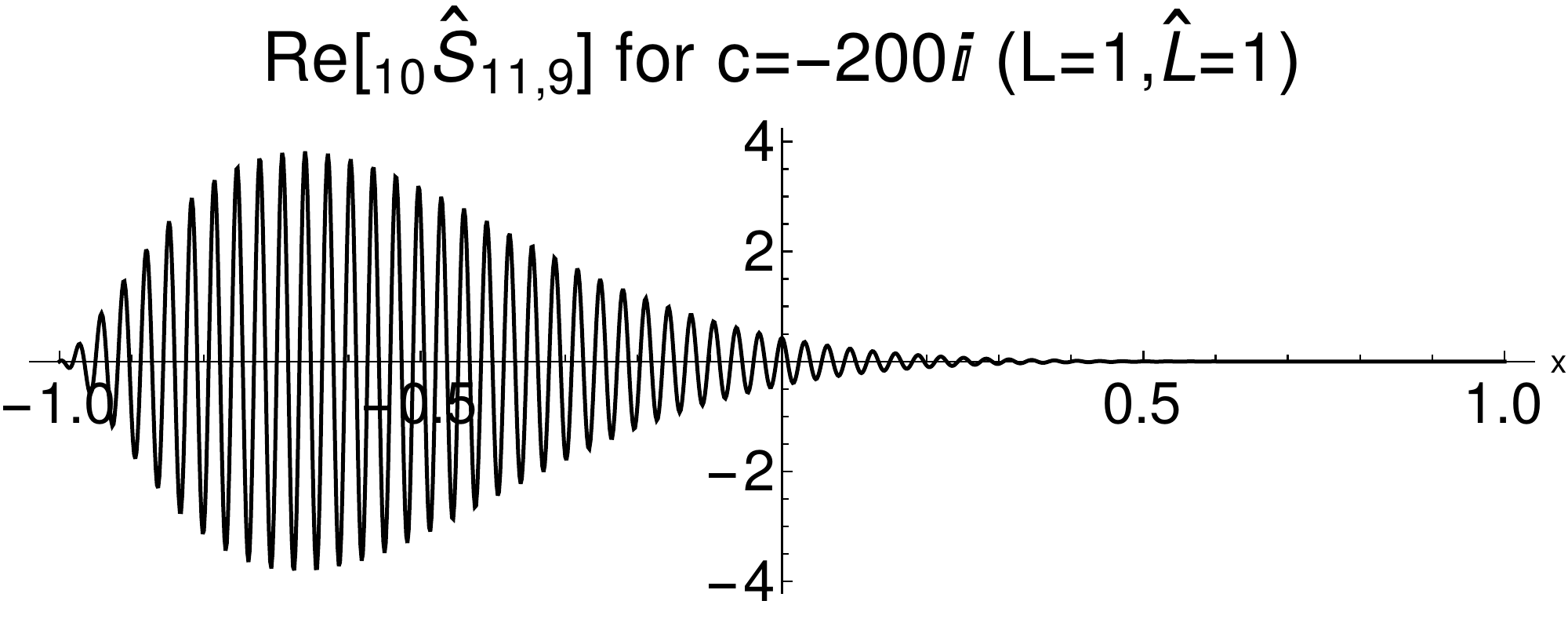} \\
\includegraphics[width=0.5\linewidth,clip]{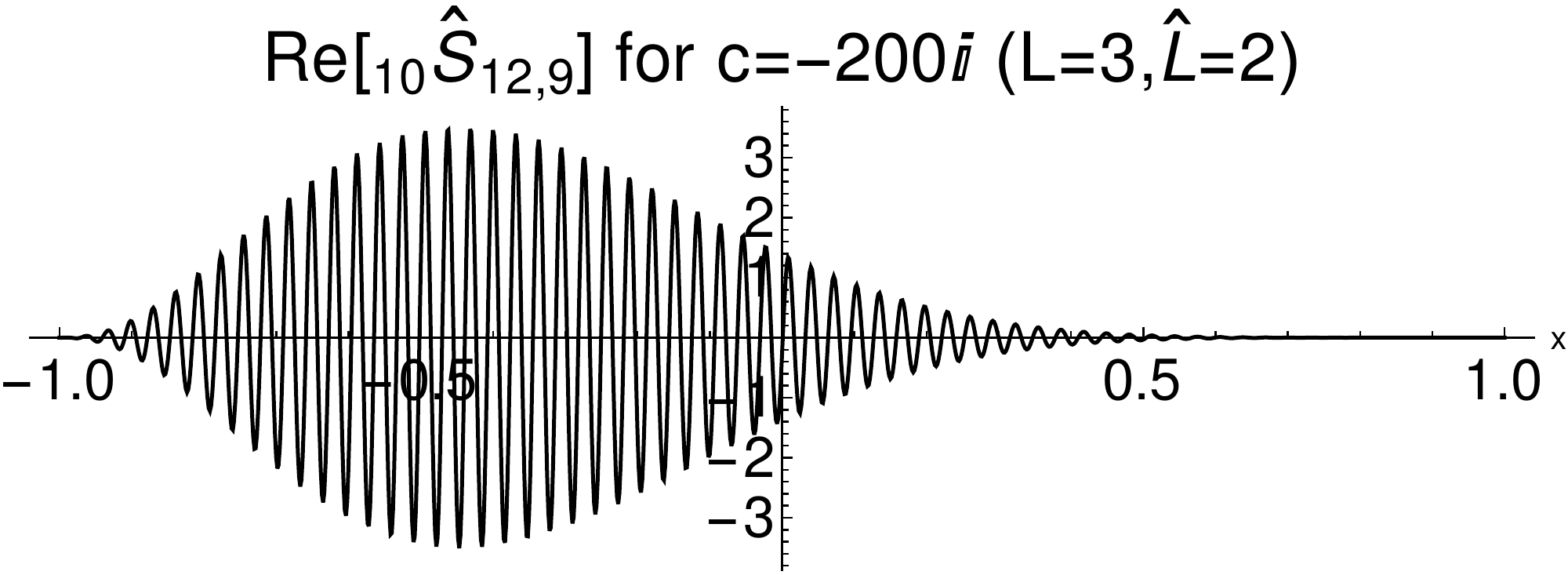} &
\includegraphics[width=0.5\linewidth,clip]{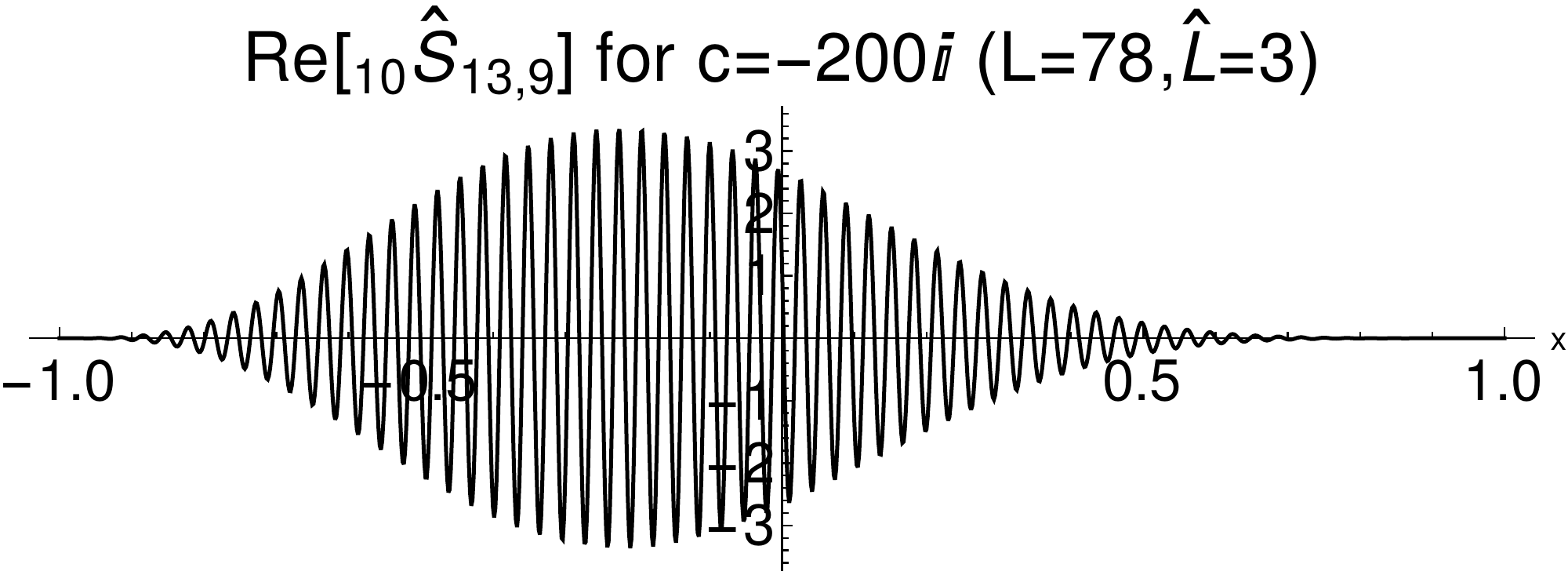}
\end{tabular}
    \caption{The real part of the $\hat{L}=0\to3$ eigenvector solution  $\swSanom{10}{\ell9}{x}{c}$ with $s\ne m$ and $c=-200i$.  The anomalous type of each decreases by $4$ from Type-$15$ for $\hat{L}=0$ to Type-$3$ for $\hat{L}=3$.}
    \label{fig:Lstar2snemeigenfunctions}
\end{figure}
And finally, Fig.~\ref{fig:Lstar2snemeigenfunctions} shows a set of representative  anomalous eigenfunctions for the case when $m=9$ and $s=10$ and $\hat{L}=0\to3$.

\section{Conclusion}
\label{sec:conclusion}

In this paper, we have explored the results of a thorough numerical investigation of solutions to the Angular Teukolsky Equation, Eq.~(\ref{eqn:Angular Teukolsky Equation}), for purely imaginary values of the oblateness parameter $c$.  Our initial goal was to use these numerical solutions to construct an analytic expression for the eigenvalues $\scA{s}{\ell m}{c}$ of the Angular Teukolsky Equation (also referred to as the separation constant), in the limit of large imaginary values for $c$, the so-called prolate asymptotic limit.  Only limited success toward this goal had been achieved previously using purely analytic techniques (see Ref.~\cite{berti2005eigenvalues} and references therein).  Our numerical solutions, however, revealed that the prolate asymptotic limit actually exhibits two distinctly different asymptotic behaviors.

To explore the prolate asymptotic limit, we constructed sequences of solutions parameterized by the magnitude of $c$.  A family of sequences is labeled by the values of the spin-weight $s$, and azimuthal index $m$ which, along with $c$, fix the free parameters of the Angular Teukolsky Equation.  The resulting eigenvalue problem has an infinity of solutions usually parameterized by the harmonic index $\ell$.  In the spherical limit, $c\to0$, the eigenfunctions are simply the spin-weighted spherical functions which are proportional to the Wigner-d functions $\swS{s}{\ell m}{x=\cos\theta}{0} \propto d^\ell_{m(-s)}(\theta)$, and the eigenvalues are simply $\scA{s}{\ell m}{0}=\ell(\ell+1)-s(s+1)$.  For fixed values of $m$ and $s$, each purely imaginary value of $c$ yields a non-degenerate set of eigenvalues which forms a smooth sequence parameterized by $c$.

As $|c|$ gets large, we found that the sequences exhibit two distinctly different leading-order asymptotic behaviors.  We labeled one type of behavior as ``normal'' because it followed the leading order behavior of linear growth in $|c|$ of the eigenvalue predicted in Ref.~\cite{berti2005eigenvalues} for the prolate asymptotic limit.  Using our high-accuracy numerical solutions for the eigenvalues, we were able to construct the analytic form for $4$ additional terms in the normal prolate asymptotic expansion for the eigenvalues.  The full expression for $\scAnorm{s}{\ell m}{c}$ is given in Eq.~(\ref{eqn:prolate normal solution}), where the over-bar denotes the result as applying to normal sequences.  The index $\ell$ labeling $\scAnorm{s}{\ell m}{c}$ is computed as $\ell=\bar{L}+\max(|m|,|s|)$, with $\bar{L}$ appearing in Eq.~(\ref{eqn:prolate normal solution}) and also being associated with the number of zero crossings of the real part of the associated eigenfunction $\swSnorm{s}{\ell m}{x}{c}$.  As described in Sec.~\ref{sec:prolate behavior}, there will be $\bar{L}$ real zero crossing of the eigenfunction within the inner region defined by $|x|<\sqrt{(2\bar{L}+1)/|c|}$.

In the spherical limit, $c\to0$, the real eigenfunctions $\swS{s}{\ell m}{x}{0}$ also have $L$ zero crossings, where $\ell=L+\max(|m|,|s|)$.  We can denote the eigenvalue and eigenfunction for any normal sequence in terms of the $\ell$ index from the spherical limit via $\scA{s}{\ell m}{c}$ and $\swS{s}{\ell m}{x}{c}$, or in terms of the $\ell$ index from the normal prolate asymptotic limit via $\scAnorm{s}{\ell m}{c}$ and $\swSnorm{s}{\ell m}{x}{c}$.  It is tempting to assume that the value of the eigenvalue index $\ell$ in these two notations is the same for a given sequence, but they can be different.  The reason for this difference is that some sequences exhibit an anomalous behavior as they approach the prolate asymptotic limit, with the eigenvalue growing as $|c|^2$ as opposed to the linear growth of normal sequences. 

For fixed values of $m$ and $s$, there exists a countably infinite set of eigensolution sequences indexed by $\ell$.  In all cases, most of the sequences have normal behavior in the prolate asymptotic limit.  For many $(m,s)$ pairs, all of the sequences have normal behavior.  But, as shown in Tables~\ref{table:anomalous data 1} and \ref{table:anomalous data 2}, there are also many $(m,s)$ pairs for which a finite set of the sequences have anomalous prolate asymptotic behavior.  The presence of even one anomalous sequence within a set of sequences with fixed $(m,s)$ means that the number of real zero crossing of the eigenfunction will be different in the spherical and asymptotic limits for most sequences.  That is, the values of $L$ and $\bar{L}$ will be different for most sequences.  In fact, if we examine the number of real zero crossings along $x$ of $\swS{s}{\ell m}{x}{c}$ as we vary $c$ along a normal sequence, the number of crossings will change regardless of the presence of anomalous sequences.  This is because of the behavior of the eigenfunction in the normal prolate asymptotic limit.  As review in Sec.~\ref{sec:prolate behavior}, there are $\bar{L}$ real zero crossings in the inner region.  But there will also be some predictable number of real zero crossings in the outer regions.  This is clearly illustrated for $\swS{2}{33}{x}{c}$ in Figs.~\ref{fig:eigenvector 3x2x0} and \ref{fig:eigenvector 3x2x0 wide}.  In Fig.~\ref{fig:eigenvector 3x2x0} with $c=-100i$, the solution is well within the asymptotic limit and the two additional zero crossings in the outer regions are exponentially suppressed though still detectable.  In Fig~\ref{fig:eigenvector 3x2x0 wide} with $c=-10i$, the solution is still in the transition region between spherical and asymptotic behavior and we see two real zero crossing for a case where $L=\bar{L}=0$.

The behavior just described above presents something of a problem if one is trying to classify some single eigensolution with known values of $m$ and $s$.  If the solution is in the transition region between spherical and asymptotic behavior, then the value of the eigenvalue cannot be compared to either asymptotic form, nor can the number of real zero crossings of the eigenfunction be used as a reliable discriminant.  But, given a single eigensolution, that solution can be extended toward either asymptotic limit and classified once it is sufficiently close to either one.  If it is extended toward the spherical limit, then the value of $L$ can be determined either from $\swS{s}{\ell m}{s}{c}$ or $\scA{s}{\ell m}{c}$.  If the sequence is normal and it is extended toward the prolate asymptotic limit, then the value of $\bar{L}$ can be determined either from $\swSnorm{s}{\ell m}{x}{c}$ or $\scAnorm{s}{\ell m}{c}$.  It is important to emphasize that a given normal sequence can be denoted either as the eigensolution pair $(\swS{s}{\ell m}{s}{c},\scA{s}{\ell m}{c})$ or as $(\swSnorm{s}{\ell m}{x}{c},\scAnorm{s}{\ell m}{c})$ {\em for any value of $c$ along the sequence}.  But, the correspondence between these two notations can only be known if the sequence has been extended to both limits.  In this case, we know both $L$, and $\bar{L}$ for the given sequence and, by definition, the value of $\sNlm{s}{\ell m}=L-\bar{L}$.  Recall from Sec.~\ref{sec:Normal Sequences} that $\sNlm{s}{\ell m}$ is the number of anomalous eigensolutions for given $m$ and $s$ with values of $\ell$ smaller than $\ell=L+\max(|m|,|s|)$.

The second type of asymptotic behavior seen in the the prolate solutions manifests in the eigenvalue as quadratic growth, $\scA{s}{\ell m}{c}\approx|c|^2$.  Generically, we have labeled eigensolutions along such a sequence as ``anomalous'', but these sequences can be sub-categorized as ``Type-$\mathcal{N}$ anomalous'' where $\mathcal{N}$ is an odd integer.  This classification is based entirely on the asymptotic behavior of the eigenvalue.  We find that the Type-$\mathcal{N}$ anomalous prolate asymptotic eigenvalue $\scAanom{s}{\ell m}{c}$ is equal to Eq.~(\ref{eqn:oblate solution}), with ${}_sq_{\ell m}$ replaced by ${}_s\hat{q}_{\ell m}$ as defined in Eq.~(\ref{eqn:anomalous sqlm}), up to but not including terms of order $c^{-\mathcal{N}}$.  In all cases, the term at order $c^{-\mathcal{N}}$ in Eq.~(\ref{eqn:oblate solution}) should be purely imaginary for the prolate case.  However, we find numerically that the term at order $c^{-\mathcal{N}}$ for Type-$\mathcal{N}$ anomalous prolate asymptotic sequences also includes a real part.  Moreover, the full behavior of the term at order $c^{-\mathcal{N}}$ seems to be a correction to the term from the oblate case, given by Eq.~(\ref{eqn:oblate solution A1}) for Type-$1$ solutions and by Eq.~(\ref{eqn:oblate solution A3}) for Type-$3$ solutions.  Such a correction term seems to behave somewhat as if it were proportional to the exponential of a purely imaginary function of $c$, but we have not been able to find a suitable form for this.  Figure~\ref{fig:loglogReImanomresiduals} shows an example of the behavior of the correction terms for a Type-$1$ anomalous sequence for $m=s=4$.  In this figure, the lines labeled by $\hat{L}=0$ correspond to a Type-$5$ sequence and the lines labeled by $\hat{L}=1$ correspond to the Type-$1$ sequence of interest.  The $\hat{L}=1$ lines in this figure clearly show that the corrections to Eq.~(\ref{eqn:oblate solution A1}) are complex and highly oscillatory while the magnitudes of both the real and the imaginary parts decay on average like $c^{-1}$.  We cannot separate out the correction term for the Type-$5$ sequence seen in these plots because we do not have an expression for the oblate sequence at order $c^{-5}$.  The upper-left plot in Fig.~\ref{fig:logloganomresiduals} shows the magnitude of the Type-$1$ correction term for the same $m=s=4$ sequence, and we can clearly see that the rapid oscillations vanish asymptotically.  This non-oscillatory behavior for the magnitude of the correction term seems to be generic and is the foundation for our conjecture that the correction terms are proportional to the exponential of a purely imaginary function of $c$.

For any anomalous sequence, its type can be determined easily as 
\begin{equation}
	\mathcal{N}\equiv-2{}_s\hat{q}_{\ell|m|}-1.  
\end{equation}
In fact, numerical evidence shows that $\mathcal{N}\ge1$ is a necessary, but not sufficient condition for a sequence to be anomalous.  Without explicitly constructing any solution sequences, given values of $m$ and $s$, and taking values of $\ell=\hat{L}+\max(|m|,|s|)$ for $\hat{L}\ge0$, any combination for which $\mathcal{N}\ge1$ is most likely Type-$\mathcal{N}$ anomalous.  As discussed in Sec.~\ref{sec:predict anomalous}, we have not been able to determine whether certain sequences with $\mathcal{N}\ge1$ are missing from the set of anomalous sequences or if they simply occur at a very large value of $L$.  Here, it is perhaps useful to remember that the value of $\ell$ in ${}_s\hat{q}_{\ell|m|}$ used to construct $\mathcal{N}$ is associated with $\hat{L}$ which takes on consecutive values for all of the anomalous sequences for given $m$ and $s$.  The value of $L$ is used to construct the $\ell$ used in $\sNlm{s}{\ell m}$ which defines the position of each anomalous sequence in the set of eigensolutions in the spherical limit($c\to0$).  In summary, $\mathcal{N}\le0$ for $\hat{L}=0$ allows us to be certain that no anomalous sequences are present for given $m$ and $s$.  The number of values of $\hat{L}$ which yield $\mathcal{N}\ge1$, we conjecture, gives us an upper limit on the number of anomalous sequences for given $m$ and $s$
\begin{equation}
	N_{anom}\le\max\left(\lceil(|s|-\bigl||m|-|s|\bigr|-1)/2\rceil,0\right),
\end{equation}
where $\lceil x\rceil$ denotes the ceiling of $x$.  While the number of anomalous sequences is not guaranteed, our evidence suggests that the minimum number is likely no more than one less than the maximum.

The condition $\mathcal{N}\ge1$ gives us insight into the possible number of anomalous sequences for give $m$ and $s$, but the allowed values for $\mathcal{N}$ provide no information about the value of $\sNlm{s}{\ell m}$ for a given anomalous sequence.  Recall that for an anomalous sequence, $\sNlm{s}{\ell m}=\hat{L}$, explicitly relating the values of $L$ and $\hat{L}$ for a specific anomalous sequence, and thus specifying the location of the asymptotic sequence in the spherical limit($c\to0$). The only constraints on the location of any prolate anomalous sequence within the set of all prolate sequences have been obtained empirically for the special cases where $L=\hat{L}$.  By explicit construction, we find that in the region $\mathcal{A}$ defined by $L=0$ and bound by $11|s|\ge7|m|+9$ and $|s|\le10|m|-24$, all sequences are anomalous with $\bar{L}=0$ if $|m|\le20$ and $|s|\le20$.  It may be true for larger magnitudes of $m$ and $s$, but we have no analytic proof of this.  Many additional anomalous sequences with $L=\hat{L}$ exist within the region $\mathcal{D}$ defined by $L\ge1$, $14L\le11|s|-7|m|-10$, and $25L\le9|m|-2|s|-15$.  Unfortunately, not all of the points within this region are guaranteed to be anomalous.  Within the limits of $|m|\le20$ and $|s|\le20$, we have found by explicit construction that $8$ points in each quadrant are within $\mathcal{D}$ but are normal.  These $8$ points are listed in Table~\ref{table:missing anomalous}.  While sequences with parameters within region $\mathcal{D}$ are not guaranteed to be anomalous, of the known $8$ know sequences that are normal, $5$ lie on one of the bounding planes and the remaining $3$ just within one of them.  So, sequences with parameters within $\mathcal{D}$ are very likely to be anomalous unless they lie on or adjacent to one of the bounding planes.

Using our high-accuracy numerical solutions to the Angular Teukolsky Equation, we also explored the behavior of the prolate anomalous eigenfunctions $\swSanom{s}{\ell m}{x}{c}$.  Most notable is that, in the asymptotic limit, these eigenfunctions behave nothing like the eigenfunctions on normal sequences.  In the asymptotic limit, the anomalous eigenfunctions display a number of real zero crossings along $x$ that is not correlated with $\hat{L}$, but instead rapidly increases as $|c|$ increases.  And, while the number of real zero crossing changes with $c$, the envelope modulating the oscillations seems to be relatively insensitive to changes in $c$ in the asymptotic regime.  The number of real zero crossings does seem to be correlated with the anomalous eigenvalue's position in the list of all eigenvalues, when sorted by the magnitude of the real part of the eigenvalue.  As the real part of each anomalous sequence's eigenvalue grows as $|c|^2$, it quickly crosses successive normal sequences as $|c|$ increases as seen in Figs.~\ref{fig:2x2 anomalous} and \ref{fig:2x2 anomalous small}.

The path of anomalous eigenvalue sequences through the family of all eigenvalue sequences for given $m$ and $s$ also seems to be correlated with the region of transition to asymptotic behavior for both normal and anomalous sequences.  Figures~\ref{fig:2x2 anomalous small} and \ref{fig:ImA9x10anom} show that, for sequences with $\hat{L}>L$, the value of $c$ at which the real part of the anomalous eigenvalue deflects from a normal sequence mark the transition into prolate anomalous asymptotic behavior.  This value of $c$ is very close to the point at which the imaginary part of the anomalous eigenvalue crosses near the peak of a normal sequence, giving further support to this conjecture.  It also seems that normal sequences, for which the real part of the eigenvalue cross that of a prolate anomalous sequence, seem to transition to normal asymptotic behavior at values of $|c|$ slightly larger than the crossing point.  As seen in Fig.~\ref{fig:realimag sample plot}, even when no anomalous sequences are present for given $m$ and $s$, it seems that the transition to prolate normal asymptotic behavior seems to be in the region where the real part of the eigenvalue is comparable to $|c|^2$.

The numerical solutions of the Angular Teukolsky Equation in the prolate limit which we have examined in this paper have provided substantial new insights into the behavior of prolate solutions in general.  We hope that these insights and various conjectures will motivate and aid future analytic studies.  We also hope that the numerically determined asymptotic expansions for both the normal and anomalous sequences will prove useful for high-accuracy approximations to the SWSHs for use in other works.

\acknowledgments Some computations were performed on the Wake Forest
University DEAC Cluster, a centrally managed resource with support
provided in part by the University.

\appendix
\section{Tables of Anomalous Sequences}
\label{sec:anomalous table appendix}
The presence of eigensolution sequences with anomalous behavior complicates the task of matching prolate sequences in the asymptotic regime with corresponding sequences in the spherical limit.  In the spherical limit, each eigensolution is conveniently labeled by $L\ge0$ which gives the number of zero crossings of the eigenfunction $\swS{s}{\ell m}{x}{0}$ and is related to the more commonly used harmonic index $\ell$ by Eq.~(\ref{eqn:L def}), and the eigenvalues are given by $\scA{s}{\ell m}{0}=\ell(\ell+1)-s(s+1)$.  Sequences of eigensolutions, parameterized by $c$, smoothly connect prolate solutions at $c=0$ to the asymptotic limit of large $|c|$, where the eigensolutions can take on either normal or anomalous behavior.  Normal asymptotic solutions can also be labeled by $\bar{L}=L-\sNlm{s}{\ell m}$, and $\sNlm{s}{\ell m}$ is the number of anomalous eigensolutions that exist for $m$ and $s$ with smaller values of $\ell$.  Anomalous asymptotic solutions  can also be labeled by $\hat{L}=\sNlm{s}{\ell m}$.

Tables~\ref{table:anomalous data 1} and \ref{table:anomalous data 2} explicitly list all of the anomalous sequences we have found for $|m|\le10$ and $|s|\le10$.  The first two columns of each table list the values of $|m|$ and $|s|$.  The third column in each table lists the harmonic index $\ell$ as defined in the spherical limit ($c\to0$) for that sequence.  These three indices are the most common set used to specify eigensolutions in the spherical limit.  The fourth column in each table lists the value of $L$ for the sequence as given by Eq.~(\ref{eqn:L def}).  The fifth column in each table lists the value of $\sNlm{s}{\ell m}$ for the anomalous sequence labeled by the common set of indices used in the spherical limit.

The sixth column lists the values of $n$ which is the slope of a log-log plot of the difference between the numerical solution for the eigenvalue and the base prolate anomalous fit given by Eq.~(\ref{eqn:oblate solution}) with $L$ replaced by $\hat{L}=\sNlm{s}{\ell m}$ and ${}_sq_{\ell m}$ replaces by ${}_s\hat{q}_{\ell m}$ as defined by Eq.~(\ref{eqn:anomalous sqlm}).  The value of $n$ gives the power in $c$ at which the base prolate anomalous fit requires correction.  Because Eq.~(\ref{eqn:oblate solution}) is only known to order $c^{-4}$, the minimum value is $n=-5$.  The seventh column lists our conjectured value, $2{}_s\hat{q}_{\ell|m|}+1$, for the power in $c$ at which the base prolate anomalous fit requires correction.  The negative of this conjectured value defines the specific type $\mathcal{N}=-2{}_s\hat{q}_{\ell|m|}-1$ of the given anomalous sequence.  Note that for $2{}_s\hat{q}_{\ell|m|}+1\ge-5$, there is agreement with $n$.

The final column in each table lists separately the slopes of log-log plots of the real and imaginary parts of the same residual whose magnitude was plotted to obtain $n$.  For $2{}_s\hat{q}_{\ell|m|}+1\ge-5$, the values of $\text{Re}[n]$ and $\text{Im}[n]$ should be equal to $n$.  For $2{}_s\hat{q}_{\ell|m|}+1\le-7$, we expect the values to be $\text{Re}[n]=-6$ and $\text{Im}[n]=-5$ because the real part of the base prolate anomalous fit vanishes at order $c^{-5}$.  We note one unusual value for the case $|m|=9$, $|s|=10$, $\ell=13$, where $\text{Re}[n]=-7$ and we should expect this to be $-6$\footnote{See Sec.~\ref{sec:anomalous form} for additional discussion.}.

The tables only give values for $\sNlm{s}{\ell m}$ for anomalous sequences.  However, these values are all that is necessary to specify $\sNlm{s}{\ell m}$ for all values of $\ell$ for given $m$ and $s$.  For given $m$ and $s$, if there are no anomalous sequences, then $\sNlm{s}{\ell m}=0$ for all $\ell$.  For example $\sNlm{2}{\ell 1}=0$ for all $\ell$.  For $|m|=2$ and $|s|=2$, we have $\sNlm{2}{\ell 2}=0$ for $\ell\le3$ while $\sNlm{2}{\ell 2}=1$ for $\ell>3$.  As a final example, for $|m|=10$ and $|s|=10$, we have $\sNlm{10}{\ell 10}=0$ for $\ell=10$, $\sNlm{10}{\ell 10}=1$ for $\ell=11$, $\sNlm{10}{\ell 10}=2$ for $12\le\ell<25$, and $\sNlm{2}{\ell 2}=3$ for $25\le\ell$.  In this last example, we would expect that a fifth anomalous sequence of Type-$1$ exists with $\sNlm{2}{\ell 2}=4$ for some large value of $\ell$.  Unfortunately, finding anomalous sequences with very large values of $L$ is extremely expensive computationally\footnote{See Sec.~\ref{sec:predict anomalous} for additional discussion.}.  The following sets of sequences may be missing a Type-$1$ anomalous sequence for the same reason:
$(|m|,|s|)\in\{(4,8\to10),(6,6\to10),(8,6\to10),(10,7\to10)\}$.

\begin{table}
 \begin{tabular}{ccc|cc|cc|c}
 $|m|$ & $|s|$ & $\ell$ & $L$ & ${}_sN_{\ell m}$ & $n$ & $2{}_s\hat{q}_{\ell|m|}+1$ & $\{\text{Re}[n],\text{Im}[n]\}$ \\
\hline
 2 & 2 & 3 & 1 & 0 & -1 & -1 & \{-1,-1\} \\
 2 & 3 & 6 & 3 & 0 & -1 & -1 & \{-1,-1\} \\
 2 & 4 & 9 & 5 & 0 & -1 & -1 & \{-1,-1\} \\
 2 & 5 & 13 & 8 & 0 & -1 & -1 & \{-1,-1\} \\
 2 & 6 & 18 & 12 & 0 & -1 & -1 & \{-1,-1\} \\
 2 & 7 & 23 & 16 & 0 & -1 & -1 & \{-1,-1\} \\
 3 & 3 & 3 & 0 & 0 & -3 & -3 & \{-3,-3\} \\
 3 & 4 & 4 & 0 & 0 & -3 & -3 & \{-3,-3\} \\
 3 & 5 & 5 & 0 & 0 & -3 & -3 & \{-3,-3\} \\
 3 & 6 & 6 & 0 & 0 & -3 & -3 & \{-3,-3\} \\
 3 & 7 & 8 & 1 & 0 & -3 & -3 & \{-3,-3\} \\
 3 & 8 & 9 & 1 & 0 & -3 & -3 & \{-3,-3\} \\
 3 & 9 & 10 & 1 & 0 & -3 & -3 & \{-3,-3\} \\
 3 & 10 & 12 & 2 & 0 & -3 & -3 & \{-3,-3\} \\
 4 & 3 & 13 & 9 & 0 & -1 & -1 & \{-1,-1\} \\
 4 & 4 & 4 & 0 & 0 & -5 & -5 & \{-5,-5\} \\
 \text{} & \text{} & 32 & 28 & 1 & -1 & -1 & \{-1,-1\} \\
 4 & 5 & 5 & 0 & 0 & -5 & -5 & \{-5,-5\} \\
 \text{} & \text{} & 60 & 55 & 1 & -1 & -1 & \{-1,-1\} \\
 4 & 6 & 6 & 0 & 0 & -5 & -5 & \{-5,-5\} \\
 \text{} & \text{} & 99 & 93 & 1 & -1 & -1 & \{-1,-1\} \\
 4 & 7 & 7 & 0 & 0 & -5 & -5 & \{-5,-5\} \\
 \text{} & \text{} & 143 & 136 & 1 & -1 & -1 & \{-1,-1\} \\
 4 & 8 & 8 & 0 & 0 & -5 & -5 & \{-5,-5\} \\
 4 & 9 & 9 & 0 & 0 & -5 & -5 & \{-5,-5\} \\
 4 & 10 & 10 & 0 & 0 & -5 & -5 & \{-5,-5\} \\
 5 & 4 & 5 & 0 & 0 & -3 & -3 & \{-3,-3\} \\
 5 & 5 & 5 & 0 & 0 & -5 & -7 & \{-6,-5\} \\
 \text{} & \text{} & 8 & 3 & 1 & -3 & -3 & \{-3,-3\} \\
 5 & 6 & 6 & 0 & 0 & -5 & -7 & \{-6,-5\} \\
 \text{} & \text{} & 11 & 5 & 1 & -3 & -3 & \{-3,-3\} \\
 5 & 7 & 7 & 0 & 0 & -5 & -7 & \{-6,-5\} \\
 \text{} & \text{} & 14 & 7 & 1 & -3 & -3 & \{-3,-3\} \\
 5 & 8 & 8 & 0 & 0 & -5 & -7 & \{-6,-5\} \\
 \text{} & \text{} & 18 & 10 & 1 & -3 & -3 & \{-3,-3\} \\
 5 & 9 & 9 & 0 & 0 & -5 & -7 & \{-6,-5\} \\
 \text{} & \text{} & 22 & 13 & 1 & -3 & -3 & \{-3,-3\} \\
 5 & 10 & 10 & 0 & 0 & -5 & -7 & \{-6,-5\} \\
 \text{} & \text{} & 27 & 17 & 1 & -3 & -3 & \{-3,-3\} \\
 6 & 4 & 43 & 37 & 0 & -1 & -1 & \{-1,-1\} \\
 6 & 5 & 6 & 0 & 0 & -5 & -5 & \{-5,-5\} \\
 \text{} & \text{} & 118 & 112 & 1 & -1 & -1 & \{-1,-1\} \\
 6 & 6 & 6 & 0 & 0 & -5 & -9 & \{-6,-5\} \\
 \text{} & \text{} & 7 & 1 & 1 & -5 & -5 & \{-5,-5\} \\
 6 & 7 & 7 & 0 & 0 & -5 & -9 & \{-6,-5\} \\
 \text{} & \text{} & 8 & 1 & 1 & -5 & -5 & \{-5,-5\} \\
 6 & 8 & 8 & 0 & 0 & -5 & -9 & \{-6,-5\} \\
 \text{} & \text{} & 10 & 2 & 1 & -5 & -5 & \{-5,-5\} \\
 6 & 9 & 9 & 0 & 0 & -5 & -9 & \{-6,-5\} \\
 \text{} & \text{} & 12 & 3 & 1 & -5 & -5 & \{-5,-5\} \\
 6 & 10 & 10 & 0 & 0 & -5 & -9 & \{-6,-5\} \\
 \text{} & \text{} & 14 & 4 & 1 & -5 & -5 & \{-5,-5\} \\
\end{tabular}
\caption{Sequences known to have anomalous behavior.  $m$, $s$, and $\ell$ designate each sequence in terms if its behavior as $c\to0$.  $L$ is the alternate labeling as defined by Eq.~(\ref{eqn:L def}) and ${}_sN_{\ell m} = \hat{L}$ is the number of anomalous sequences that exist for $m$ and $s$ with smaller $\ell$.  See the text in Appendix~\ref{sec:anomalous table appendix} for descriptions of the last 3 columns of data.}
\label{table:anomalous data 1}
\end{table}

\begin{table}
 \begin{tabular}{ccc|cc|cc|c}
 $|m|$ & $|s|$ & $\ell$ & $L$ & ${}_sN_{\ell m}$ & $n$ & $2{}_s\hat{q}_{\ell|m|}+1$ & $\{\text{Re}[n],\text{Im}[n]\}$ \\
\hline
7 & 5 & 10 & 3 & 0 & -3 & -3 & \{-3,-3\} \\
 7 & 6 & 7 & 0 & 0 & -5 & -7 & \{-6,-5\} \\
 \text{} & \text{} & 15 & 8 & 1 & -3 & -3 & \{-3,-3\} \\
 7 & 7 & 7 & 0 & 0 & -5 & -11 & \{-6,-5\} \\
 \text{} & \text{} & 8 & 1 & 1 & -5 & -7 & \{-6,-5\} \\
 \text{} & \text{} & 23 & 16 & 2 & -3 & -3 & \{-3,-3\} \\
 7 & 8 & 8 & 0 & 0 & -5 & -11 & \{-6,-5\} \\
 \text{} & \text{} & 9 & 1 & 1 & -5 & -7 & \{-6,-5\} \\
 \text{} & \text{} & 31 & 23 & 2 & -3 & -3 & \{-3,-3\} \\
 7 & 9 & 9 & 0 & 0 & -5 & -11 & \{-6,-5\} \\
 \text{} & \text{} & 10 & 1 & 1 & -5 & -7 & \{-6,-5\} \\
 \text{} & \text{} & 41 & 32 & 2 & -3 & -3 & \{-3,-3\} \\
 7 & 10 & 10 & 0 & 0 & -5 & -11 & \{-6,-5\} \\
 \text{} & \text{} & 11 & 1 & 1 & -5 & -7 & \{-6,-5\} \\
 \text{} & \text{} & 53 & 43 & 2 & -3 & -3 & \{-3,-3\} \\
 8 & 6 & 8 & 0 & 0 & -5 & -5 & \{-5,-5\} \\
 8 & 7 & 8 & 0 & 0 & -5 & -9 & \{-6,-5\} \\
 \text{} & \text{} & 10 & 2 & 1 & -5 & -5 & \{-5,-5\} \\
 8 & 8 & 8 & 0 & 0 & -5 & -13 & \{-6,-5\} \\
 \text{} & \text{} & 9 & 1 & 1 & -5 & -9 & \{-6,-5\} \\
 \text{} & \text{} & 13 & 5 & 2 & -5 & -5 & \{-5,-5\} \\
 8 & 9 & 9 & 0 & 0 & -5 & -13 & \{-6,-5\} \\
 \text{} & \text{} & 10 & 1 & 1 & -5 & -9 & \{-6,-5\} \\
 \text{} & \text{} & 16 & 7 & 2 & -5 & -5 & \{-5,-5\} \\
 8 & 10 & 10 & 0 & 0 & -5 & -13 & \{-6,-5\} \\
 \text{} & \text{} & 11 & 1 & 1 & -5 & -9 & \{-6,-5\} \\
 \text{} & \text{} & 19 & 9 & 2 & -5 & -5 & \{-5,-5\} \\
 9 & 6 & 17 & 8 & 0 & -3 & -3 & \{-3,-3\} \\
 9 & 7 & 9 & 0 & 0 & -5 & -7 & \{-6,-5\} \\
 \text{} & \text{} & 29 & 20 & 1 & -3 & -3 & \{-3,-3\} \\
 9 & 8 & 9 & 0 & 0 & -5 & -11 & \{-6,-5\} \\
 \text{} & \text{} & 10 & 1 & 1 & -5 & -7 & \{-6,-5\} \\
 \text{} & \text{} & 44 & 35 & 2 & -3 & -3 & \{-3,-3\} \\
 9 & 9 & 9 & 0 & 0 & -5 & -15 & \{-6,-5\} \\
 \text{} & \text{} & 10 & 1 & 1 & -5 & -11 & \{-6,-5\} \\
 \text{} & \text{} & 12 & 3 & 2 & -5 & -7 & \{-6,-5\} \\
 \text{} & \text{} & 64 & 55 & 3 & -3 & -3 & \{-3,-3\} \\
 9 & 10 & 10 & 0 & 0 & -5 & -15 & \{-6,-5\} \\
 \text{} & \text{} & 11 & 1 & 1 & -5 & -11 & \{-6,-5\} \\
 \text{} & \text{} & 13 & 3 & 2 & -5 & -7 & \{-7,-5\} \\
 \text{} & \text{} & 88 & 78 & 3 & -3 & -3 & \{-3,-3\} \\
 10 & 7 & 11 & 1 & 0 & -5 & -5 & \{-5,-5\} \\
 10 & 8 & 10 & 0 & 0 & -5 & -9 & \{-6,-5\} \\
 \text{} & \text{} & 15 & 5 & 1 & -5 & -5 & \{-5,-5\} \\
 10 & 9 & 10 & 0 & 0 & -5 & -13 & \{-6,-5\} \\
 \text{} & \text{} & 11 & 1 & 1 & -5 & -9 & \{-6,-5\} \\
 \text{} & \text{} & 20 & 10 & 2 & -5 & -5 & \{-5,-5\} \\
 10 & 10 & 10 & 0 & 0 & -5 & -17 & \{-6,-5\} \\
 \text{} & \text{} & 11 & 1 & 1 & -5 & -13 & \{-6,-5\} \\
 \text{} & \text{} & 12 & 2 & 2 & -5 & -9 & \{-6,-5\} \\
 \text{} & \text{} & 25 & 15 & 3 & -5 & -5 & \{-5,-5\}
\end{tabular}
\caption{Sequences known to have anomalous behavior.  $m$, $s$, and $\ell$ designate each sequence in terms if its behavior as $c\to0$.  $L$ is the alternate labeling as defined by Eq.~(\ref{eqn:L def}) and ${}_sN_{\ell m} = \hat{L}$ is the number of anomalous sequences that exist for $m$ and $s$ with smaller $\ell$.  See the text in Appendix~\ref{sec:anomalous table appendix} for descriptions of the last 3 columns of data.}
\label{table:anomalous data 2}
\end{table}

\section{Correcting and Extending the Oblate Asymptotic Expansion}
\label{sec:Oblate appendix}
The oblate asymptotic expansion obtained by Breuer, Ryan, and Waller\cite{AnalyticOblate} did not agree well with our oblate numerical results.  To correct and improve this expansion, we used the same approach outline in Sec.~\ref{sec:Normal Sequences} for finding the prolate normal asymptotic expansion to explicitly fit our numerical data for the coefficients in the oblate asymptotic expansion.  As mentioned in Sec.~\ref{sec:Oblate data sets}, we modified the recurrence relation of Eq.~(\ref{eqn:cook recursion}) to move the leading order $-c^2$ asymptotic behavior from the eigenvalue to the matrix coefficients.  Removing this rapid growth in the eigenvalue improved the accuracy of the numerical solutions.  We then fit the asymptotic expansion given in Eq.~(\ref{eqn:oblate solution}) to obtain the coefficients $A_1$, $A_2$, $A_3$, and $A_4$ given by Eqs.~(\ref{eqn:oblate solution A1}--\ref{eqn:oblate solution A4}).  We also fit explicitly for the two terms that are linear and constant in $c$ within Eq.~(\ref{eqn:oblate solution}).

The coefficients were obtained using a greedy algorithm, first obtaining the form for the coefficient linear in $c$, then obtaining the forms for each subsequent coefficient at successively smaller powers of $c$.  At each step, we fit for the coefficients of the first $4$ unknown terms in the asymptotic expansion.  Once each coefficient was determined, it was included in the asymptotic expansion before fitting for the next coefficient.  Fitting each coefficient is a two-step process.  First the value of the coefficient was obtained at various values of $m$, $s$, and $L$ by directly fitting to the last 40 (largest values of $|c|$) data points from each of the numerically generate oblate sequences described in Sec.~\ref{sec:Oblate data sets}.  In the second step, this set of data points describing the coefficients as functions of $m$, $s$, and $L$ was fit to find the coefficient as an explicit function of $m$, $s$, and ${}_sq_{\ell m}$.  It is necessary to fit each coefficient as a function of ${}_sq_{\ell m}$ as defined in Eqs.~(\ref{eqn:sqlm all}) and (\ref{eqn:L def}) because of the complicated dependence of ${}_sq_{\ell m}$ on $m$, $s$, and $L$.

Each coefficient was fit using data from all sequences with values of $|m|\le10$, $|s|\le10$, and ${}_sq_{\ell m}\le14$.  The results from fitting for the linear coefficient are displayed in Table~\ref{table:oblate Am1}, and the results for the constant term are shown in Table~\ref{table:oblate A0}.  Finally, the results for the coefficients $A_{1\to4}$ are displayed respectively in Tables~\ref{table:oblate A1}--\ref{table:oblate A4}.
\begin{table}[h]
 \begin{tabular}{c| d{9} d{7}}
 &\multicolumn{1}{c}{Estimate} & \multicolumn{1}{c}{$\sigma$}  \\
\hline
$1$ & -5.0\times10^{-9} & 3.7\times10^{-10} \\
$q$ &  2.00000000 & 4.0\times10^{-11}
\end{tabular}
\caption{Oblate sequence linear fit results for the term linear in $c$.  The fit for the linear term is given in Eq.~(\ref{eqn:oblate solution}).}
\label{table:oblate Am1}
\end{table}
\begin{table}[h]
 \begin{tabular}{c| d{9} d{7}}
 &\multicolumn{1}{c}{Estimate} & \multicolumn{1}{c}{$\sigma$}  \\
\hline
  $1$ & -1.00000010 & 8.2\times10^{-9} \\
  $s$ & -2.00000000 & 1.0\times10^{-9} \\
$m^2$ &  1.00000000 & 1.6\times10^{-10} \\
$q^2$ & -1.00000000 & 6.4\times10^{-11}
\end{tabular}
\caption{Oblate sequence linear fit results for twice the constant term.  The fit for the constant term is given in Eq.~(\ref{eqn:oblate solution}).}
\label{table:oblate A0}
\end{table}
\begin{table}[h]
 \begin{tabular}{c| d{9} d{7}}
 &\multicolumn{1}{c}{Estimate} & \multicolumn{1}{c}{$\sigma$}  \\
\hline
   $1$ &  4.0\times10^{-6} & 5.6\times10^{-7} \\
   $q$ & -1.00000178 & 1.1\times10^{-7} \\
 $q^3$ & -0.99999998 & 4.6\times10^{-10} \\
$m^2q$ &  1.00000000 & 9.3\times10^{-10} \\
$ms^2$ &  1.99999999 & 1.6\times10^{-9} \\
$qs^2$ &  1.99999999 & 1.1\times10^{-9}
\end{tabular}
\caption{Oblate sequence linear fit results for $8A_1$.  The fit for $A_1$ is given in Eq.~(\ref{eqn:oblate solution A1}).}
\label{table:oblate A1}
\end{table}
\begin{table}[h]
 \begin{tabular}{c| d{9} d{7}}
 &\multicolumn{1}{c}{Estimate} & \multicolumn{1}{c}{$\sigma$}  \\
\hline
     $1$ &  -0.99974542 & 0.000076 \\
   $m^2$ &   2.00001612 & 3.3\times10^{-6} \\
   $s^2$ &   4.00000290 & 1.6\times10^{-6}\\
   $q^2$ & -10.00002405 & 1.4\times10^{-6} \\
   $m^4$ &  -1.00000013 & 3.1\times10^{-8} \\
   $q^4$ &  -4.99999977 & 5.8\times10^{-9} \\
$m^2s^2$ &   3.99999990 & 4.0\times10^{-8} \\
$m^2q^2$ &   5.99999992 & 1.6\times10^{-8} \\
$q^2s^2$ &  11.99999984 & 1.7\times10^{-8} \\
 $mqs^2$ &  15.99999979 & 2.7\times10^{-8}
\end{tabular}
\caption{Oblate sequence linear fit results for $64A_2$.  The fit for $A_2$ is given in Eq.~(\ref{eqn:oblate solution A2}).}
\label{table:oblate A2}
\end{table}
\begin{table}[h]
 \begin{tabular}{c| d{9} d{7}}
 &\multicolumn{1}{c}{Estimate} & \multicolumn{1}{c}{$\sigma$}  \\
\hline
      $1$ &   -0.00217492 & 0.0015 \\
      $q$ &  -36.99794124 & 0.00054 \\
    $q^3$ & -114.00010838 & 8.9\times10^{-6} \\
   $m^2q$ &   50.00005295 & 0.000013 \\
   $ms^2$ &   52.00005913 & 0.000022 \\
   $qs^2$ &  100.00008913 & 0.000019 \\
    $q^5$ &  -32.99999885 & 4.6\times10^{-8} \\
   $m^4q$ &  -13.00000037 & 1.0\times10^{-7} \\
   $ms^4$ &   -8.00000048 & 2.4\times10^{-7} \\
   $qs^4$ &   -8.00000093 & 2.1\times10^{-7} \\
 $m^3s^2$ &   -4.00000025 & 1.7\times10^{-7} \\
 $q^3s^2$ &   91.99999883 & 1.3\times10^{-7} \\
 $m^2q^3$ &   45.99999958 & 9.2\times10^{-8} \\
$m^2qs^2$ &   35.99999887 & 3.7\times10^{-7} \\
$mq^2s^2$ &  131.99999783 & 2.7\times10^{-7}
\end{tabular}
\caption{Oblate sequence linear fit results for $512A_3$.  The fit for $A_3$ is given in Eq.~(\ref{eqn:oblate solution A3}).}
\label{table:oblate A3}
\end{table}
\begin{table}[h]
 \begin{tabular}{c| d{9} d{7}}
 &\multicolumn{1}{c}{Estimate} & \multicolumn{1}{c}{$\sigma$}  \\
\hline
        $1$ &  -14.00498229 & 0.012 \\
      $q^2$ & -238.99817258 & 0.00084 \\
      $s^2$ &   59.99902632 & 0.00087\\
      $m^2$ &   29.99967476 & 0.0012 \\
      $q^4$ & -340.00012316 & 0.000018 \\
      $s^4$ & - 15.99998345 & 0.000012 \\
      $m^4$ &  -18.00003108 & 0.000034 \\
   $q^2s^2$ &  460.00022510 & 0.000039 \\
   $m^2q^2$ &  230.00010283 & 0.000030 \\
    $mqs^2$ &  372.00025561 & 0.000066 \\
   $m^2s^2$ &   40.00007593 & 0.000063 \\
      $q^6$ &  -62.99999840 & 1.1\times10^{-7} \\
      $m^6$ &    2.00000051 & 2.9\times10^{-7} \\
   $q^4s^2$ &  199.99999753 & 3.2\times10^{-7} \\
   $m^2q^4$ &   99.99999905 & 2.4\times10^{-7} \\
  $mq^3s^2$ &  291.99999484 & 7.5\times10^{-7} \\
   $m^4q^2$ &  -39.00000077 & 2.8\times10^{-7} \\
   $q^2s^4$ &  -48.00000368 & 6.5\times10^{-7} \\
$m^2q^2s^2$ &   59.99999575 & 1.3\times10^{-6} \\
    $mqs^4$ &  -72.00000403 & 1.4\times10^{-6} \\
  $m^3qs^2$ &  -36.00000207 & 1.1\times10^{-6} \\
   $m^2s^4$ &  -24.00000103 & 1.0\times10^{-6} \\
   $m^4s^2$ &   -4.00000064 & 6.9\times10^{-7}
\end{tabular}
\caption{Oblate sequence linear fit results for $1024A_4$.  The fit for $A_4$ is given in Eq.~(\ref{eqn:oblate solution A4}).}
\label{table:oblate A4}
\end{table}

\section{Additional Anomalous Figures}
\label{sec:additional anom figs append}
This Appendix provides a few additional figures illustrating the behavior of the normal and anomalous sequences.  Figures~\ref{fig:ReIm4x4}--\ref{fig:Re9x10} include both the normal and anomalous sequences for several cases not fully explored in the main text.  Figures~\ref{fig:TypeNAnomRe} and \ref{fig:TypeNAnomIm} compare the behavior of the anomalous sequences grouped by anomalous type.

\begin{widetext}

\begin{figure}[t]
\begin{tabular}{cc}
\includegraphics[width=0.5\linewidth,clip]{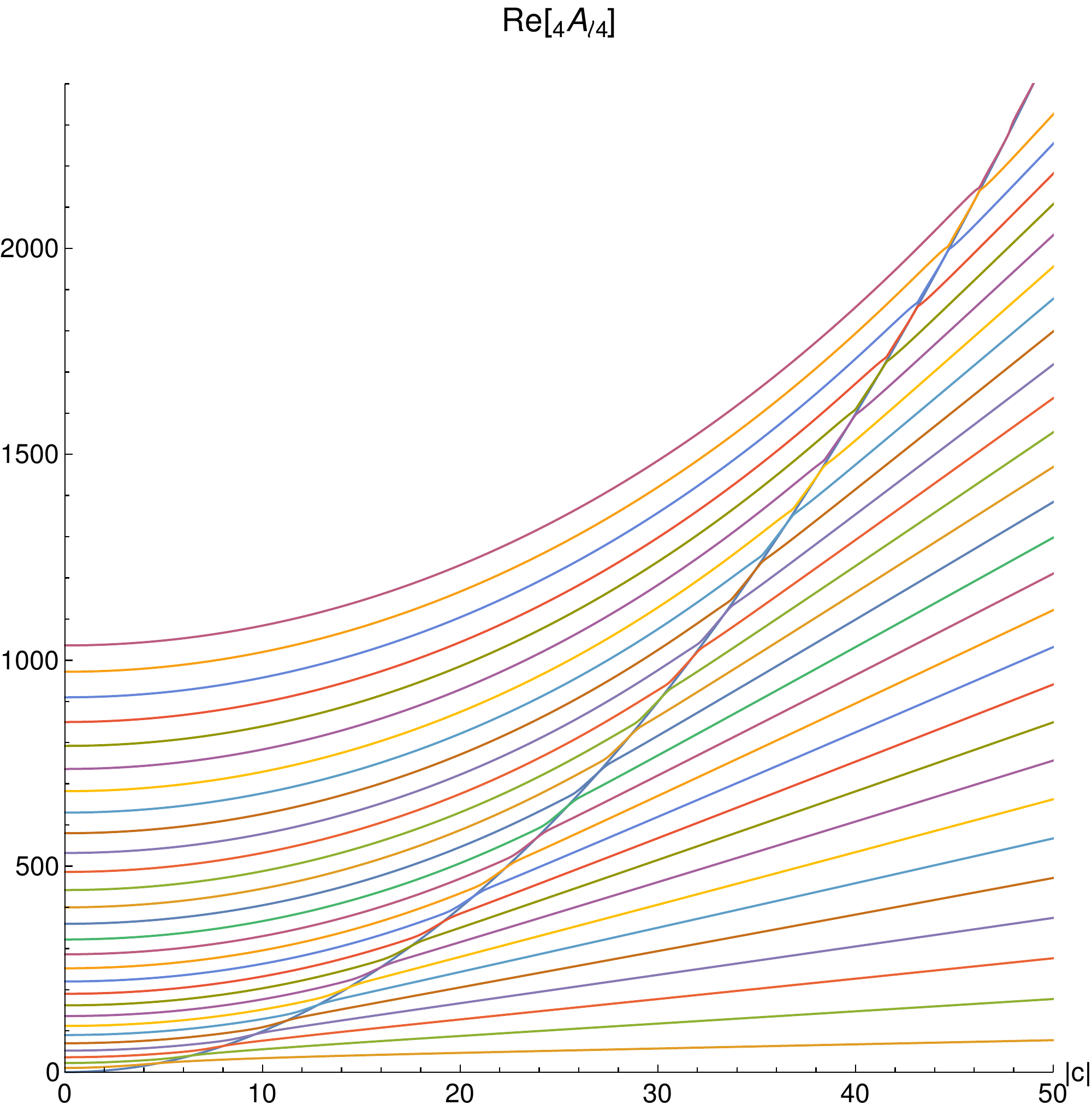} &
\includegraphics[width=0.5\linewidth,clip]{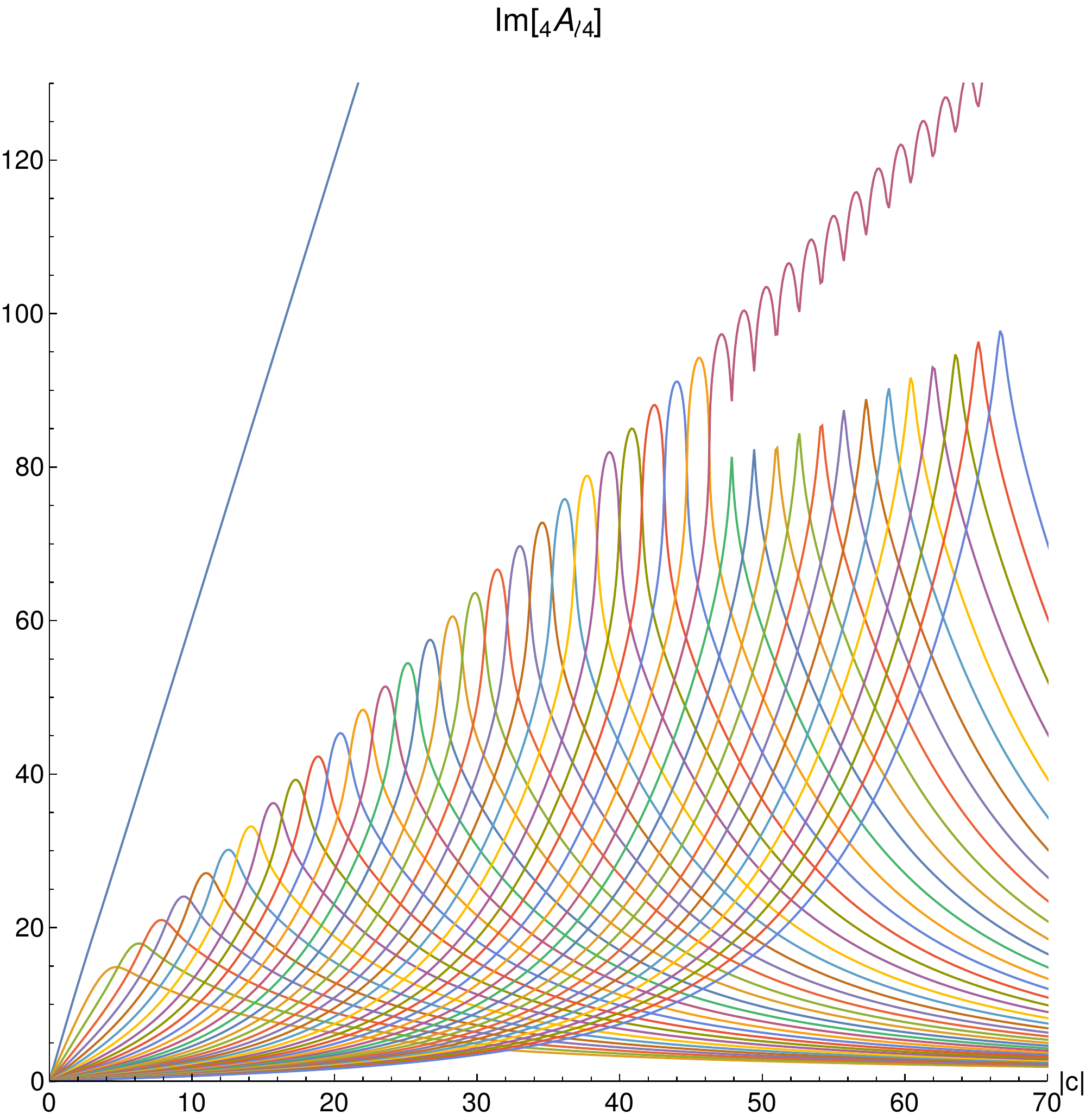}
\end{tabular}
\caption{Eigenvalue sequences for $\scA{4}{\ell4}{-i|c|}$. Note the unusual behavior of the $L=0$ and $L=28$ sequences which have quadratic leading-order behavior for the real components and linear leading-order behavior for the imaginary components.  The left plot includes the sequences with $0\le L\le28$, so the upper-most sequence is anomalous.  The right plot includes $13$ additional sequences to illustrate the behavior of these sequences following the transition of the $L=28$ sequence to anomalous prolate asymptotic behavior.}
\label{fig:ReIm4x4}
\end{figure}

\begin{figure}[b]
\begin{tabular}{cc}
\includegraphics[width=0.5\linewidth,clip]{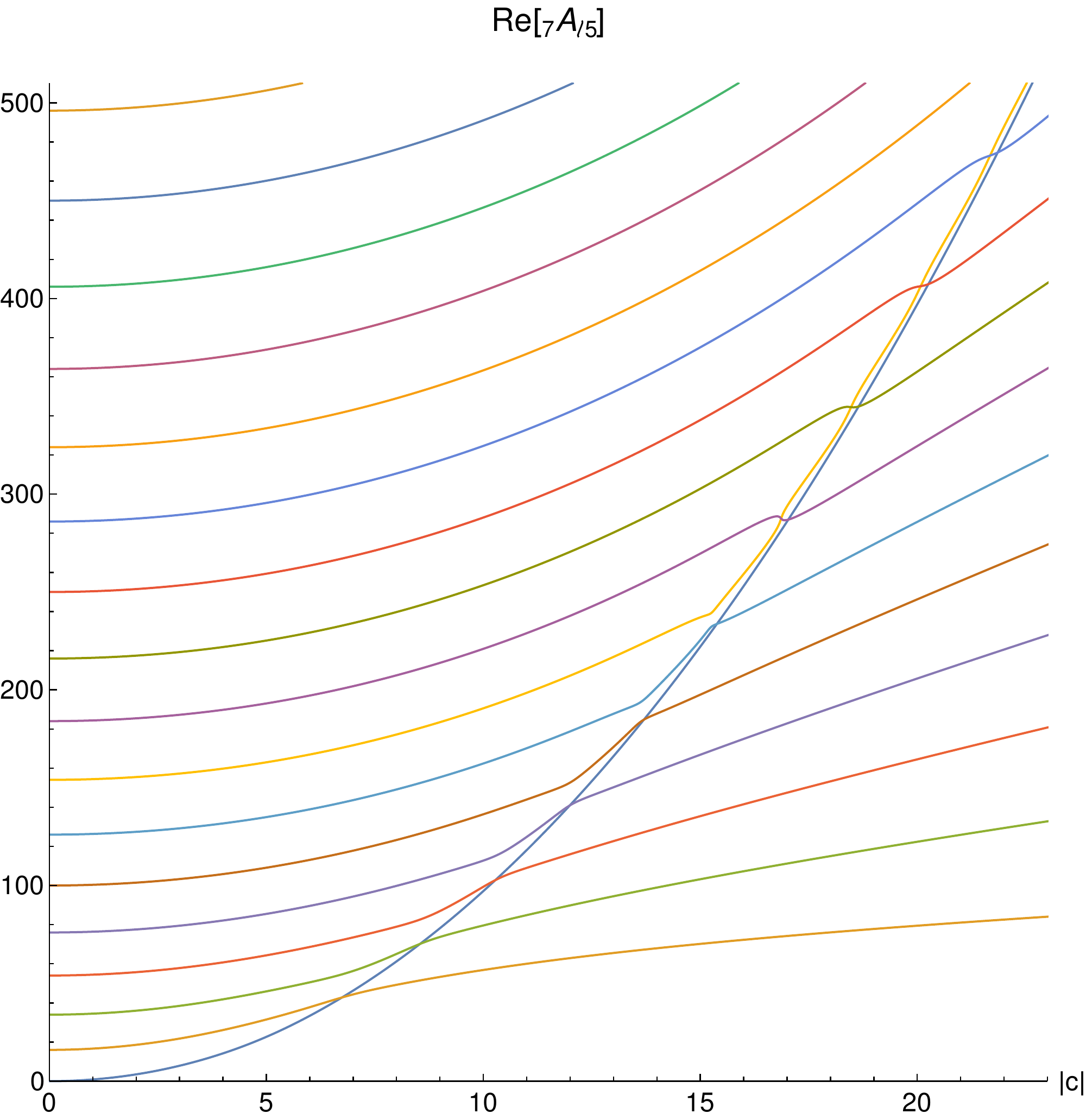} &
\includegraphics[width=0.5\linewidth,clip]{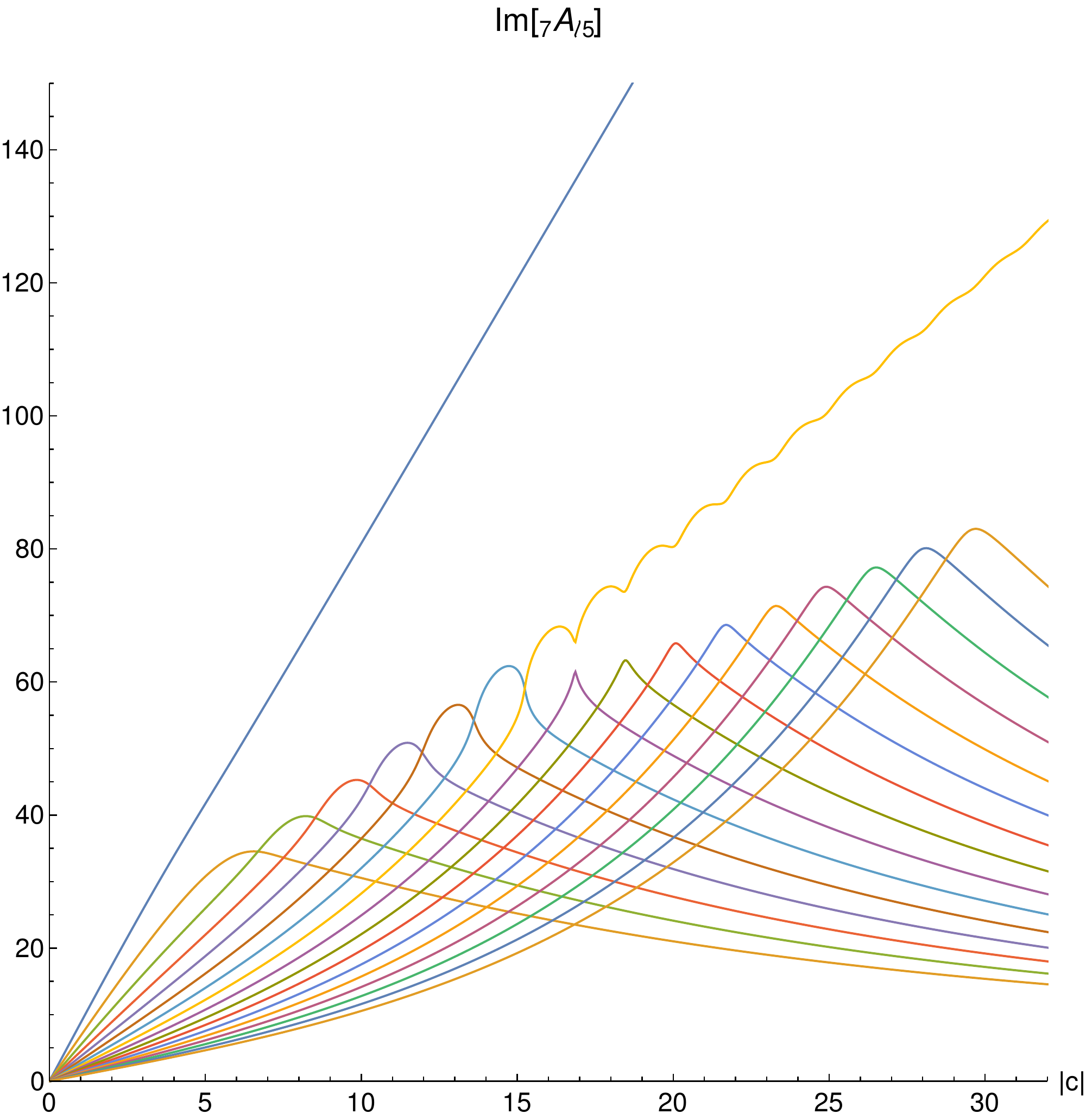}
\end{tabular}
\caption{Eigenvalue sequences for $\scA{7}{\ell5}{-i|c|}$. Note the unusual behavior of the $L=0$ and $L=7$ sequences which have quadratic leading-order behavior for the real components and linear leading-order behavior for the imaginary components.  Like Fig.~\ref{fig:ReIm4x4}, this set of sequences has two anomalous sequences with the first at $L=0$.  In this case, however, the second anomalous sequence appears sooner and we can zoom in on a smaller region on the plots.}
\label{fig:ReIm5x7}
\end{figure}

\begin{figure}
\begin{tabular}{cc}
\includegraphics[width=0.5\linewidth,clip]{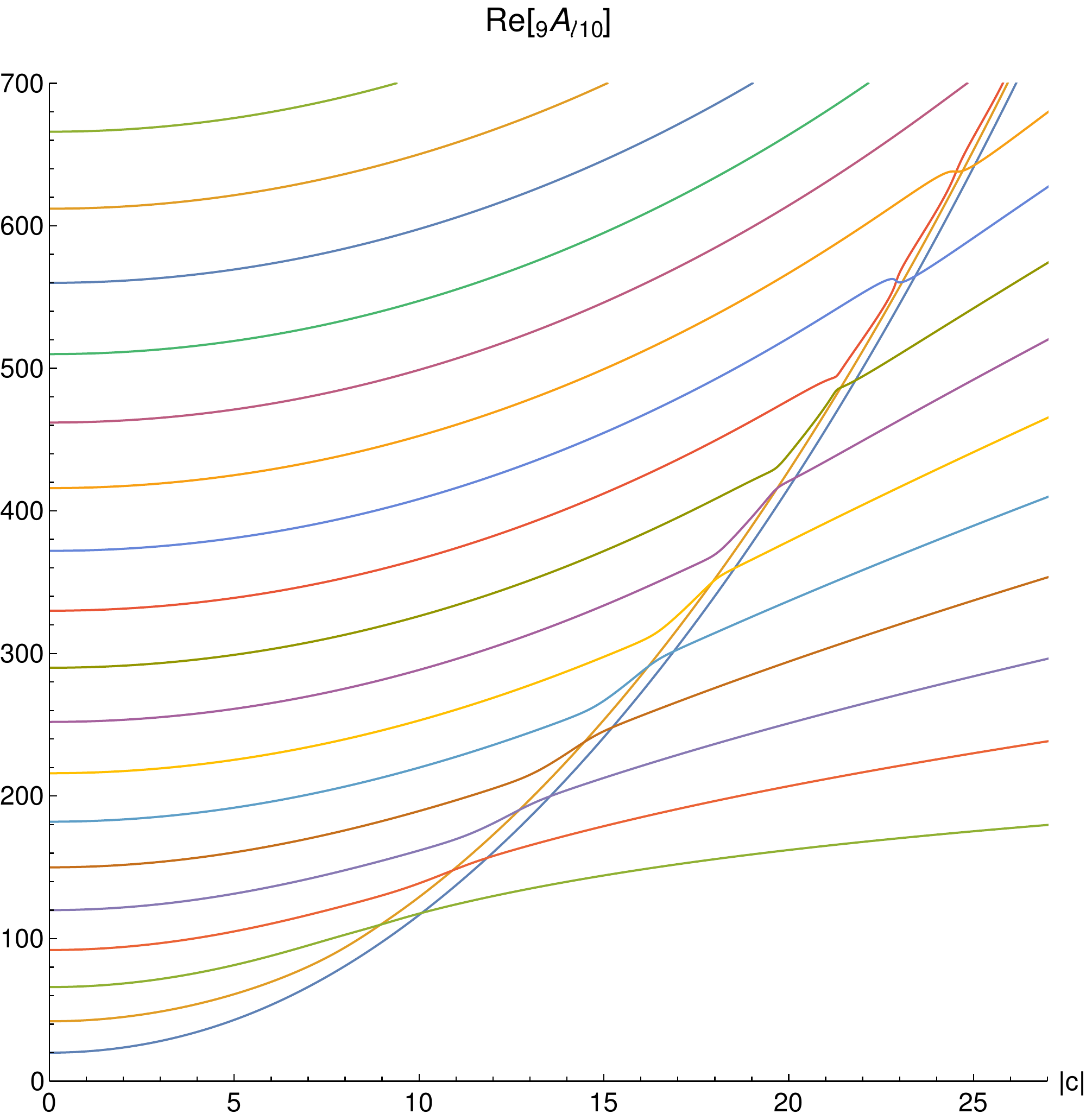} &
\includegraphics[width=0.5\linewidth,clip]{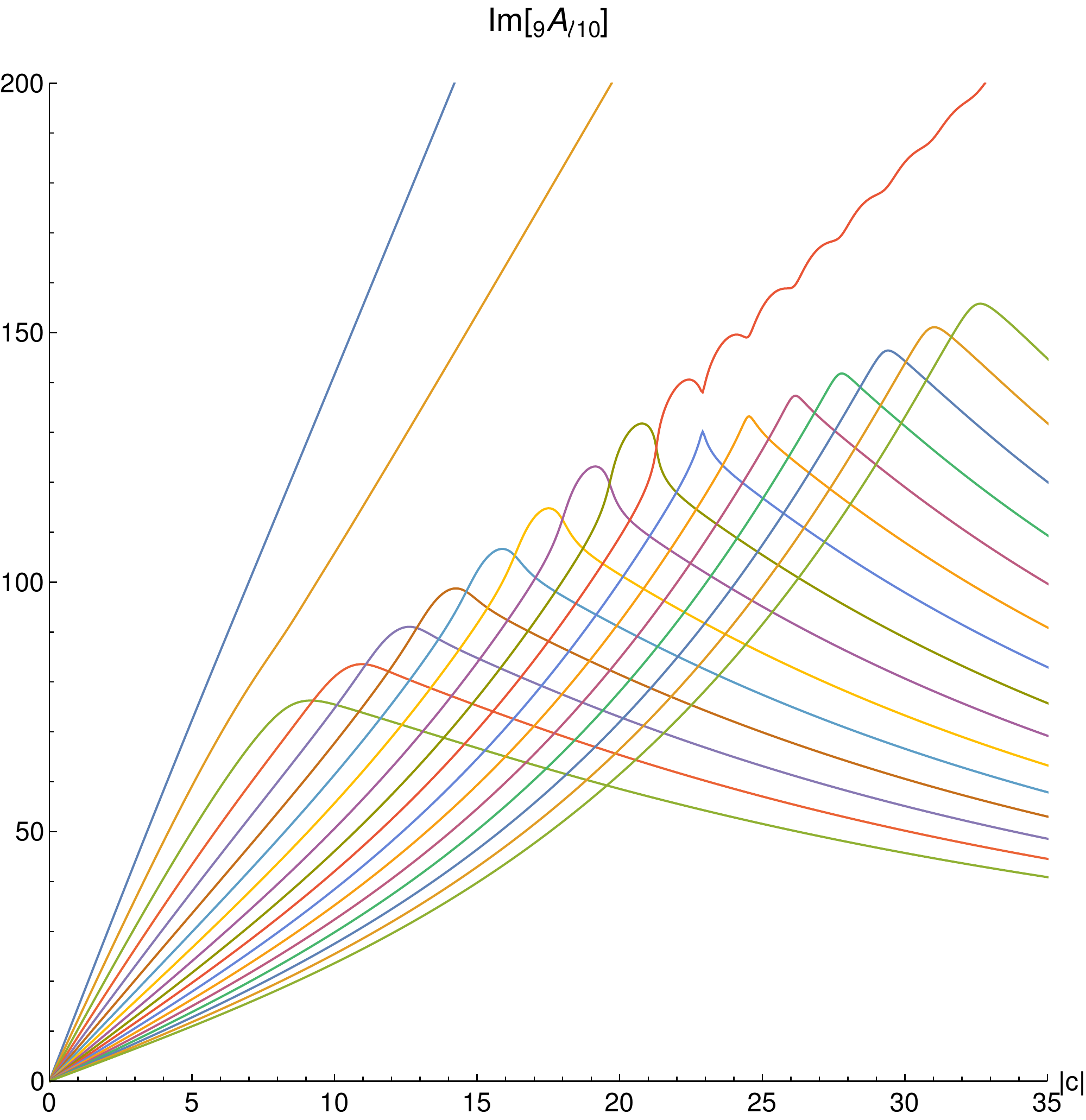}
\end{tabular}
\caption{Eigenvalue sequences for $\scA{9}{\ell10}{-i|c|}$. Note the unusual behavior of the $L=0$, $L=1$, and $L=10$ sequences which have quadratic leading-order behavior for the real components and linear leading-order behavior for the imaginary components.  This case is similar to  Fig.~\ref{fig:ReIm5x7}, but include 3 anomalous sequences.}
\label{fig:ReIm10x9}
\end{figure}

\begin{figure}
\begin{tabular}{cc}
\multirow[b]{2}{*}[125pt]{\includegraphics[width=0.525\linewidth,clip]{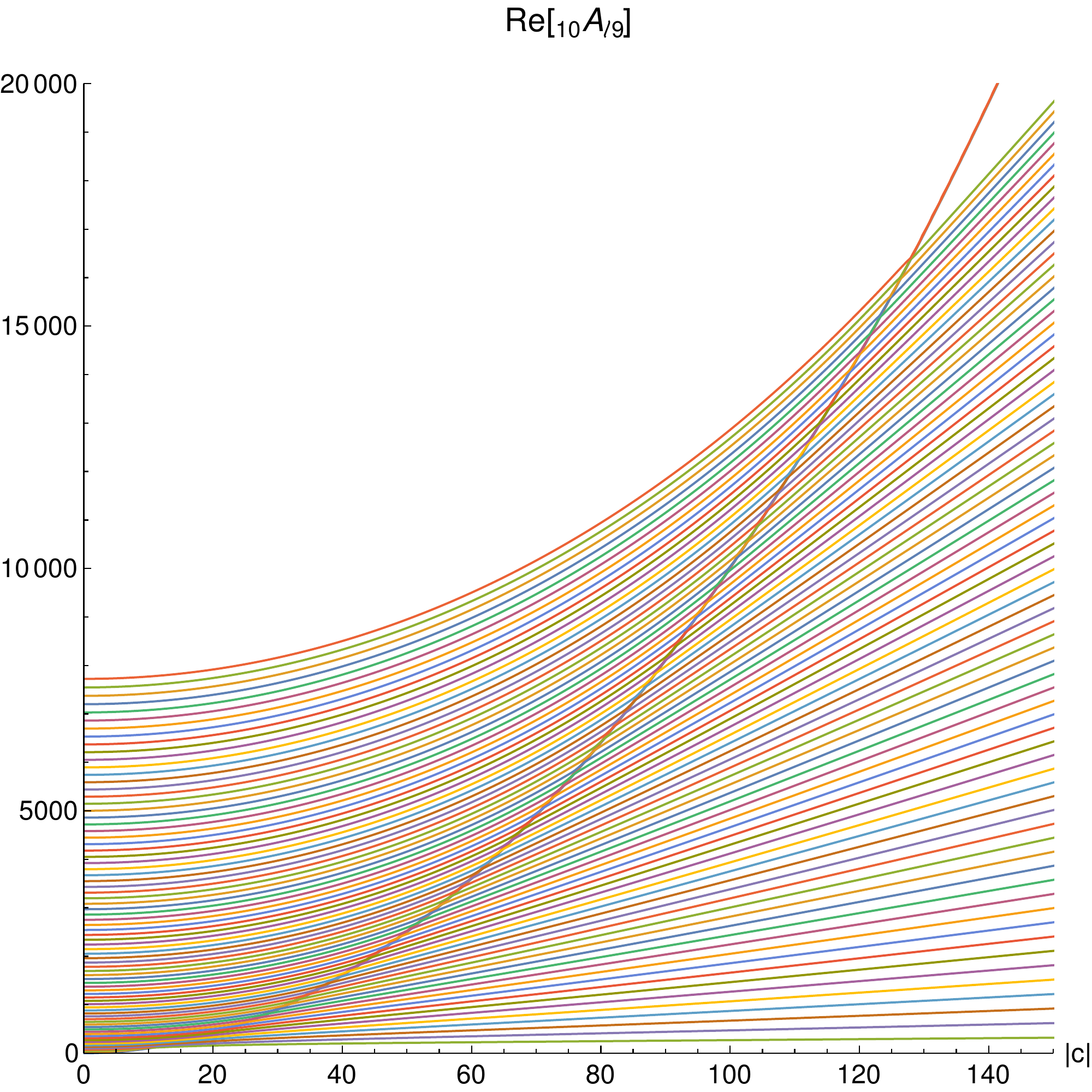}} &
\includegraphics[width=0.475\linewidth,clip]{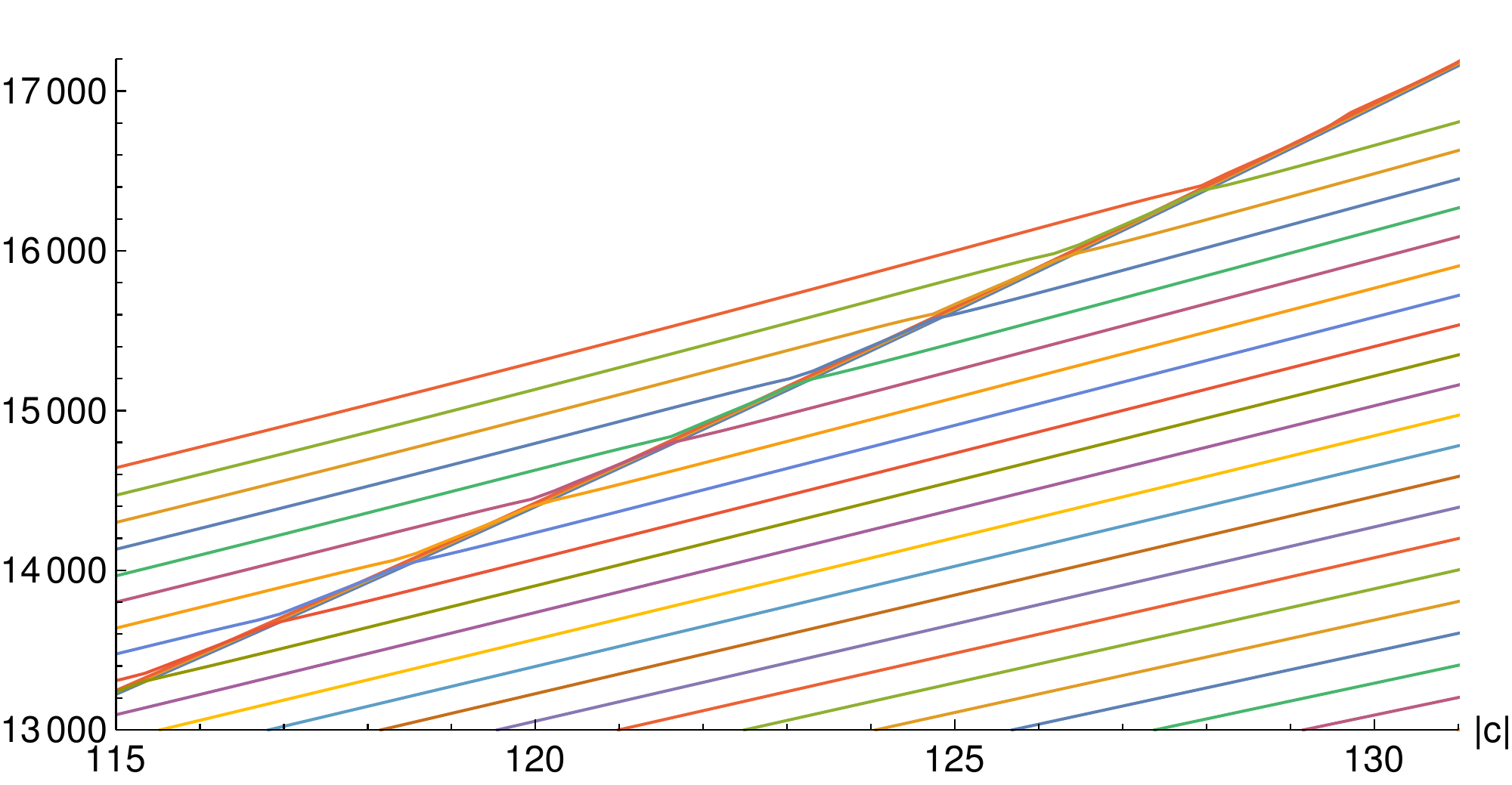} \\
&\includegraphics[width=0.475\linewidth,clip]{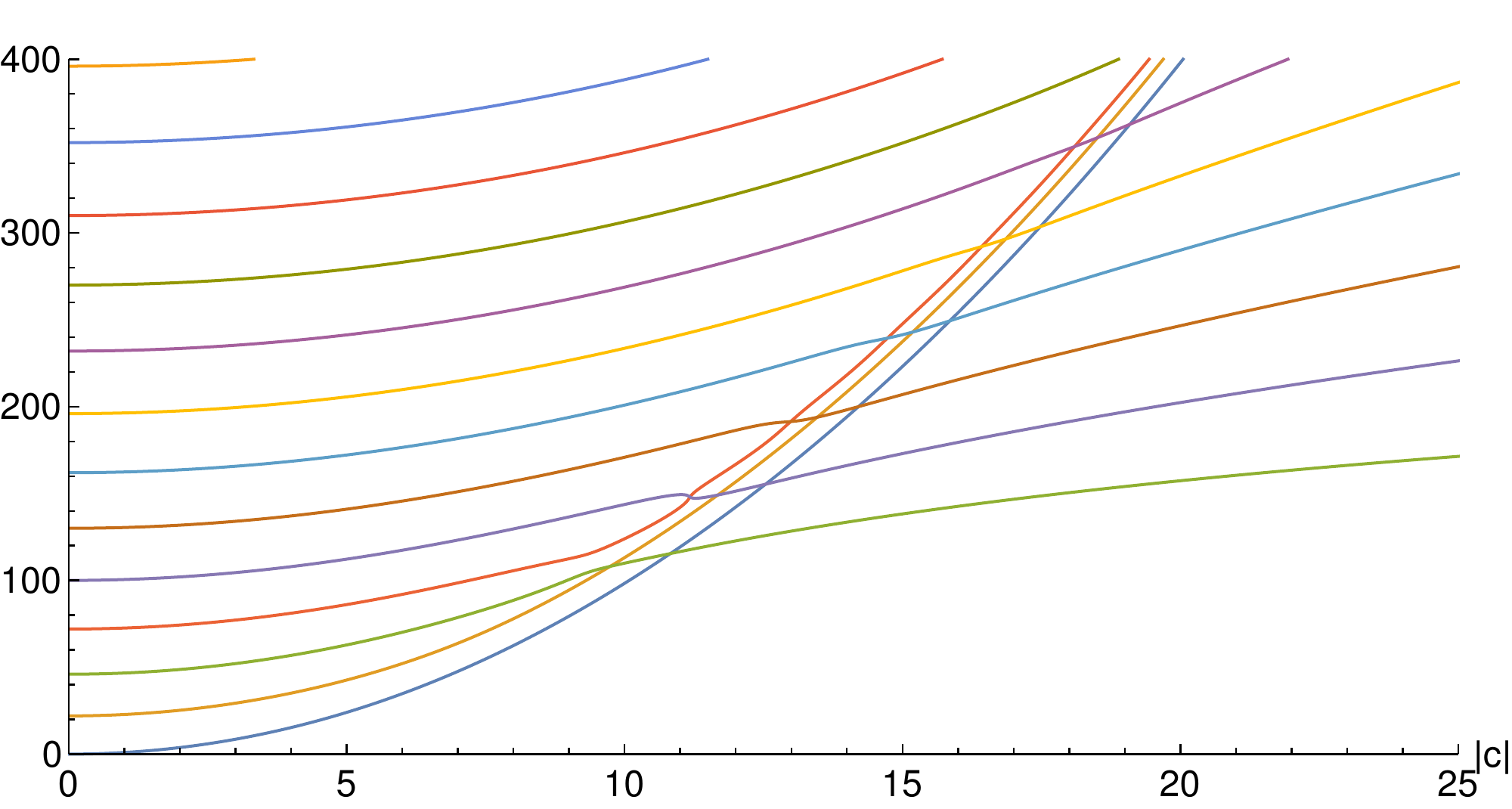}
\end{tabular}
\caption{Eigenvalue sequences for the real part of $\scA{10}{\ell9}{-i|c|}$. The behavior of the imaginary part can be seen in Fig.~\ref{fig:ImA9x10anom}.  Note the unusual behavior of the $L=0$, $L=1$, $L=3$, and $L=78$ sequences which have quadratic leading-order behavior for the real components.  The left plot includes the sequences with $0\le L\le78$, so the upper-most sequence is anomalous.  The upper-right plot focuses on the region of the plot were the $L=28$ anomalous sequence transitions to anomalous prolate asymptotic behavior.  The lower-right plot focuses on the behavior of the first $3$ anomalous sequences.}
\label{fig:Re9x10}
\end{figure}

\begin{figure}
\begin{tabular}{cc}
\includegraphics[width=0.5\linewidth,clip]{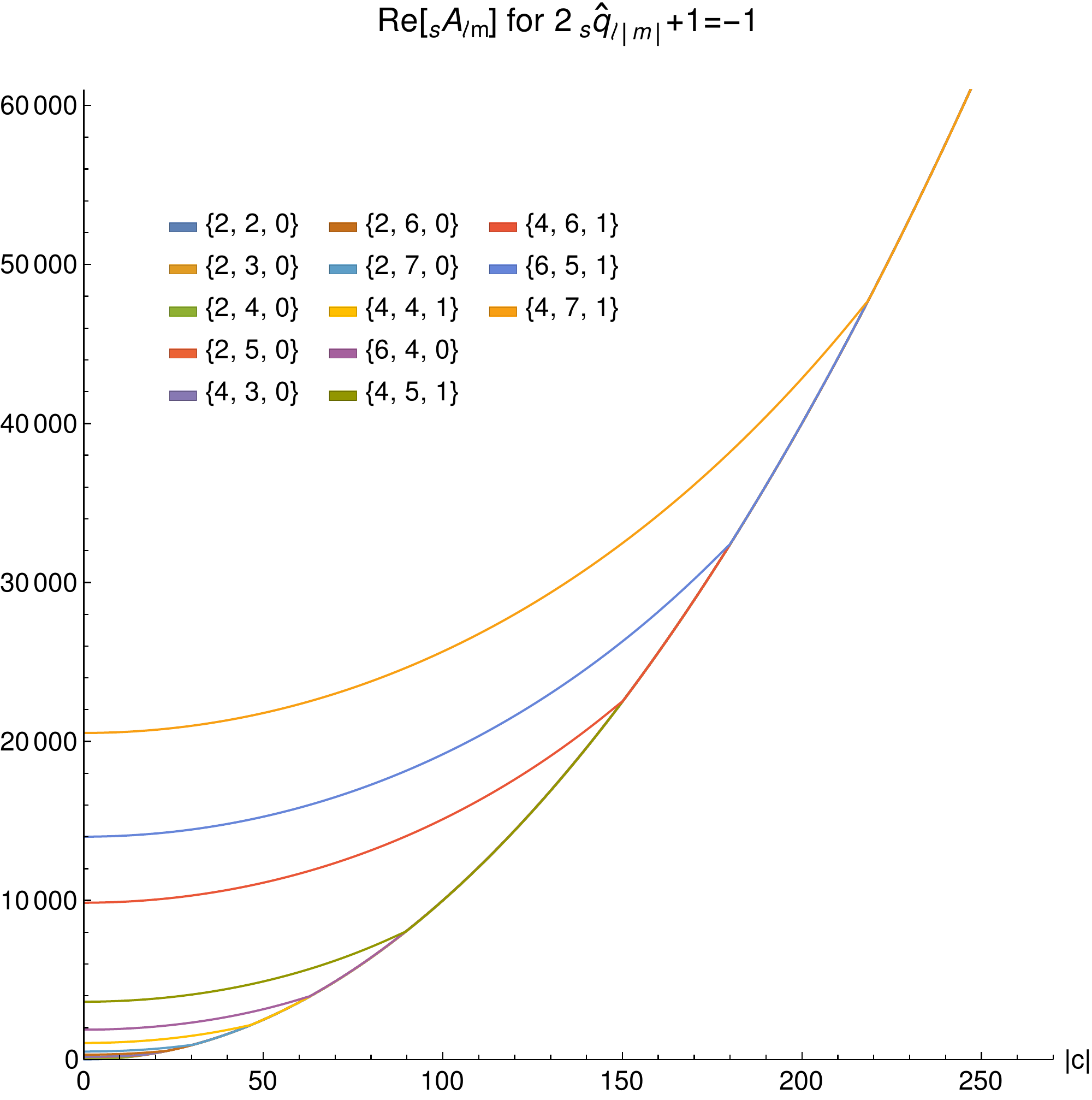} &
\includegraphics[width=0.5\linewidth,clip]{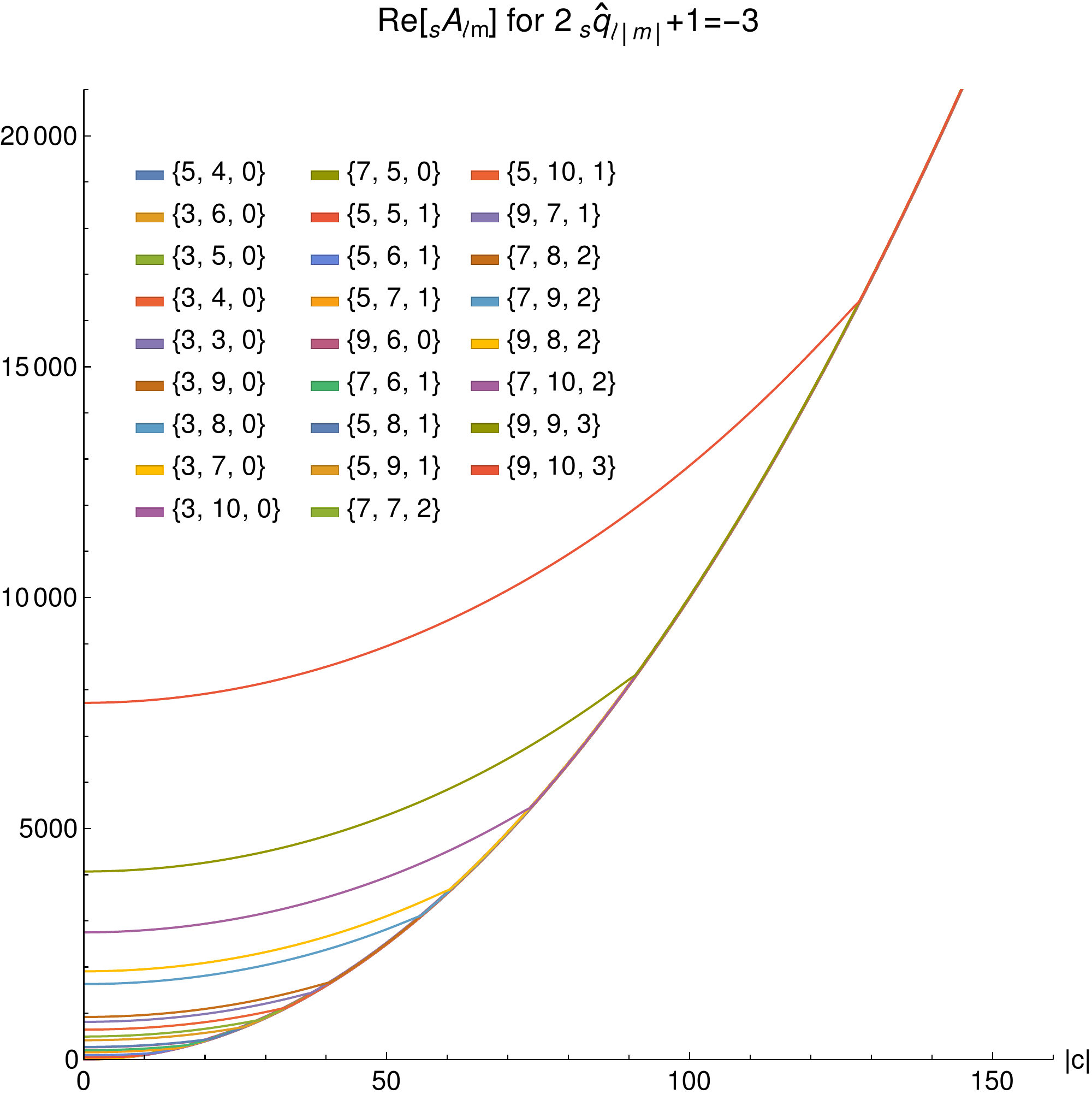} \\
\includegraphics[width=0.5\linewidth,clip]{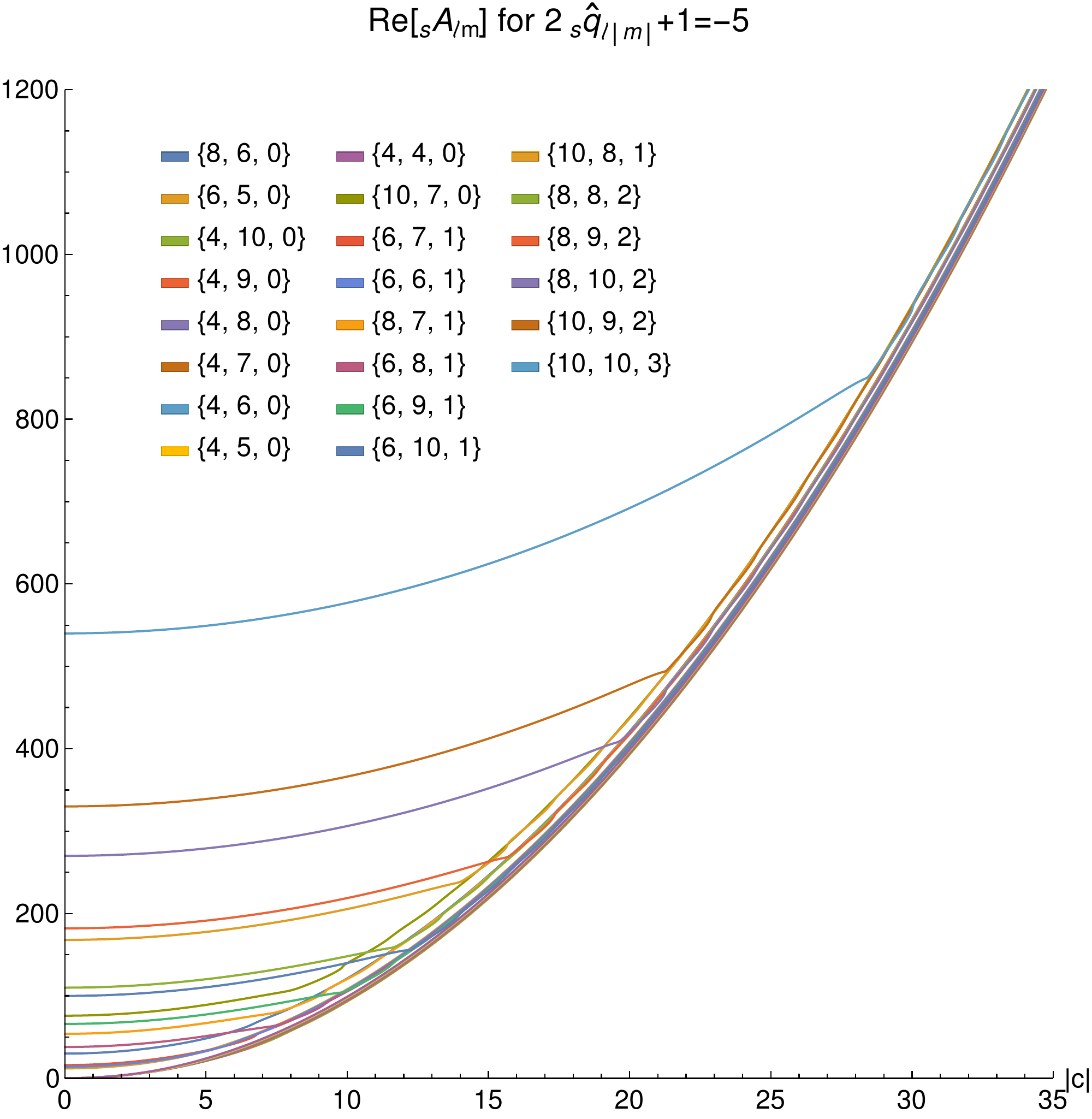} &
\includegraphics[width=0.5\linewidth,clip]{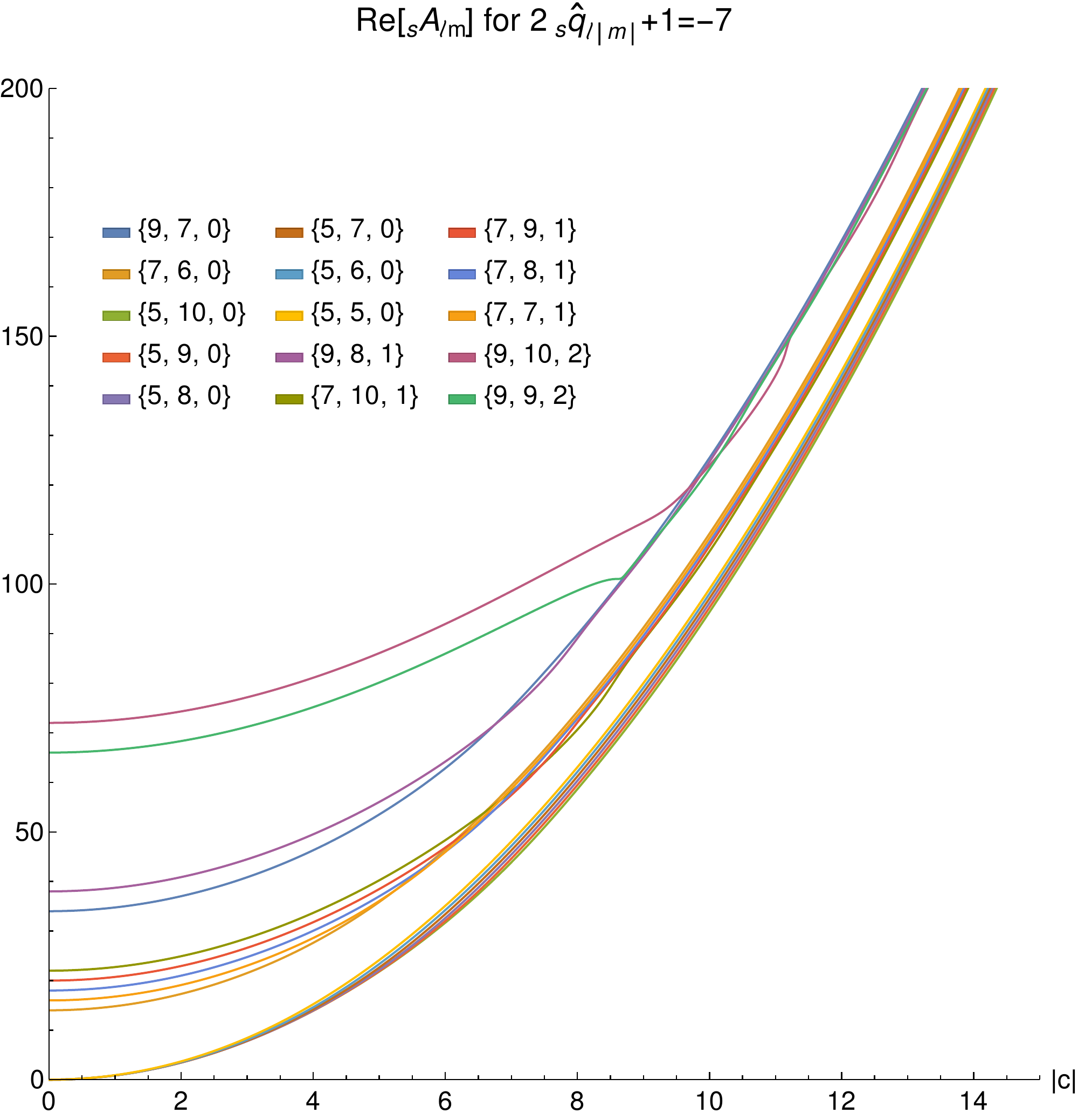}
\end{tabular}
\caption{Eigenvalue sequences for the real part of $\scA{s}{\ell m}{-i|c|}$ showing only the real parts of anomalous sequences grouped by anomalous type.  Each plot includes all sequences of a given type found with $0\le m\le10$ and $0\le s\le10$.  The upper-left plot displays the Type-1 anomalous sequences.  Note that several additional sequences may exist, but may occur for such large values of $L$ that we have not found them.  The upper-right plot displays the Type-3 anomalous sequences.  The lower-left and -right plots show, respectively the plots for Type-5 and Type-7 anomalous sequences.}
\label{fig:TypeNAnomRe}
\end{figure}

\begin{figure}
\begin{tabular}{cc}
\includegraphics[width=0.5\linewidth,clip]{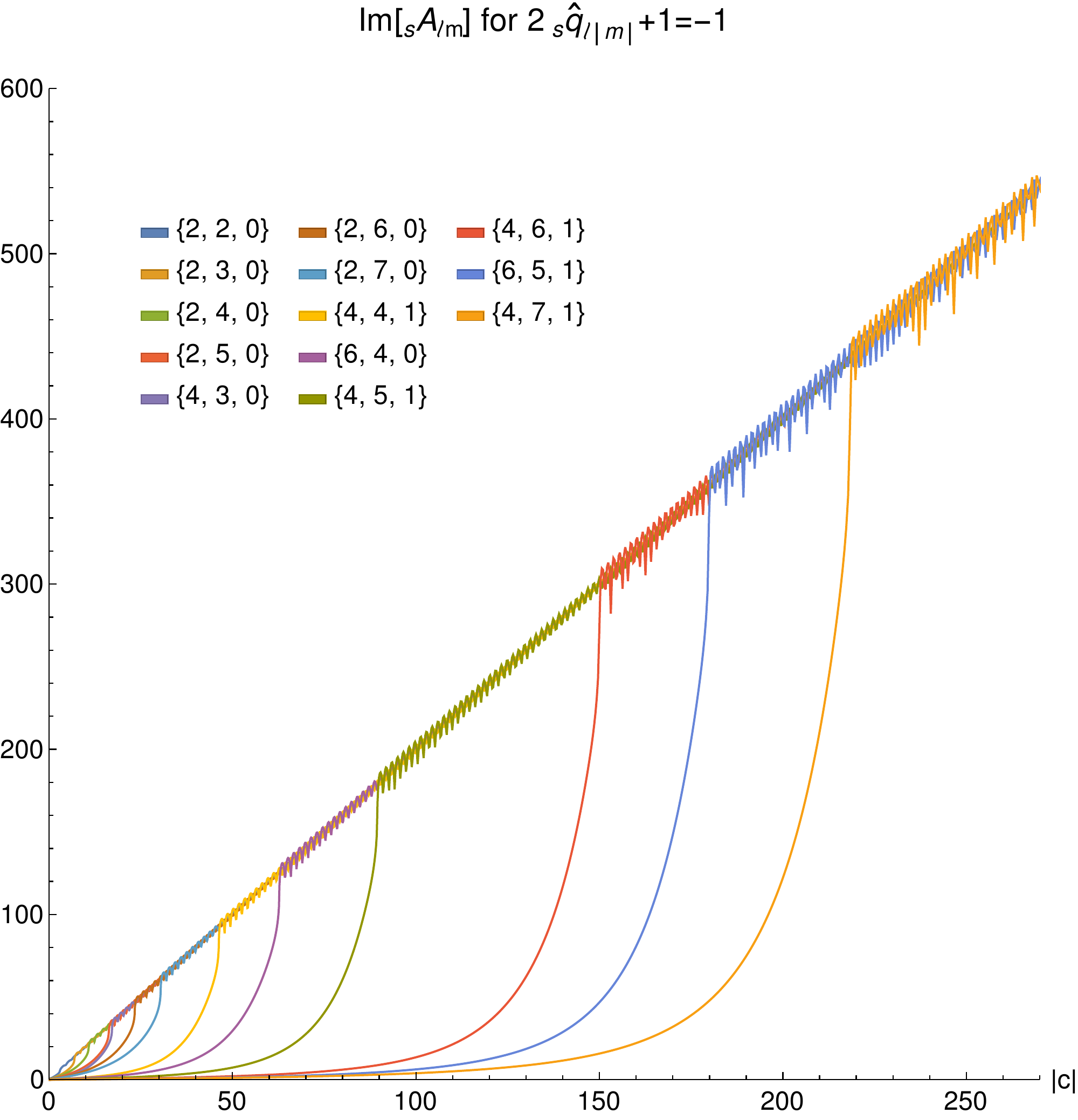} &
\includegraphics[width=0.5\linewidth,clip]{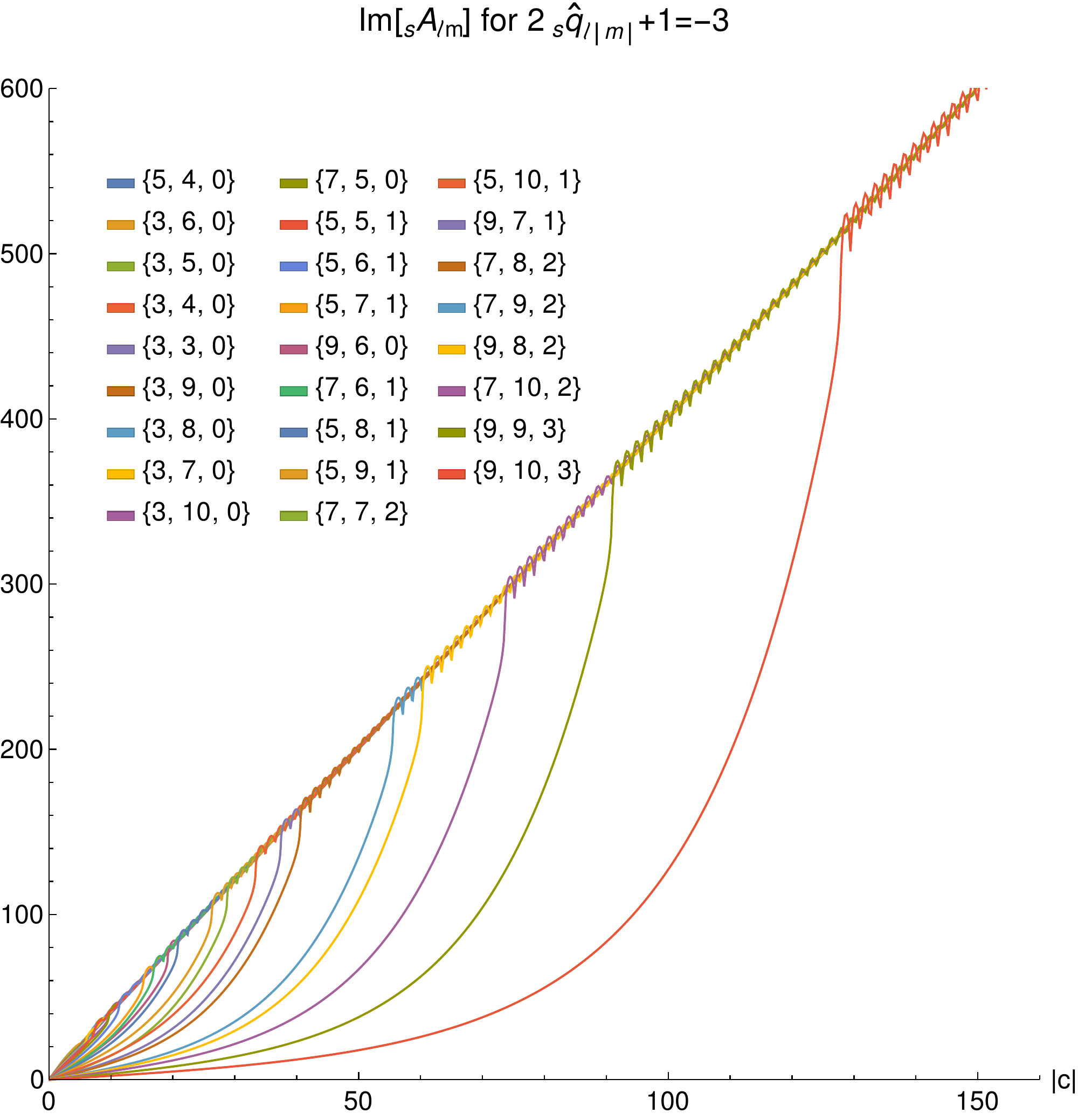} \\
\includegraphics[width=0.5\linewidth,clip]{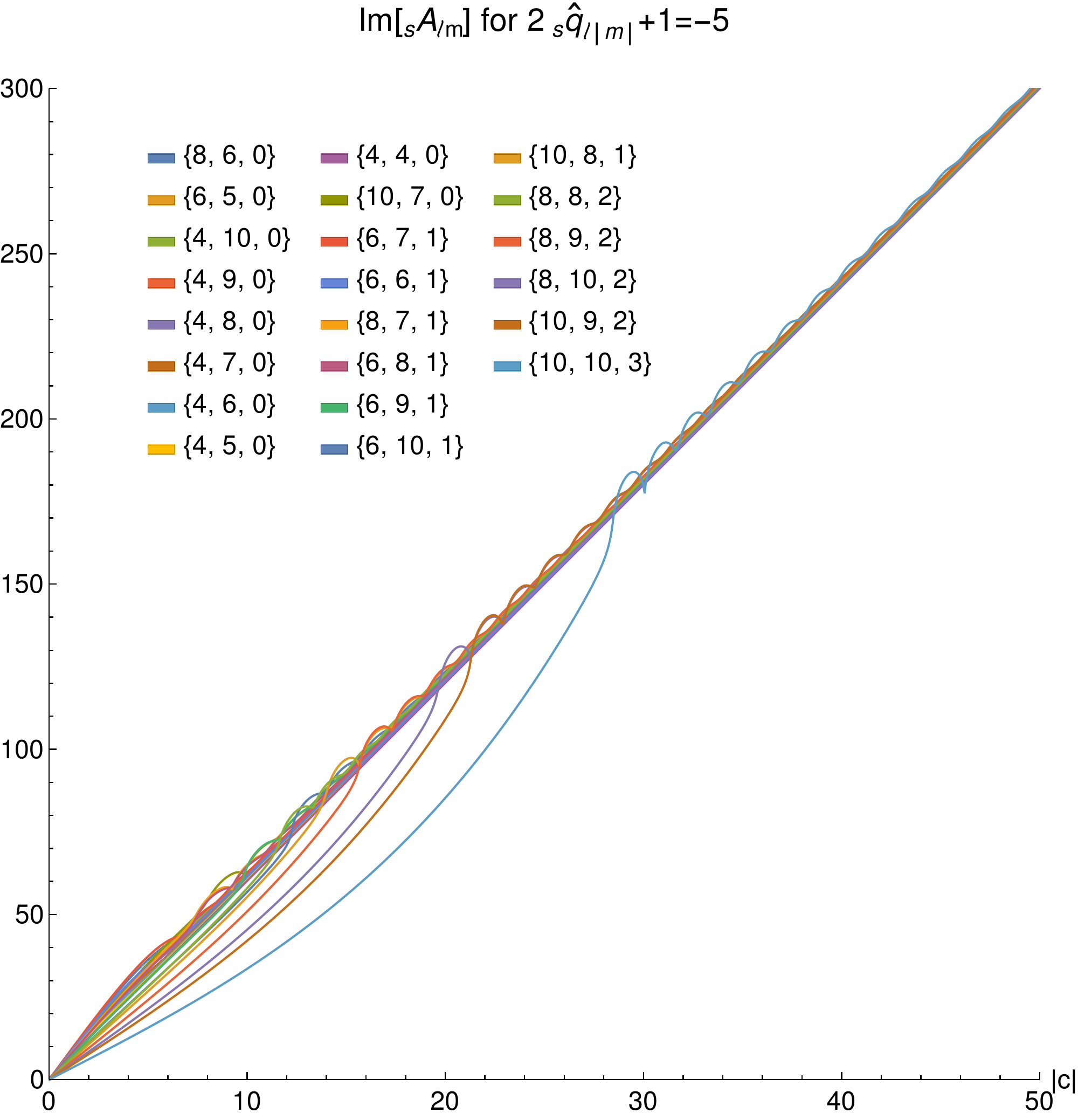} &
\includegraphics[width=0.5\linewidth,clip]{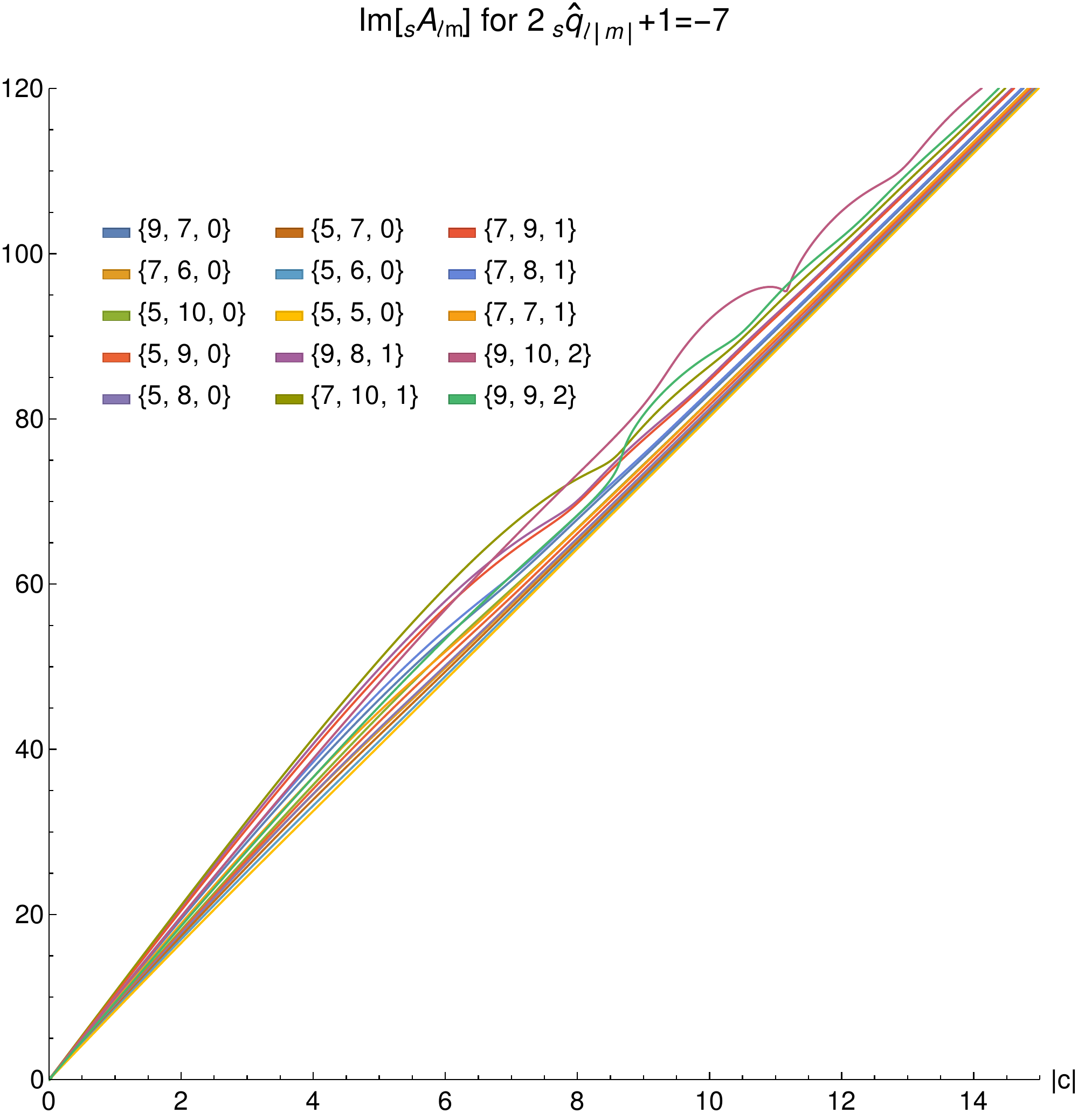}
\end{tabular}
\caption{Eigenvalue sequences for the imaginary part of $\scA{s}{\ell m}{-i|c|}$ showing only the imaginary parts of anomalous sequences grouped by anomalous type.  Each plot includes all sequences of a given type found with $0\le m\le10$ and $0\le s\le10$.  The upper-left plot displays the Type-1 anomalous sequences.  Note that several additional sequences may exist, but may occur for such large values of $L$ that we have not found them.  The upper-right plot displays the Type-3 anomalous sequences.  The lower-left and -right plots show, respectively the plots for Type-5 and Type-7 anomalous sequences.}
\label{fig:TypeNAnomIm}
\end{figure}
\end{widetext}

\end{document}